\documentclass[aps,pra,floatfix,notitlepage,groupedaddress,10pt]{revtex4-1}
\usepackage{natbib}
\usepackage{amsmath}
\usepackage{amssymb}
\usepackage{amsfonts}
\usepackage{graphicx}
\usepackage{dcolumn}
\usepackage{bm}
\usepackage{multirow}
\usepackage{array}
\newcolumntype{P}[1]{>{\centering\arraybackslash}p{#1}}
\usepackage{hyperref}
\usepackage{slashed}
\usepackage{float}
\floatstyle{plaintop}
\restylefloat{table}
\usepackage{caption}
\usepackage{ragged2e}
\allowdisplaybreaks
\graphicspath{{/home/mmh10d/Dropbox/Computational/Latex/FiguresPdf/}}

\begin{document}

\title{$\Lambda_c$ Semileptonic Decays in a Quark Model}

\author{Md Mozammel Hussain}
\email[]{mozammelhussain@gmail.com}
\author{Winston Roberts}
 \email[]{wroberts@fsu.edu}
\affiliation{Department of Physics,
Florida State University, Tallahassee, 32306}
\begin{abstract}
Hadronic form factors for semileptonic decay of the $\Lambda_c$ are calculated in a nonrelativistic quark 
model. The full quark model wave functions are employed to numerically calculate the form factors to all relevant orders in ($1/m_c$, $1/m_s$). The form factors obtained satisfy relationships expected from the heavy quark effective theory (HQET). The differential decay rates and branching fractions are calculated for transitions to the ground state and a number of excited states of $\Lambda$. The branching fraction of the semileptonic decay width to the total width of $\Lambda_c$ has been calculated and compared with other theoretical estimates and experimental results. The branching fractions for
$\Lambda_c\to\Lambda^{*} l^{+}\nu_l\to\Sigma\pi l^{+}\nu_l$ and $\Lambda_c\to\Lambda^{*} l^{+}\nu_l\to N\bar{K} l^{+}\nu_l$ are also calculated. Apart from decays to the ground state $\Lambda$(1115), it is found that decays through the $\Lambda(1405)$ provide a significant portion of the branching fraction
$\Lambda_c\to X_s l\nu_l$. A new estimate for $f=B(\Lambda_c^{+}\to\Lambda l^{+}\nu_l)/B(\Lambda_c^{+}\to X_s l^{+}\nu_l)$ is obtained.
\end{abstract}
\maketitle
\nopagebreak
\section{Introduction and Motivation} 

Semileptonic decays of hadrons are the main sources for precise 
knowledge on Cabibo-Kobayashi-Maskawa (CKM) matrix elements \cite{glashow}. The form factors that parametrize 
the non-perturbative QCD effects in these transitions play a crucial role in the extraction of CKM matrix elements and 
the precision depends on how well the form factors are calculated. 

A great deal of work has been done on semileptonic decay processes to calculate and improve the modeling of the form factors.
For example,  monopole type form factors were used to study semileptonic decay of heavy mesons by Wirbel, Stech and Bauer \cite{wb}.
Isgur, Scora, Grinstein and Wise caculated the semileptonic $B$ and $D$ meson decays in a non-relativistic quark model \citep{isgw}. 
Lattice QCD calculations of semileptonic decay form factors have been done in ref \citep{okamoto}. 
These are a very few out of a huge number of articles. More work has been done on semileptonic meson decays than baryon decays. 
Pervin, Roberts and Capstick worked on semileptonic baryon  decays of $\Lambda_Q$ \citep{PRC} and $\Omega_Q$ \citep{PRC1}) in a constituent quark model.
Some baryon decays have also been addressed in QCD sum rules \citep{Huang}, perturbative lattice QCD \citep{UKQCD} and a number of other approaches \citep{Migliozzi}.

The description of the weak decays of heavy hadrons are somewhat simplified because of the so-called heavy quark symmetry.
This was first pointed out by Isgur and Wise \citep{IW1}. 
Hadrons containing one heavy quark $Q$ (with $m_Q>>\Lambda_{\text{QCD}}$) possess this symmetry, which has been formalized into the heavy quark effective theory (HQET).
In HQET the properties of the hadrons are governed by the light degrees of freedom and are independent of the heavy quark degrees of freedom.
For semileptonic decays of heavy hadrons, HQET reduces the number of independent form factors needed to describe the decays.

In this paper, we examine the semileptonic decays of the $\Lambda_c^{+}$ to a number of $\Lambda$s, including the ground state. Because it is the lightest charmed baryon, $\Lambda_c^{+}$ plays an important role in understanding charm and bottom baryons.
The lowest-lying bottom baryon is most often detected through its weak decay to $\Lambda_c^{+}$.
In addition, the study of all of the $\Lambda_c^{+}$-type and $\Sigma_c$-type baryons are directly linked to the understanding of the ground state of $\Lambda_c^{+}$, as these baryons eventually decay into a $\Lambda_c^{+}$. 

Among the  branching fractions of the $\Lambda_c$, $\mathcal{B}(\Lambda_c^{+}\to pK^{-}\pi^{+})$ is used to normalize most of its other branching fractions.
The Particle Data Group (PDG), in their previous version \citep{PDG} reported that there was no model independent measurement of $\mathcal{B}(\Lambda_c^{+}\to pK^{-}\pi^{+})$. 
Two model-dependent measurements were reported, with two different results obtained from different assumptions. 
The model that calculated branching fractions $\mathcal{B}(\Lambda_c^{+}\to pK^{-}\pi^{+})$ from semileptonic decays, estimated that
\begin{equation}
  \mathcal{B}(\Lambda_{c}^{+}\to pK^{-}\pi^{+})=RfF 
  \frac{B(D\to Xl^{+}\nu_l)}{1+|\frac{V_{cd}}{V_{cs}}|^2}\tau(\Lambda_c^{+}),
\end{equation}
\noindent
where, 
\begin{eqnarray}
R&=&B(\Lambda_{c}^{+}\to pK^{-}\pi^{+})/B(\Lambda_c^{+}\to\Lambda l^{+}\nu_l),\nonumber\\
f&=&B(\Lambda_c^{+}\to\Lambda l^{+}\nu_l)/B(\Lambda_c^{+}\to X_s l^{+}\nu_l),\nonumber\\
F&=&B(\Lambda_c^{+}\to X_s l^{+}\nu_l)/B(D\to X_sl^{+}\nu_l).\nonumber
\end{eqnarray}
They estimated  $  B(\Lambda_{c}^{+}\to pK^{-}\pi^{+}) = (7.3\pm 1.4)\%$ with the theoretical estimate of $f=F=1.0$ with significant uncertainties.
 
However, in their most recent release, PDG \citep{PDG16} reports a model independent measurement of $\mathcal{B}(\Lambda_c^{+}\to pK^{-}\pi^{+})$.
A. Zupanc {\it et al.} (Belle Collaboration) \citep{ZUPANC} measured it to be $6.84{}^{+0.32}_{-0.40}\%$, while M. Ablikim {\it et al.} (BESIII Collaboration) \citep{BESIII2} measured it to be $5.84\pm 0.27\pm 0.23\%$.
The PDG fit is $6.35\pm 0.33\%$ that leads to a new estimate of
\begin{equation}
f=B(\Lambda_c^{+}\to\Lambda l^{+}\nu_l)/B(\Lambda_c^{+}\to X_s l^{+}\nu_l)=0.87{}^{+0.13}_{-0.17},\nonumber
\end{equation}
with the assumption of $F=1.0$. 
Pervin, Roberts and Capstick (PRCI) \citep{PRC} estimated the value of $f$ to be $0.85\pm 0.04$.
Mott and Roberts \citep{Mott} later estimated the rare decay branching fractions of the $\Lambda_b$ using two different methods. 
Their results indicated that the results were sensitive to the precision with which the form factors were estimated, and this further implied that $f$ could be even smaller than $0.85$.
The semileptonic branching fraction, $\mathcal{B}(\Lambda_c^{+}\to\Lambda l^{+} \nu_l)$ is reported to be $2.8\pm 0.5\%$ 
with the assumption that the $\Lambda_c^{+}$ decays only to the  ground state $\Lambda(1115)$.  
No semileptonic decays to excited $\Lambda$ have been reported.
This provides the motivation for our work.  

There have been a number of theoretical articles on the semileptonic decay of $\Lambda_c^{+}$ in recent years. 
Gutsche {\it et al.} used a covariant quark model to estimate the branching fraction for $\Lambda_c\to\Lambda l^{+} \nu_l$ \citep{CQM}. Liu {\it et al.} used QCD light cone sum rules to examine this decay \citep{LCSR}, while Ikeno and Oset have examined the semileptonic decay to the $\Lambda(1405)$, treating that state as a dynamically generated molecular state \citep{Oset}.

In the work presented herein, we work in the framework of a constituent quark model. 
Such models have been quite successful in explaining the main features of hadron phenomenology. In computing the form factors for $\Lambda_c\to\Lambda^{*}$, 
we have deployed two approximations. In the first approximation, single component wave functions are used to 
compute the analytic form factors for $\Lambda_c\to\Lambda^{*}$ transitions. As in PRCI \citep{PRC} a variational 
diagonalization of a quark model Hamiltonian was used to extract the single component wave functions and the 
quark operators were reduced to their non-relativistic Pauli form.
In the second method we keep the full relativistic form of 
the quark spinors and use the full quark model wave functions. We believe that this second method provides more reliable numerical 
values of the form factors as it uses fewer approximations. 

We calculate the decay widths and branching fractions for decays to 
ground state and a number of excited $\Lambda^{(*)}$. We also study the decay widths and branching fractions of
two other decay channels, namely $\Lambda_c^{+}\to \Sigma\pi l^{+}\nu_l$
and $\Lambda_c^{+}\to N\bar{K} l^{+}\nu_l$, via a set of $\Lambda$ resonances.

The rest of this paper is organized as follows: in section II, we discuss the hadronic matrix elements and
decay rates. Section III presents a concise overview of HQET and the relationships predicted by HQET among the form factors for the transitions we study. 
In section IV we describe the model we employ to calculate the form factors. Section V is devoted to discussing
the numerical results such as form factors, decay rates and branching fractions. 
Section VI presents our conclusions and outlook. A number of details of the calculation are shown in the appendices.
 
\section{Matrix Elements and Decay Widths}

\subsection{Semileptonic decay \texorpdfstring{($\Lambda_c^{+}\to\Lambda l^{+} \nu_l $)}{}}

\subsubsection{Matrix Elements}\label{semime}

\begin{figure}
\caption{Semileptonic decay $\Lambda_c^{+}\to\Lambda^{(*)} l^{+} \nu_l$}
\label{semilamdec}
\includegraphics[width = 0.4\textwidth]{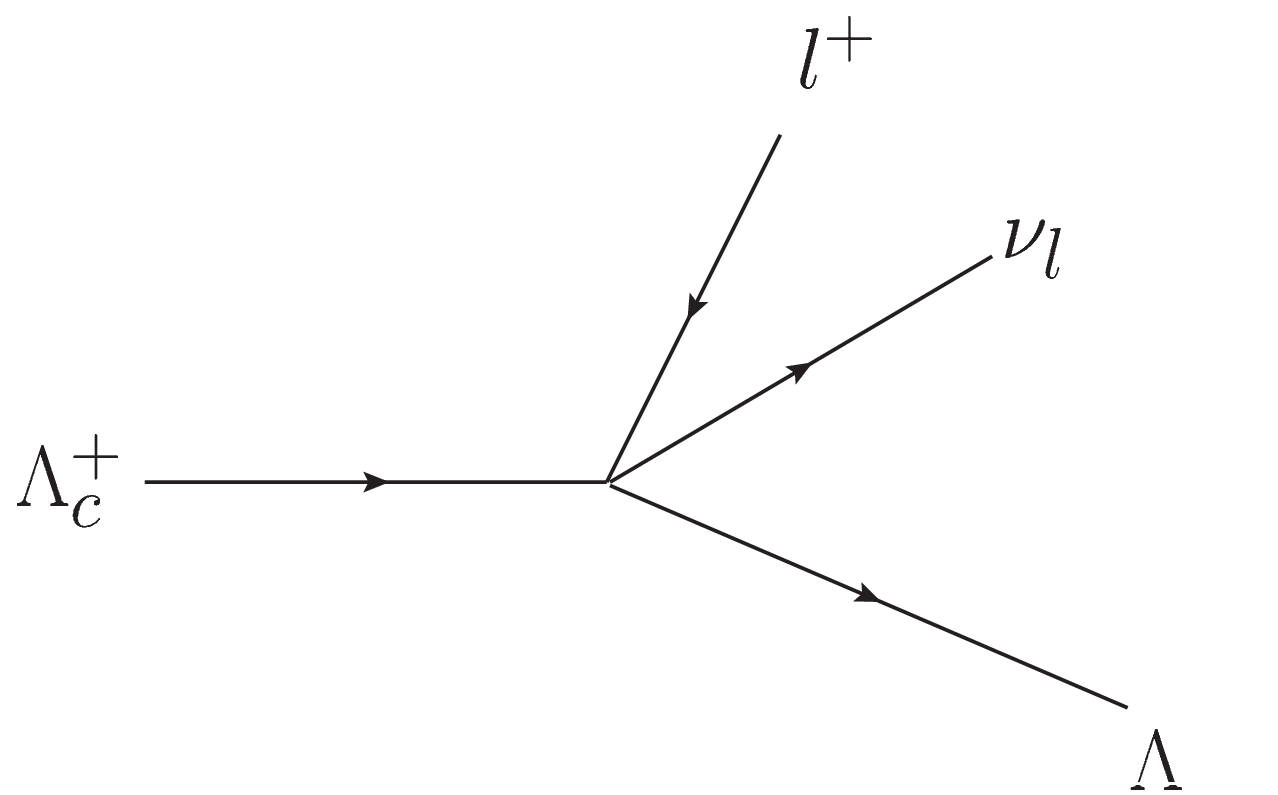}
\end{figure}

\noindent
Fig.\ref{semilamdec} depicts the semileptonic decay 
$\Lambda_c^{+}\to\Lambda^{(*)} l^{+} \nu_l$. We work in the rest frame of the parent $\Lambda_c$. 
The transition matrix element for the decay  is
\begin{equation}
 \mathcal{M}=\frac{G_F}{\sqrt{2}}V_{cs}L^\mu \langle\Lambda(p^\prime,s^\prime)|J_{\mu}|\Lambda_c(p,s)\rangle,
\end{equation}
\noindent
where $V_{cs}$ is the CKM matrix element, $L^\mu=\bar{u}_{\nu_l}\gamma^{\mu}(1-\gamma_5)v_l$ is the lepton current and 
$J_{\mu}=\bar{s}\gamma_{\mu}(1-\gamma_5)c$ is the hadronic current. The momenta of the $\Lambda_c$, $\Lambda$, $l$, $\nu_l$ are 
labeled as $p$, $p^{\prime}$, $p_l$ and $p_{\nu_l}$, respectively.
The hadronic matrix element is defined as
\begin{equation}
H_\mu=\langle\Lambda^{(*)}|J_\mu|\Lambda_c\rangle.
\end{equation}
\noindent
The hadronic matrix elements are parametrized in terms of a number of form factors. For transitions from the ground state 
$\Lambda_c$ ($J^P=\frac{1}{2}^{+}$) to the ground state $\Lambda$ ($J^P=\frac{1}{2}^{+}$), 
the matrix elements for the vector ($V_\mu$) and axial-vector ($A_\mu$) currents are, respectively,
\begin{align}\label{had.mat.el}
\langle\Lambda(p^{\prime},s^{\prime})|V_{\mu}|\Lambda_c(p,s)\rangle
& =\bar{u}(p^{\prime},s^\prime)
 \biggl(\gamma_{\mu}F_1+\frac{p_{\mu}}{m_{\Lambda_c}}F_2
 +\frac{p_{\mu}^{\prime}}{m_{\Lambda}}F_3\biggr)u(p,s),\\
 \langle\Lambda(p^{\prime},s^{\prime})|A_{\mu}|\Lambda_c(p,s)\rangle
 & =\bar{u}(p^{\prime},s^\prime)\biggl(\gamma_{\mu}G_1 + \frac{p_{\mu}}{m_{\Lambda_c}}G_2
 +\frac{p_{\mu}^{\prime}}{m_{\Lambda}}G_3\biggr)\gamma_5 u(p,s),
\end{align}
where the $F_i$'s and $G_i$'s are the form factors and $s(s^\prime)$ is the spin of $\Lambda_c(\Lambda)$. 
The matrix elements for transitions to a daughter baryon with $J^P = \frac{3}{2}^{-}$ are
\begin{align}
\langle\Lambda(p^\prime,s^\prime)|V_{\mu}|\Lambda_c(p,s)\rangle 
& =\bar{u}^{\alpha}(p^{\prime},s^\prime)
\biggl[\frac{p_{\alpha}}{m_{\Lambda_c}}\biggr(\gamma_{\mu}F_1
+\frac{p_{\mu}}{m_{\Lambda_c}}F_2
+\frac{p^{\prime}_{\mu}}{m_{\Lambda}}F_3\biggr)+g_{\alpha\mu}F_4\biggr]u(p,s),\\
\langle\Lambda(p^\prime,s^\prime)|A_{\mu}|\Lambda_c(p,s)\rangle 
& =\bar{u}^{\alpha}(p^{\prime},s^\prime)
\biggl[\frac{p_{\alpha}}{m_{\Lambda_c}} \biggl(\gamma_{\mu}G_1 + \frac{p_{\mu}}{m_{\Lambda_c}}G_2 +\frac{p_{\mu}^{\prime}}{m_{\Lambda}}G_3\biggr)+g_{\alpha\mu}G_4\biggr]\gamma_5u(p,s).
\end{align}
The Rarita-Schwinger spinor $\bar{u}^{\alpha}$ satisfies the conditions
 \begin{equation}
  p_{\alpha}^{\prime}\bar{u}^{\alpha}(p^{\prime},s^{\prime})=0,\,\,\,\,
  \bar{u}^{\alpha}(p^{\prime},s^{\prime})\gamma_{\alpha}=0,\,\,\,\,
   \bar{u}^{\alpha}(p^{\prime},s^{\prime})\slashed{p}^{\prime}=m_{\Lambda^{3/2}}\bar{u}^{\alpha}(p^{\prime},s^{\prime}).
 \end{equation}
The corresponding matrix elements for transitions to a daughter baryon with $J^p = \frac{5}{2}^{+}$ are
\begin{align*}
\langle\Lambda(p^\prime,s^\prime)|V_{\mu}|\Lambda_c(p,s)\rangle 
& =\bar{u}^{\alpha\beta}(p^{\prime},s^\prime)\frac{p_\alpha}{m_{\Lambda_c}}
\biggl[\frac{p_{\beta}}{m_{\Lambda_c}}\biggr(\gamma_{\mu}F_1
+\frac{p_{\mu}}{m_{\Lambda_c}}F_2
+\frac{p^{\prime}_{\mu}}{m_{\Lambda}}F_3\biggr)+g_{\beta\mu}F_4\biggr]u(p,s).\\
\langle\Lambda_c(p^\prime,s^\prime)|A_{\mu}|\Lambda_c(p,s)\rangle
& =\bar{u}^{\alpha\beta}(p^{\prime},s^\prime)
\frac{p_\alpha}{m_{\Lambda_c}}\biggl[\frac{p_{\beta}}{m_{\Lambda_c}} \biggl(\gamma_{\mu}G_1 + \frac{p_{\mu}}{m_{\Lambda_c}}G_2
 +\frac{p_{\mu}^{\prime}}{m_{\Lambda}}G_3\biggr)+g_{\beta\mu}G_4\biggr]\gamma_5u(p,s).
\end{align*}
\noindent
The spinor $\bar{u}^{\alpha\beta}$ satisfies the conditions

\begin{align*}
p_{\alpha}^{\prime}\bar{u}^{\alpha\beta}(p^{\prime},s^{\prime})=p_{\beta}^{\prime}
\bar{u}^{\alpha\beta}(p^{\prime},s^{\prime})=0,\,\,\,\,
\bar{u}^{\alpha\beta}(p^{\prime},s^{\prime})\gamma_{\alpha}&=\bar{u}^{\alpha\beta}(p^{\prime},s^{\prime})\gamma_{\beta}=0,\\
\bar{u}^{\alpha\beta}(p^{\prime},s^{\prime})\slashed{p}^{\prime}=m_{\Lambda^{5/2}}
\bar{u}^{\alpha\beta}(p^{\prime},s^{\prime}),\,\,\,\,
\bar{u}^{\alpha\beta}(p^{\prime},s^{\prime})g_{\alpha\beta}&=0.
\end{align*}
\noindent
Here we have shown the hadronic transition matrix elements for the decays to daughter baryons with natural parity.
For decays to states with unnatural parity, the matrix elements are constructed by switching $\gamma_5$ from the equations defining 
the $G_i$ to the equations defining the $F_i$.

\subsubsection{Decay Width}

\noindent
The differential decay rate for the transition $\Lambda_c\to\Lambda^{(*)}l^{+}\nu_l$ is
\begin{equation}
 d\Gamma=\frac{1}{2m_{\Lambda_c}}\overline{|\mathcal{M}|^2}\frac{d^3p_ld^3p_{\nu_l}d^3p^{\prime}}
 {2E_l2E_{\nu_l}2E^{\prime}}\frac{(2\pi)^4\delta^4(p-p^{\prime}-p_l-p_{\nu_l})}
 {(2\pi)^3(2\pi)^3(2\pi)^3},
\end{equation}
\noindent
where 
\begin{align}
\overline{|\mathcal{M}|^2} &=\frac{G_F^2}{2}|V_{cs}|^2\frac{1}{2}\sum_{\text{spins}}H_{\mu}^{\dagger}
H_{\nu}{L^{\mu}}^{\dagger}L^{\nu}\nonumber ,\\
& =\frac{G_F^2}{4}|V_{cs}|^2H_{\mu\nu}L^{\mu\nu}.
\end{align}
\noindent
$ \overline{|\mathcal{M}|^2}$ is the squared amplitude averaged over the initial spins (the factor of $\frac{1}{2}$) and summed over the final spins.

\noindent
The most general Lorentz form of the hadronic tensor can be written as
\begin{align}{\label{hadrontensor}}
H_{\mu\nu}&={\alpha} g_{\mu\nu} +\beta_{PP} P_{\mu}P_{\nu}
 +\beta_{PL}P_{\mu}L_{\nu} +\beta_{LP}L_{\mu}P_{\nu} +\beta_{LL}L_{\mu}L_{\nu}
 +i\gamma\epsilon_{\mu\nu\rho\sigma}P_{\rho}L_{\sigma},
\end{align}
\noindent
where we have defined $P =p^\prime$ and $L = p-p^{\prime}$. The lepton tensor is
\begin{equation}\label{leptontensor}
 L^{\mu\nu}=8[p_l^{\mu}p_{\nu_l}^{\nu}+p_{\nu_l}^{\mu}p_l^{\nu}-g^{\mu\nu}(p_l.p_{\nu_l})
 +i\epsilon^{\mu\nu\alpha\beta}p_{l\alpha}p_{{\nu_l}\beta}].
\end{equation}
\noindent
Integrating over the lepton momenta allows us to write the lepton tensor 
as
\begin{equation}\label{lepten2}
 \int \frac{d^3p_l d^3p_{\nu_l}}{(2\pi)^3 (2\pi)^3 2E_l 2E_{\nu_l}} L^{\mu\nu}= \int d\Omega_l (Ag^{\mu\nu}+A^{\prime} L^{\mu} L^{\nu}),
\end{equation}
where $q^2=(p-p^\prime)^2$ and
\begin{equation}
 A=-\frac{(q^2-m_l^2)^2(2q^2+m_l^2)}{384\pi^6q^4},\,\,\,\,A^{\prime}=\frac{(q^2-m_l^2)^2(q^2+2m_l^2)}{192\pi^6q^6}.
\end{equation}
\noindent
The complete expression for the differential decay rate becomes
\begin{align}
\frac{d\Gamma}{dq^2} =  \frac{|V_{cs}|^2}{192} 
\frac{ G_F^2}{\pi^3 m_{\Lambda_c}^3}\lambda^{1/2}(m_{\Lambda_c}^2,m_{\Lambda}^2,q^2)\frac{(q^2-m_l^2)^2}{4q^4}
&\bigg(-6\alpha q^2+ \beta_{PP}\Big[2q^2\big((P\cdot L)-m_{\Lambda}^2\big)+m_l^2\big(4(P\cdot L)-m_{\Lambda}^2\big)\Big]\\
&
+\Big[\beta_{LP}(P\cdot L) +\beta_{PL}(P\cdot L)+ \beta_{LL}q^2\Big]3m_l^2 \bigg),\nonumber 
\end{align}
where $P\cdot L = \frac{1}{2}(m_{\Lambda_c}^2 - m_{\Lambda}^2-q^2)$ and 
$\lambda^{1/2}(x,y,z) = (x^2+y^2+z^2-2xy-2yz-2zx)^{1/2}$. When contracted with the lepton tensor, all of the $\beta$s (except $\beta_{PP}$) are proportional
to powers of the lepton mass $m_l$ and thus give small contributions to the decay rate.
The complete form of $\beta_{PP}$ is given in appendix \ref{HadTen1}.

\subsection{\texorpdfstring{$\Lambda_c\to\Lambda^{*} l^{+} \nu_l \to \Sigma\pi l^{+} \nu_l/N\bar{K}l^{+}\nu_l$}{}}

\begin{figure}[t!]
\caption{(a) shows the semileptonic decay $\Lambda^{+}_c\to\Lambda^{*}l^{+}\nu_l$ followed by the strong decay $\Lambda\to\Sigma\pi$;
(b) shows the semileptonic decay $\Lambda^{+}_c\to\Lambda^{*}l^{+}\nu_l$ followed by the strong decay $\Lambda\to N\bar{K}$;
(c) shows the strong decay $\Lambda^{+}_c\to \Sigma_c^{*}\pi$  followed by the semileptonic decay  
$\Sigma^{*}_c\to\Sigma l^{+}\nu_l$; 
(d) shows the strong decay $\Lambda^{+}_c\to D^{*}N$  followed by the semileptonic decay $D^{*}\to \bar{K} l^{+}\nu_l$.}
\label{decaymodes}
\includegraphics[width=0.6\textwidth]{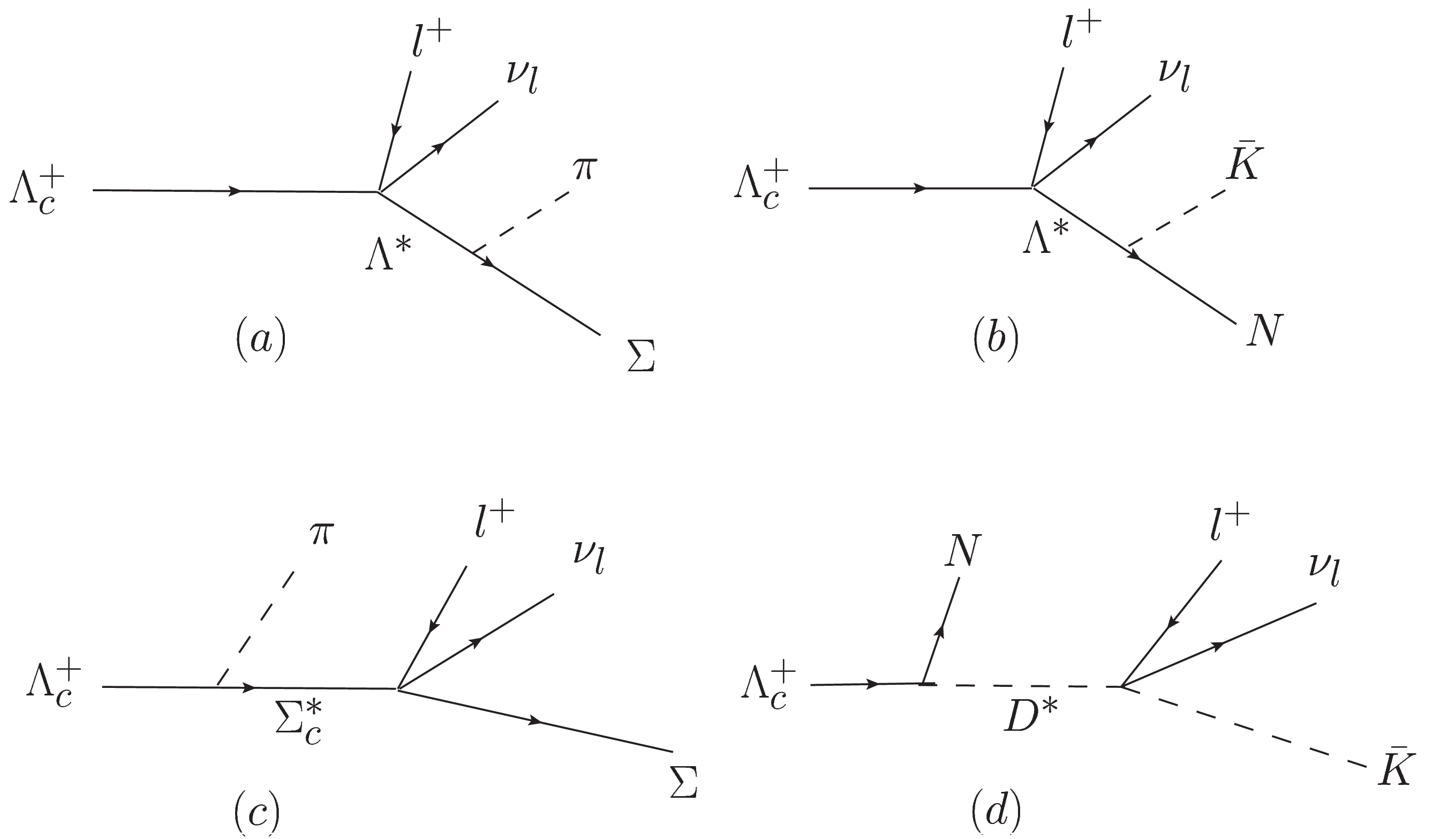}
\end{figure}

We include six $\Lambda^{(*)}$ in this calculation. We denote these as $\Lambda_i, \,\,i=1\dots 6$. In this notation, $\Lambda_1=\Lambda(1115)\,\,1/2^+$;  $\Lambda_2=\Lambda(1600)\,\,1/2^+$; $\Lambda_3=\Lambda(1405)\,\,1/2^-$; $\Lambda_4=\Lambda(1520)\,\,3/2^-$; $\Lambda_5=\Lambda(1890)\,\,3/2^+$; $\Lambda_6=\Lambda(1820)\,\,5/2^+$.
With the exception of $\Lambda_1$, these excited $\Lambda_i$ are not stable particles and will decay strongly to $\Sigma\pi$ or $N\bar{K}$. Thus we study the 
four-body decays, $\Lambda_c\to\Lambda_i l^{+} \nu_l \to \Sigma\pi l^{+} \nu_l$ and 
$\Lambda_c\to\Lambda_i l^{+} \nu_l \to N\bar{K} l^{+} \nu_l$ as shown in Fig. \ref{decaymodes}(a, b). There are other contributions to each of these four-body final states, two of which are shown in Fig. \ref{decaymodes} (c, d). However, in each case, the intermediate resonance is very heavy and very far from the mass shell. Thus, we expect these contributions to be small.

\subsubsection{Kinematics}
\begin{figure}
\caption{Kinematics for the process $\Lambda_c\to\Sigma\pi l\nu$. The lepton momenta define the lepton plane, while the momenta of the hadrons define the hadron plane.}\label{kinematic}
\includegraphics[width=0.5\textwidth]{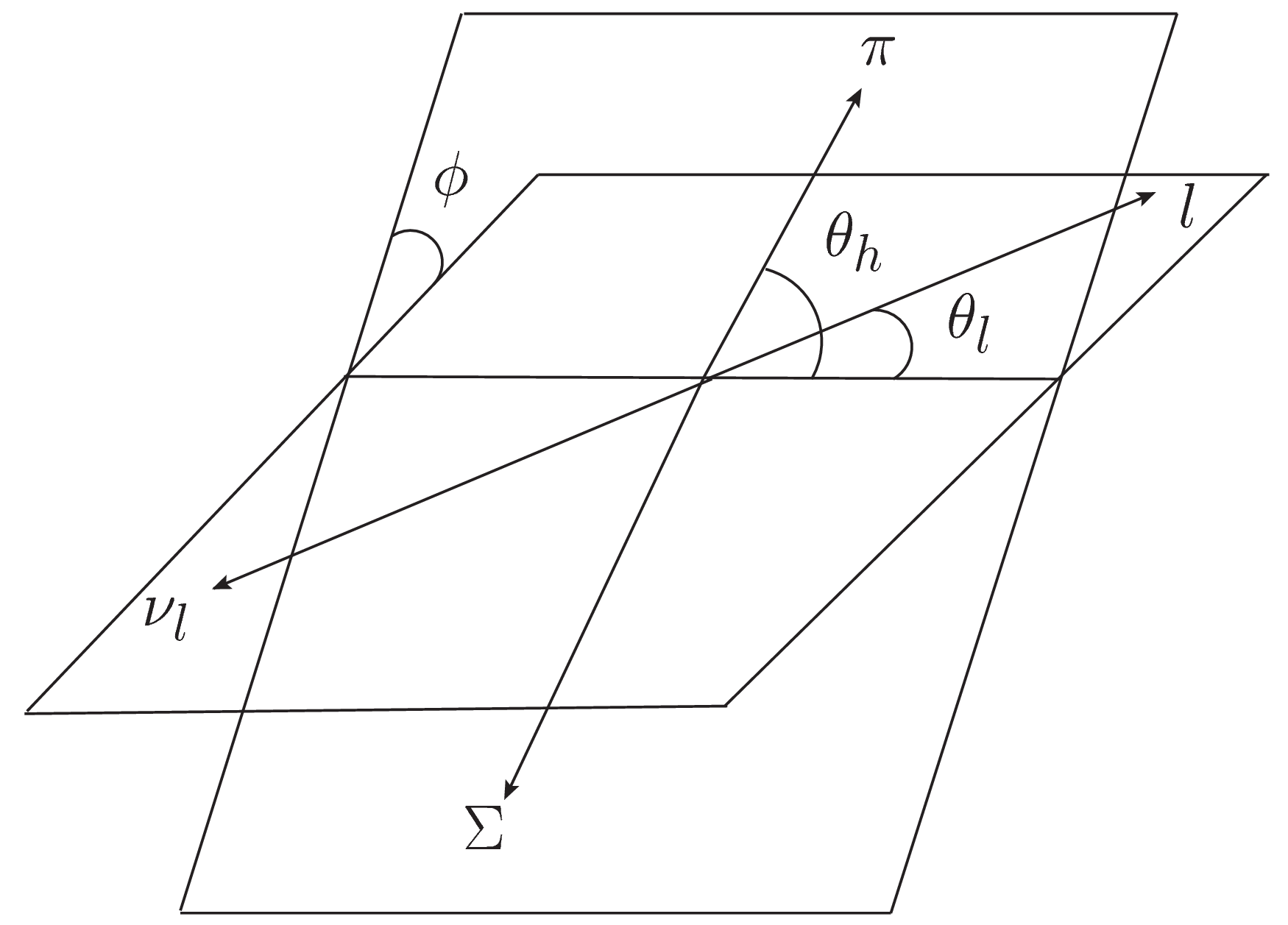}
\end{figure}

Fig \ref{kinematic} shows the kinematic diagram for the four-particle decay 
$\Lambda_c\to\Sigma\pi l^{+}\nu_l$.
We define 
\begin{equation}\label{pql}
P  \equiv  p_\Sigma + p_\pi,\,\,\,\, Q \equiv p_\Sigma-p_\pi,\,\,\,\,L \equiv p_l+p_\nu,
\end{equation}
so that $p_{\Lambda_c}=P+L$.
In the rest frame of the $\Lambda_c$, the back-to back momenta $\vec{P}$ and $\vec{L}$ define a common $z$-axis. In the rest frame of the daughter hadrons, $\theta_h^*$ is the polar angle between the pion momentum and $\vec{P}$. Similarly, in the rest frame of the lepton pair, $\theta_l^*$ is the polar angle between the lepton momentum and $\vec{L}$. $\phi^*$ is then the angle between the lepton and hadron planes. 

In the overall rest frame of $\Lambda_c$, the momenta $P$ and $L$ are
\begin{align*}
P & = \Big(\frac{1}{2m_{\Lambda_c}}(m_{\Lambda_c}^2+S_{\Sigma\pi}-q^2),\,\,0,\,\, 0,\,\,
 \frac{1}{2m_{\Lambda_c}}\lambda^{1/2}(m_{\Lambda_c}^2,S_{\Sigma\pi},q^2)\Big),\\
L & = \Big(\frac{1}{2m_{\Lambda_c}}(m_{\Lambda_c}^2-S_{\Sigma\pi}+q^2),\,\, 0,\,\, 0,\,\,
 -\frac{1}{2m_{\Lambda_c}}\lambda^{1/2}(m_{\Lambda_c}^2,S_{\Sigma\pi},q^2)\Big).
\end{align*}

In the rest frame of the daughter hadrons, the momenta $p_\Sigma$ and $p_\pi$ are
\begin{align*}
p_\pi & = \Big(\frac{1}{2\sqrt{S_{\Sigma\pi}}}(S_{\Sigma\pi}-m_\Sigma^2+m_\pi^2),\,\, \frac{1}{2\sqrt{S_{\Sigma\pi}}}\lambda^{1/2}(S_{\Sigma\pi},m_\Sigma^2,m_\pi^2)\sin{\theta_h^*},\,\, 0,\,\, \frac{1}{2\sqrt{S_{\Sigma\pi}}}\lambda^{1/2}(S_{\Sigma\pi},m_\Sigma^2,m_\pi^2)\cos{\theta_h^*}\Big),\\
p_\Sigma & = \Big(\frac{1}{2\sqrt{S_{\Sigma\pi}}}(S_{\Sigma\pi}+m_\Sigma^2-m_\pi^2),\,\, -\frac{1}{2\sqrt{S_{\Sigma\pi}}}\lambda^{1/2}(S_{\Sigma\pi},m_\Sigma^2,m_\pi^2)\sin{\theta_h^*},\,\,0,\,\,
 -\frac{1}{2\sqrt{S_{\Sigma\pi}}}\lambda^{1/2}(S_{\Sigma\pi},m_\Sigma^2,m_\pi^2)\cos{\theta_h^*}\Big).
\end{align*}

In the rest frame of the lepton pair, the lepton momenta are 
\begin{align*}
p_l & = \Big(\frac{1}{2\sqrt{q^2}}(q^2+m_l^2),\,\, \frac{1}{2\sqrt{q^2}}(q^2-m_l^2)\sin{\theta_l^*},\,\, 0,\,\,
 \frac{1}{2\sqrt{q^2}}(q^2-m_l^2)\cos{\theta_l^*}\Big),\\
p_\nu & = \Big(\frac{1}{2\sqrt{q^2}}(q^2-m_l^2),\,\,-\frac{1}{2\sqrt{q^2}}(q^2-m_l^2)\sin{\theta_l^*},\,\, 0,\,\,
 -\frac{1}{2\sqrt{q^2}}(q^2-m_l^2)\cos{\theta_l^*}\Big).
\end{align*}

\subsubsection{Matrix Elements}

The hadron matrix elements for the decays $\Lambda_c\to\Lambda_i l \nu_l \to BM l \nu_l$, where $B$ is a baryon with $J^P=1/2^+$ and $M$ is a pseudoscalar meson, can be written as
\begin{equation}
\langle (B(p_B)M(p_M))_i|J_{\mu}|\Lambda_c(p_{\Lambda_c})\rangle = \bar{u}(p_B)\Upsilon^s R(P){\cal J}^i_\mu u(P+L).
\end{equation}
In this expression, $\Upsilon^s$ represents the strong decay vertex, $p_B$ and $p_M$ are the momenta of the daughter baryon $B$ and meson $M$, respectively, $R(P)$ is the propagator with momentum $P$. $J_\mu$ is the weak current leading to the weak decay, while ${\cal J}_\mu^i$ is the matrix element for the semileptonic decay $\Lambda_c\to\Lambda_i$, written in terms of the form factors of section \ref{semime}. In this notation, the momenta of eqn. \ref{pql} are more generally written as
\begin{equation}\label{pqlp}
P  \equiv  p_B + p_M,\,\,\,\, Q \equiv p_B-p_M,\,\,\,\,L \equiv p_l+p_\nu.
\end{equation}

When the intermediate baryon has $J^P=1/2^+$, the hadron matrix elements are
\begin{align}\label{spin1}
 \langle B(p_B)M(p_M)|V^i_{\mu}|\Lambda_c(P+L)\rangle &=g_{\Lambda_i BM}\bar{u}(p_B)\gamma_5 
(\slashed{P}+M^i_{\Gamma}) \nabla^i 
 \biggl[\gamma_{\mu}F^i_1+\frac{(P+L)_\mu}{m_{\Lambda_c}}F^i_2 +\frac{P_{\mu}}{m_{\Lambda_i}}F^i_3\biggr]u(P+L),\nonumber\\
 \langle B(p_B)M(p_M)|A^i_{\mu}|\Lambda_c(P+L)\rangle &=g_{\Lambda_i BM}\bar{u}(p_B)\gamma_5 
(\slashed{P}+M^i_{\Gamma})\nabla^i \biggl[\gamma_{\mu}G^i_1 + \frac{(P+L)_{\mu}}{m_{\Lambda_c}}G^i_2
 +\frac{P_{\mu}}{m_{\Lambda_i}}G^i_3\biggr]\gamma_5u(P+L),
\end{align}
where $\nabla^i =1/(P^2-{M^i_\Gamma}^2)$
and $M^i_\Gamma = m_{\Lambda_i}- i\Gamma_i/2$, with $m_{\Lambda_i}$ and $\Gamma_i$ the mass and total
decay width of the $\Lambda_i$, respectively. $g_{\Lambda_i BM}$ is 
the strong coupling constant for the decay $\Lambda_i\to BM$.

For an intermediate state with $J^P=3/2^-$, the hadron matrix elements are
\begin{align}\label{spin3}
\langle B(p_B)M(p_M)|V^i_{\mu}|\Lambda_c(P+L)\rangle= &g_{\Lambda_i BM}\bar{u}(p_B)\gamma_5 \frac{p_{M_\alpha}}{m_M} R^{\alpha\beta}(P)\nabla^i
\biggl[\frac{(P+L)_{\beta}}{m_{\Lambda_c}}\biggr(\gamma_{\mu}F^i_1
+\frac{(P+L)_\mu}{m_{\Lambda_c}}F^i_2\nonumber\\&
+\frac{P_{\mu}}{m_{\Lambda_i}}F^i_3\biggr)+g_{\beta\mu}F^i_4\biggr]u(P+L),\nonumber\\
\langle B(p_B)M(p_M)|A^i_{\mu}|\Lambda_c(P+L)\rangle &= g_{\Lambda_i BM}\bar{u}(p_B) \gamma_5\frac{p_{M_\alpha}}{m_M}R^{\alpha\beta}(P)\nabla^i 
\biggl[\frac{(P+L)_{\beta}}{m_{\Lambda_c}} \biggl(\gamma_{\mu}G^i_1 + \frac{(P+L)_{\mu}}{m_{\Lambda_c}}G^i_2 \nonumber\\&+\frac{p_{\mu}}{m_{\Lambda_i}}G^i_3\biggr)+g_{\beta\mu}G^i_4\biggr]\gamma_5u(P+L),
\end{align}
where $R^{\alpha\beta}(P)$ is the Rarita-Schwinger tensor for a
massive spin $3/2$ propagator, which takes the form
\begin{align*}
R^{\alpha\beta}(P) = -(\slashed{P}+M^i_{\Gamma}) \bigg[g^{\alpha\beta} - \frac{1}{3}\gamma^\alpha\gamma^\beta - \frac{2P^{\alpha} P^{\beta}}{3m^2_{\Lambda_i}} 
- \frac{\gamma^\alpha P^{\beta} - \gamma^\beta P^{\alpha}}{3m_{\Lambda_i}}\bigg].
\end{align*} 
\noindent
For an intermediate state with $J^P=5/2^+$, the hadronic matrix elements are
\begin{align}\label{spin5}
\langle B(p_B)M(p_M)|V^i_{\mu}|\Lambda_c(p_{\Lambda_c})\rangle &= g_{\Lambda_i BM}\bar{u}(p_B)\gamma_5 
\frac{p_M^{\alpha^\prime}p_M^{\beta^\prime}}{m_M^2}R^{\alpha\beta}_{\alpha^\prime\beta^\prime}(P)\nabla^i
\frac{(P+L)_\alpha}{m_{\Lambda_c}}\biggl[\frac{(P+L)_{\beta}}{m_{\Lambda_c}}\biggr(\gamma_{\mu}F^i_1
+\frac{(P+L)_{\mu}}{m_{\Lambda_c}}F^i_2\nonumber\\&+\frac{P_{\mu}}{m_{\Lambda_i}}F^i_3\biggr)+g_{\beta\mu}F^i_4\biggr]u(p),\nonumber\\
\langle B(p_B)M(p_M)|A^i_{\mu}|\Lambda_c(P+L)\rangle &=  g_{\Lambda_i BM}\bar{u}(p_B)\gamma_5 
\frac{p_M^{\alpha^\prime}p_M^{\beta^\prime}}{m_M^2} R^{\alpha\beta}_{\alpha^\prime\beta^\prime}(P)\nabla^i
\frac{(P+L)_\alpha}{m_{\Lambda_c}}
\biggl[\frac{(P+L)_{\beta}}{m_{\Lambda_c}} \biggl(\gamma_{\mu}G^i_1 + \frac{(P+L)_{\mu}}{m_{\Lambda_c}}G^i_2\nonumber\\&+\frac{P_{\mu}}{m_{\Lambda_i}}G^i_3\biggr)+g_{\beta\mu}G^i_4\biggr]\gamma_5u(P+L),
\end{align}
where $R^{\alpha\beta}_{\alpha^\prime\beta^\prime}(P)$ is the Rarita-Schwinger propagator tensor for a massive particle with total angular momentum $5/2$ \citep{Spin5half}.

\vspace{0.2cm}

\noindent
We need to cast the matrix elements from the previous three equations into a more general form that makes it easier to organize the calculation. The most general form of the contribution of the $i$th state to the matrix element for the four-body decay $\Lambda_c^{+}\to \Lambda_il^{+}\nu_l\to BMl^{+}\nu_l$ can be written
\begin{equation*}
M^i_\nu =\bar{u}(p_B)\left(\sum_{j=1}^{16}c^i_j {\cal O}_j\right) u(P+L),
\end{equation*}
where the Lorentz-Dirac operators ${\cal O}_i$ are
\begin{align*}
{\cal O}_1=&\gamma_\nu ,\,\,\,\, {\cal O}_2=\slashed{P}\gamma_\nu,\,\,\,\, {\cal O}_3= P_\nu,\,\,\,\, {\cal O}_4=\slashed{P}P_\nu,\,\,\,\,
{\cal O}_5= L_\nu,\,\,\,\, {\cal O}_6=\slashed{P}L_\nu,\,\,\,\, {\cal O}_7= Q_\nu,\,\,\,\, {\cal O}_8=\slashed{P}Q_\nu,\nonumber\\
{\cal O}_9=&\gamma_\nu\gamma_5 ,\,\,\,\, {\cal O}_{10}=\slashed{P}\gamma_\nu\gamma_5,\,\,\,\, {\cal O}_{11}= P_\nu\gamma_5,\,\,\,\, {\cal O}_{12}=\slashed{P}P_\nu\gamma_5,\,\,\,\,
{\cal O}_{13}= L_\nu\gamma_5,\,\,\,\, {\cal O}_{14}=\slashed{P}L_\nu\gamma_5,\,\,\,\, {\cal O}_{15}= Q_\nu\gamma_5,\,\,\,\, {\cal O}_{16}=\slashed{P}Q_\nu\gamma_5.
\end{align*}
Because of the forms of the propagators in eqns. \ref{spin1} - \ref{spin5}, there are no terms containing $\slashed{L}$ or $\slashed{Q}$ among the ${\cal O}_i$. For the cases of eqns. \ref{spin3} and \ref{spin5}, the meson momentum $p_M$ can be replaced by $P-p_B$, and any factors of $\slashed{p}_B$ can be commuted leftward until they are adjacent to the spinor $\bar{u}(p_B)$. The Dirac equation can then be used to write this as the scalar $m_B$. 

The $c^i_j$ can be written
\begin{equation}
c^i_j =\frac{g_{\Lambda_i BM}}{P^2-{M^i_{\Gamma}}^2}\sum_{k}(C^{i,F}_{jk}F_k+C^{i,G}_{jk}G_k).
\end{equation}
where $k$ runs from $1$ to $3$ for spin $\frac{1}{2}$ states and from $1$ to $4$ for states with higher spin.

In the above, we have shown the forms for the $\Lambda_i$ states with natural parity. For the states with unnatural parity, the weak and strong vertices each acquire an extra multiplicative factor of $\gamma_5$.
\vspace{1cm}

\subsubsection{Decay Width}

The differential decay rate for the decay $\Lambda_c^{+} \to BM l^{+} \nu_l$ is,
\begin{equation}
d\Gamma = \frac{1}{2m_{\Lambda_c}} \frac{G_F^2}{2} |V_{cs}|^2\frac{d^3p_B d^3p_M d^3p_l d^3p_{\nu_l}}{2E_B 2E_M 2E_l 2E_{\nu_l}}\delta^4(p_{\Lambda_c}-p_B - p_M- p_l - p_{\nu_l})H_{\mu\nu}L^{\mu\nu},
\end{equation}

\noindent
The hadron tensor that arises from each intermediate state $i$ can be written as,
\begin{align*}
H^i_{\mu\nu} &= \sum_{\text{spins}}M^{i\dagger}_\mu M^i_\nu
= \alpha^i g_{\mu\nu} +\beta_{PP}^iP_\mu P_\nu +\beta_{PQ}^iP_\mu Q_\nu
				+\beta_{QP}^iQ_\mu P_\nu +\beta_{QQ}^iQ_\mu Q_\nu\\
&				+\beta_{QL}^iQ_\mu L_\nu
				+\beta_{LQ}^iL_\mu Q_\nu 
				+\beta_{LL}^iL_\mu L_\nu +\beta_{PL}^iP_\mu L_\nu
				+\beta_{PL}^iP_\mu L_\nu\\
&				+ i\gamma_a^i\epsilon^{\mu\nu\rho\delta} P_\rho Q_\delta
				+ i\gamma_b^i\epsilon^{\mu\nu\rho\delta} L_\rho P_\delta
				+ i\gamma_c^i\epsilon^{\mu\nu\rho\delta} L_\rho Q_\delta
				+ i\gamma_d^i\epsilon^{\sigma\mu\rho\delta}L_\sigma P_\rho Q_\delta P_\nu
				+ i\gamma_e^i\epsilon^{\sigma\mu\rho\delta}L_\sigma P_\rho Q_\delta Q_\nu\\
&				+ i\gamma_f^i\epsilon^{\sigma\mu\rho\delta}L_\sigma P_\rho Q_\delta L_\nu				
				+ i\gamma_g^i\epsilon^{\sigma\nu\rho\delta}L_\sigma P_\rho Q_\delta P_\mu
				+ i\gamma_h^i\epsilon^{\sigma\nu\rho\delta}L_\sigma P_\rho Q_\delta Q_\mu
				+ i\gamma_k^i\epsilon^{\sigma\nu\rho\delta}L_\sigma P_\rho Q_\delta L_\mu.
\end{align*}
In this expression, 
\begin{equation}
\alpha^i=\sum_{j,k=1}^{16}a^i_{jk}c^{i\dag }_jc^i_k,
\end{equation}
with similar forms for all of the other coefficients. The terms in $\gamma_i$ do not contribute to the decay rates that we consider, due to the symmetry of the lepton tensor. 

For the process $\Lambda_c\to BMl\nu_l$ we examine the contribution from each $\Lambda_i$ individually, as well as the coherent contribution of all the $\Lambda_i$. For the coherent sum, we write
\begin{equation}
M_\nu=\bar{u}(p_B)\left(\sum_{j=1}^{16}{\cal C}_j {\cal O}_j\right)u(p_{\Lambda_c})=\bar{u}(p_B)\sum_{i=1}^6\left(\sum_{j=1}^{16}c^i_j {\cal O}_j\right) u(p_{\Lambda_c}),
\end{equation}
which ultimately leads to
\begin{equation}
{\cal C}_j=\sum_{i=1}^6c^i_j.
\end{equation}

Integrating over the lepton momenta, and making use of eqn. (\ref{lepten2}) leads to
\begin{align}
\frac{d\Gamma}{dS_{BM}dq^2d\theta_hd\theta_ld\phi}=&  \frac{|V_{cs}|^2}{2} 
\frac{4\pi G_F^2}{128 m_{\Lambda_c}^3S_{BM}}\sin{\theta_h}\lambda^{1/2}(m_{\Lambda_c}^2,S_{BM}^2,q^2)
\lambda^{1/2}(S_{BM}^2,m_B^2,m_M^2)\bigg( \alpha\Big[4A + A^{\prime}q^2\Big]\nonumber\\
&
+ \beta_{PP}\Big[AS_{BM}+A^{\prime}(P\cdot L)^2 \Big] 
+ \Big[\beta_{PQ}+ \beta_{QP}\Big]\Big[A(P\cdot Q) + A^{\prime}(P\cdot L)(Q\cdot L)\Big]\nonumber \\  
&
+ \beta_{QQ}\Big[A(Q\cdot Q)+A^{\prime}(Q\cdot L)^2\Big]\nonumber\\
&+\Big[(\beta_{LP} +\beta_{PL})(P\cdot L)
+ (\beta_{LQ} + \beta_{QL})(Q\cdot L) + \beta_{LL}q^2\Big]\Big[A + A^{\prime}q^2\Big] \bigg),
\end{align}
where
\begin{align*}
P\cdot P & = p_{\Lambda^{*}}^2 \equiv S_{BM}, \\
P\cdot Q & = m_B^2-m_M^2,\\
P\cdot L & = (m_{\Lambda_c}^2-S_{BM} - q^2)/2,\\
L\cdot L & = q^2,\\
Q\cdot Q & = 2m_B^2 + 2m_M^2 - S_{BM},\\
Q\cdot L & = \frac{1}{2S_{BM}}\left[(m_B^2 - m_M^2)(m_{\Lambda_c}^2 - S_{BM} - q^2) 
+ \cos{\theta_h}\lambda^{1/2}(m_{\Lambda_c}^2,S_{BM},q^2)\lambda^{1/2}(S_{BM},m_B^2,m_M^2)\right].\\
\end{align*}

\section{Heavy Quark Effective Theory }

The heavy quark effective theory (HQET) has been a very useful tool in the study of the electroweak decays of hadrons containing one heavy quark.
In this effective theory, the matrix elements are expanded in increasing orders of $1/m_Q$, where $m_Q$ is the mass of the heavy quark. This expansion has facilitated the extraction of CKM matrix elements with decreasing model dependence.

Hadrons containing a single charm or beauty quark are considered to be heavy hadrons as the mass $m_Q>>\Lambda_{\text{QCD}}$. For such hadrons, HQET reduces the number of independent form factors required to describe the transitions mediated by electroweak transitions that change a heavy quark of one flavor into a heavy quark of different flavor. At leading order in the $1/m_Q$ expansion, such heavy to heavy transitions require a single form factor, the so-called Isgur-Wise function. This is the case independent of the total angular momentum of the daughter hadron (we assume that the parent hadron is a ground-state hadron), integer (meson) or half-integer (baryon). For transitions between a ground-state heavy hadron and a light one, HQET is not as powerful. However, for transitions between a heavy baryon (ground state) and a light one, HQET indicates that a pair of form factors is all that is needed to describe the transition, independent of the angular momentum of the daughter baryon. 

The semileptonic decays $\Lambda_c\to\Lambda^{*}$ fall into this second category, and are therefore described by two independent form factors. We may represent one of these light baryons of angular momentum $J$  
by a generalized Rarita-Schwinger field $u^{\mu_1\dots\mu_n}(p)$ where $n=J-1/2$. This field is symmetric under exchange of any pair of its Lorentz indices, and satisfies the conditions
\begin{align*}
 \slashed{p}u^{\mu_1...\mu_n}(p)&=m_{\Lambda}u^{\mu_1...\mu_n}(p),\gamma_{\mu_1}u^{\mu_1...\mu_n}(p)=0,\\
p_{\mu_1}u^{\mu_1...\mu_n}(p)&=0,u_{\mu}^{\mu...\mu_n}(p)=0.
 \end{align*}
The matrix element we are interested in is
 \begin{equation}
  \langle\Lambda^{*}(p^{\prime})|\bar{s}\Gamma c|\Lambda_c^{+}(p)\rangle=\bar{u}^{\mu_1...\mu_n}\mathit{M}_{\mu_1...\mu_n}\Gamma u(p),
 \end{equation}
where $\Gamma = \gamma^\mu \text{ or } \gamma^\mu\gamma_5$ defines
vector or axial vector current and $\mathit{M}_{\mu_1\mu_2...\mu_n}$ is a tensor. 
The most general tensor can be constructed as
\begin{equation}
 \mathit{M}_{\mu_1\mu_2...\mu_n}=\mathit{v}_{\mu_1}...\mathit{v}_{\mu_n}\mathit{A}_n,
\end{equation}
\noindent
where $\mathit{A_n}$ is the most general Lorentz scalar that can be constructed. This takes the form
\begin{equation}{\label{superscript1}}
 \mathit{A}_n=\xi_1^{(n)}+\slashed{\mathit{v}}\xi_2^{(n)},
\end{equation}
where $v=p/m_{\Lambda_c}$ is the velocity of the parent baryon.
For the transitions to daughter baryons with  unnatural parity, $M_{\mu_1\mu_2...\mu_n}$ must be a pseudo-tensor. This is easily constructed by including a factor of $\gamma_5$, so that
\begin{equation}{\label{superscript2}}
 \mathit{M}_{\mu_1\mu_2...\mu_n}=\mathit{v}_{\mu_1}...\mathit{v}_{\mu_n}
 \biggl(\zeta_1^{(n)}+\slashed{\mathit{v}}\zeta_2^{(n)}\biggr)\gamma_5.
\end{equation}

\subsection{Form Factors}\label{HQETformfactors}
 The matrix elements can be written in terms of six general form factors for spin $\frac{1}{2}^{\pm}$, or eight general form factors for
 spin $\frac{3}{2}^{\pm}$ and $\frac{5}{2}^{+}$, as shown in Section \ref{semime}.  
Comparing the predictions of HQET with the most general form of the matrix elements leads to a number of relations among the 
general form factors $F_i/G_i$ and the HQET form factors $\xi_i/\zeta_i$.

For spin $\frac{1}{2}^{+}$, these relationships are
\begin{equation}\label{xixi:1}
 F_1=\xi_1^{(0)}-\xi_2^{(0)}, 
 G_1=\xi_1^{(0)}+\xi_2^{(0)}, 
 F_2=G_2=2\xi_2^{(0)}, 
 F_3=G_3=0.
\end{equation}
For spin $\frac{1}{2}^{-}$, they are 
\begin{equation}\label{xixi:2}
 F_1=-(\zeta_1^{(0)}+\zeta_2^{(0)}), G_1=-(\zeta_1^{(0)}-\zeta_2^{(0)}), 
 F_2=G_2=-2\zeta_2^{(0)}, 
 F_3=G_3=0.
\end{equation}
For spin $\frac{3}{2}^{-}$, they are 
\begin{equation}\label{xixi:3}
 F_1=\xi_1^{(1)}-\xi_2^{(1)}, 
 G_1=\xi_1^{(1)}+\xi_2^{(1)}, 
 F_2=G_2=2\xi_2^{(1)}, 
 F_3=G_3=0, 
 F_4=G_4=0.
\end{equation}
\noindent
For spin $\frac{3}{2}^{+}$, the relationships are
\begin{equation}\label{xixi:4}
 F_1=-(\zeta_1^{(1)}+\zeta_2^{(1)}),
 G_1=-(\zeta_1^{(1)}-\zeta_2^{(1)}),
 F_2=G_2=-2\zeta_2^{(1)},
 F_3=G_3=0,
 F_4=G_4=0.
\end{equation}
\noindent
 For spin $\frac{5}{2}^{+}$, they are
\begin{equation}\label{xixi:5}
 F_1=\xi_1^{(2)}-\xi_2^{(2)},
 G_1=\xi_1^{(2)}+\xi_2^{(2)}),
 F_2=G_2=2\xi_2^{(2)},
 F_3=G_3=0,
 F_4=G_4=0.
\end{equation}
\noindent
\subsection{Decay Width}
At leading order in HQET, the differential decay rates take simple forms for all the excited states we discuss. 
This general form is
\begin{equation}{\label{HQETdecaywidth}}
 \frac{d\Gamma}{dq^2}=
 \Phi^J X \bigg[\left(A_1+A_2\frac{m_l^2}{q^2}\right)\xi_1^2 + \frac{m_{\Lambda}}{m_{\Lambda_c}}
 \left(B_1+B_2\frac{m_l^2}{q^2}\right)\xi_1\xi_2 
 + \frac{1}{m_{\Lambda_c}^2}\left(C_1+C_2\frac{m_l^2}{q^2}\right)\xi_2^2\bigg],
\end{equation}
\noindent
where $\Phi^J$ is a dimensionless quantity that 
depends on the angular momentum of the daughter baryon. $X,A_i,B_i,C_i$ are, respectively,
\begin{align*}
X & = \frac{4\pi^3}{m_{\Lambda_c}^3}\frac{G_F^2}{2}|v_{cs}|^2\lambda^{1/2}(m_{\Lambda_c}^2,m_{\Lambda}^2,q^2),\\
A_1 & = \bigg[m_{\Lambda_c}^4+m_{\Lambda_c}^2
\bigg(q^2-2m_{\Lambda}^2\bigg)+m_{\Lambda}^4+m_{\Lambda}^2 q^2-2 q^4\bigg],\\
A_2 & = \bigg[2m_{\Lambda_c}^4-m_{\Lambda_c}^2\bigg(4 m_{\Lambda}^2+q^2\bigg)+2
m_{\Lambda}^4-m_{\Lambda}^2 q^2-q^4\bigg],\\
B_1 & =2\bigg[m_{\Lambda_c}^4-2
m_{\Lambda_c}^2\bigg(m_{\Lambda}^2-2
q^2\bigg)+\bigg(m_{\Lambda}^2-q^2\bigg)^2\bigg],\\
B_2 & = 4\bigg[m_{\Lambda_c}^4+m_{\Lambda_c}^2\bigg(q^2-2
m_{\Lambda}^2\bigg)+\bigg(m_{\Lambda}^2-q^2\bigg)^2\bigg],\\
C_1 & = \bigg[m_{\Lambda_c}^4
\bigg(m_{\Lambda}^2+2 q^2\bigg)-m_{\Lambda_c}^2\bigg(2 m_{\Lambda}^4-3
m_{\Lambda}^2 q^2+q^4\bigg)+\bigg(m_{\Lambda}^2-q^2\bigg)^3\bigg],\\
C_2 & = \bigg[m_{\Lambda_c}^4
\bigg(2 m_{\Lambda}^2+q^2\bigg)+m_{\Lambda_c}^2\bigg(-4 m_{\Lambda}^4+3
m_{\Lambda}^2 q^2+q^4\bigg)+2\bigg(m_{\Lambda}^2-q^2\bigg)^3\bigg].
\end{align*}

The decay width for states with total spin $J$ does not depend on parity. The $\Phi^J$ for states with angular momentum $J$ are
\begin{align*}
\Phi^{1/2}&  =4,\\
\Phi^{3/2}& =\frac{2}{3m_{\Lambda_c}^2 m_{\Lambda}^2}\lambda\big(m_{\Lambda_c}^2,m_\Lambda^2,q^2\big),\\
\Phi^{5/2}& = \frac{1}{10m_{\Lambda_c}^4 m_{\Lambda}^4}\lambda^2\big(m_{\Lambda_c}^2,m_\Lambda^2,q^2\big).
\end{align*}

\section{The Model}

\subsection{Wave Function Components}

\noindent
In our model, a baryon state has the form
\begin{equation}\label{statefunction}
 |\Lambda_Q(\vec{p},s)\rangle = 3^{-3/2}\int d^3p_{\rho}d^3p_{\lambda}C^A\Psi^S_{\Lambda_Q}|q_1(\vec{p}_1,s_1)
 q_2(\vec{p}_2,s_2)q_3(\vec{p}_3,s_3)\rangle,
\end{equation}
\noindent
where $\Lambda_Q$ is a flavored baryon ($\Lambda_c^{+}$ or $\Lambda$) having a flavored quark ($c$ or $s$) 
$Q$, which may or may not be considered heavy. $q_i(\vec{p}_i,s_i)$ is the creation operator for quark $q_i$ with momentum $\vec{p}_i$ and spin $s_i$. $|q_1(\vec{p}_1,s_1) q_2(\vec{p}_2,s_2)q_3(\vec{p}_3,s_3)\rangle$ is the three quark state with quarks $q_i$ 
 having momenta and spins $(\vec{p}_i,s_i)$.
$\vec{p}_{\rho}=\frac{1}{\sqrt{2}}(\vec{p}_1-\vec{p}_2) $ and
$\vec{p}_\lambda=\frac{1}{\sqrt{6}}(\vec{p}_1+\vec{p}_2-2\vec{p}_3) $ are the Jacobi momenta.
$C^A$ is the antisymmetric color wave function and $\Psi_{\Lambda_Q}^S=\phi_{\Lambda_Q}\psi_{\Lambda_Q}\chi_{\Lambda_Q}$ is a symmetric 
combination of flavor, momentum and spin wave functions.
For $\Lambda_Q$ the flavor wave function is
\begin{equation}
 \phi_{{\Lambda}_Q}=\frac{1}{\sqrt{2}}(ud-du)Q.
\end{equation}
This is antisymmetric under the exchange of the first two quarks, so the spin-space wave function must also 
be antisymmetric under such exchange.

The total spin of a system of three spin-$\frac{1}{2}$ particles can be 
either $\frac{3}{2}$ or $\frac{1}{2}$. The maximally stretched spin states are
\begin{align*}
\chi_{3/2}^S(+3/2)&=|\uparrow\uparrow\uparrow\rangle,\\
\chi_{1/2}^{\rho}(+1/2)&=\frac{1}{\sqrt{2}}(|\uparrow\downarrow\uparrow\rangle-\downarrow\uparrow\uparrow\rangle),\\
\chi_{1/2}^{\lambda}(+1/2)&=-\frac{1}{\sqrt{6}}(|\uparrow\downarrow\uparrow\rangle
+|\downarrow\uparrow\uparrow\rangle-2|\uparrow\uparrow\downarrow\rangle),
\end{align*}
where the superscript $S$ indicates that the state is totally symmetric under the exchange of any pair of quarks,
while $\rho$, $\lambda$ denote the mixed-symmetric states that are
antisymmetric and symmetric under the exchange of first two spins, respectively.
 
The momentum-space wave function $\psi_{\Lambda_Q}$ can be constructed from the Clebsch-Gordan sum of the product of wave functions of the two jacobi 
momenta $p_{\rho}$, $p_{\lambda}$ with total angular momentum $\vec{L}=\vec{l}_{\rho}+\vec{l}_{\lambda}$,
\begin{equation}
 \psi_{LM_Ln_{\rho}l_{\rho}n_{\lambda}l_{\lambda}}(p_{\lambda},p_{\rho})=\sum_{m}C^{LM_L}_{l_{\rho}m,l_{\lambda}M_L-m}
 \psi_{n_{\rho}l_{\rho}m}(p_{\rho})\psi_{n_{\lambda}l_{\lambda}M_L-m}(p_{\lambda}).
\end{equation}
This wave function is then coupled to the spin wave function $\chi_{\Lambda_Q}$ to give a spin-momentum wave function
of total spin $J$ and parity $(-)^{(l_{\rho}+l_{\lambda})}$,
\begin{equation}
 \Psi_{JM}=\sum_{M_L}C^{JM}_{LM_L,SM-M_L}
 \psi_{LM_Ln_{\rho}l_{\rho}n_{\lambda}l_{\lambda}}\chi_{S}(M-M_L).
\end{equation}
The full wave function is then constructed as
\begin{align}
\Psi_{{\Lambda_Q},{J^P},M} = \phi_{\Lambda_Q}\sum_i\eta_i\Psi_{JM}^i
\end{align}
The $\eta_i$ are the coefficients determined by diagonalizing the Hamiltonian in the basis of the states $\Psi_{JM}^i$  \citep{PRC}.
In this model the expansion is restricted to $N\leq2$, where $N=2(n_{\rho}+n_{\lambda})+l_{\rho}+l_{\lambda}$.

In the notation introduced above, the wave functions for states with $J^P=\frac{1}{2}^{+}$ are written as
\begin{align}
 \Psi_{\Lambda_{Q},\frac{1}{2}^{+}M}=\phi_{\Lambda_{Q}}\biggl(&\biggl[\eta_{1}\psi_{000000}(\vec{p}_{\rho},\vec{p}_{\lambda})
 +\eta_{2}\psi_{001000}(\vec{p}_{\rho},\vec{p}_{\lambda})+\eta_{3}\psi_{000010}(\vec{p}_{\rho},\vec{p}_{\lambda})\biggr]
 \chi_{\frac{1}{2}}^{\rho}(M)\nonumber \\
 &+\eta_{4}\psi_{000101}(\vec{p}_{\rho},\vec{p}_{\lambda}) \chi_{\frac{1}{2}}^{\lambda}(M)
 +\eta_{5}\biggl[\psi_{1M_{L}0101}(\vec{p}_{\rho},\vec{p}_{\lambda}) \chi_{\frac{1}{2}}^{\lambda}(M-M_L)\biggr]_{1/2,M}\nonumber \\
 &+\biggl[\eta_{6}\psi_{1M_{L}0101}(\vec{p}_{\rho},\vec{p}_{\lambda})\chi_{\frac{3}{2}}^{S}
 (M-M_L)\biggr]_{1/2,M}+\eta_{7}\biggl[\psi_{2M_{L}0101}(\vec{p}_{\rho},\vec{p}_{\lambda})\chi_{\frac{3}{2}}^{S}
 (M-M_L)\biggr]_{1/2,M}\biggr),
\end{align}
where we have used $[\psi_{LM_Ln_{\rho}l_{\rho}n_{\lambda}l_{\lambda}}(\vec{p}_\rho,\vec{p}_\lambda)\chi_S(M-M_L)]_{J,M}$ as a 
shorthand notation for the Clebsch-Gordan sum  
$\sum_{M_L}C^{JM}_{LM_L,SM-M_L} \psi_{LM_Ln_\rho l_\rho n_\lambda l_\lambda}(\vec{p}_\rho,\vec{p}_\lambda)\chi_S(M-M_L)$. 

In our analytic calculation of the form factors we have used the following single component representation of 
the $\Lambda$ states with different $J^P$:
\begin{align}\label{scwf}
\Psi_{1/2^{+},M} & = \phi_\Lambda\biggl[\psi_{000000}(\vec{p}_\rho,\vec{p}_\lambda)\chi_{1/2}^\rho(M-M_L) \biggr]_{1/2,M};
\nonumber\\
\Psi_{1/2^{+}_1,M} & = \phi_\Lambda\biggl[\psi_{000010}(\vec{p}_\rho,\vec{p}_\lambda)\chi_{1/2}^\rho(M-M_L) \biggr]_{1/2,M};
\nonumber\\
\Psi_{1/2^{-},M} & = \phi_\Lambda\biggl[\psi_{1M_L0001}(\vec{p}_\rho,\vec{p}_\lambda)\chi_{1/2}^\rho(M-M_L) \biggr]_{1/2,M};
\nonumber\\
\Psi_{3/2^{-},M} & = \phi_\Lambda\biggl[\psi_{1M_L0001}(\vec{p}_\rho,\vec{p}_\lambda)\chi_{1/2}^\rho(M-M_L) \biggr]_{3/2,M};
\nonumber\\
\Psi_{3/2^{+},M} & = \phi_\Lambda\biggl[\psi_{2M_L0002}(\vec{p}_\rho,\vec{p}_\lambda)\chi_{1/2}^\rho(M-M_L) \biggr]_{3/2,M};
\nonumber\\
\Psi_{5/2^{+},M} & = \phi_\Lambda\biggl[\psi_{2M_L0002}(\vec{p}_\rho,\vec{p}_\lambda)\chi_{1/2}^\rho(M-M_L) \biggr]_{5/2,M}.
\end{align}
For details of the construction of the wave functions, see appendix \ref{wavefunctions}.

$\psi_{nlm}$ is expanded in the harmonic oscillator basis, whose wave functions in momentum space are,
\begin{equation}
\psi_{nlm}(\vec{p}) = \left[\frac{2n!}{(n+l+\frac{1}{2})!}\right]^{\frac{1}{2}}(i)^l(-1)^n\frac{1}{\alpha^{l+\frac{3}{2}}}
e^{-\frac{p^2}{(2\alpha^2)}}L_n^{l+\frac{1}{2}}(p^2/\alpha^2)\mathcal{Y}_{lm}(\vec{p}),
\end{equation}
where, $L_n^\beta(x)$ are the generalized Laguerre polynomials with $p=|\vec{p}|$ and $\mathcal{Y}_{lm}(\vec{p})$ are the 
solid harmonics. 
 
\subsection{Extraction of Form Factors}

The hadron matrix elements for any arbitrary current $\bar{s}\Gamma c$ take the form
\begin{align}
\langle \Lambda(p_{\Lambda},s^\prime)|\bar{s}\Gamma c|\Lambda_c(0,s) \rangle = & \int d^3p^\prime_\lambda d^3p^\prime_\rho
d^3p_\lambda d^3p_\rho C^{A*}C^A\Psi_{\Lambda}^{*}(s^\prime)\nonumber\\
& \times \langle q_1^\prime q_2^\prime s|\bar{s}\Gamma c|q_1 q_2 c \rangle \Psi_{\Lambda_c}(s),{\label{Eff}}
\end{align}
where $ \langle q_1^\prime q_2^\prime s|\bar{s}\Gamma c|q_1 q_2 c \rangle = 
\langle q_1^\prime q_2^\prime |q_1 q_2 \rangle \langle s|\bar{s}\Gamma c|c \rangle$. 
In our spectator approximation $\langle q_1^\prime q_2^\prime |q_1 q_2 \rangle$ gives delta functions in spin, momentum and flavor.

The analytic expressions for the form factors shown in appendix \ref{AnalyticFormFactors} are obtained using the single-component wave functions of eq. \ref{scwf}. We also calculate the form factors numerically using the full multi-component wave functions extracted from the diagonalization of the Hamiltonian. For this, we adapted the semi-analytic approach used by Mott and Roberts \citep{Mott} in their calculation of the rare dileptonic decay of $\Lambda_b$. In this method some of the calculation is done analytically, leaving a couple of integrations to be done numerically.
 
In the rest frame of the parent $\Lambda_c$, we write the initial quark momenta in terms of the Jacobi momenta as
\begin{align*}
\vec{p}_1 = \frac{1}{\sqrt{2}}\vec{p}_\rho + \frac{1}{\sqrt{6}}\vec{p}_\lambda, \hspace{0.5cm} 
\vec{p}_2 =- \frac{1}{\sqrt{2}}\vec{p}_\rho + \frac{1}{\sqrt{6}}\vec{p}_\lambda, \hspace{0.5cm}
\vec{p}_3\equiv\vec{p} = -\sqrt{\frac{2}{3}}\vec{p}_\lambda.
\end{align*}
We use the spectator approximation in which the first two quarks are unaffected by the transition. This allows us to integrate over the Jacobi momenta separately and write the matrix element as 
\begin{equation}
\langle \Lambda|\bar{s}\Gamma c|\Lambda_c \rangle =  \sum_{b^\prime,b} h_{b^\prime}^{\Lambda^{*}}
h_b^{\Lambda_c}\delta_{s_1^\prime s_1}\delta_{s_2^\prime s_2}(-1)^{l_{\lambda^\prime}+l_\lambda}\times B_{n_\rho l_\rho m_\rho}^{n_{\rho^\prime}l_{\rho^\prime}m_{\rho^\prime}}(\alpha_\rho,\alpha_{\rho^\prime})D_{\Gamma;n_\lambda l_\lambda m_\lambda s_q}^{n_{\lambda^\prime} l_{\lambda^\prime} m_{\lambda^\prime} s_{q^\prime}}(\alpha_\lambda,\alpha_{\lambda^\prime}),
\end{equation}
where the coefficients $h_{b(b^\prime)}$ are the products of the normalization of the baryon states, the expansion coefficients 
$\eta_i$, and the various Clebsch-Gordan coefficients that appear in the parent (daughter) baryon wave function.
The indices $b(b^\prime)$ contain all the relevant quantum numbers being summed over for the parent (daughter) baryon state. 
$B_{n_\rho l_\rho m_\rho}^{n_{\rho^\prime}l_{\rho^\prime}m_{\rho^\prime}}$ is the spectator overlap,
\begin{equation}
B_{n_\rho l_\rho m_\rho}^{n_{\rho^\prime}l_{\rho^\prime}m_{\rho^\prime}}(\alpha_\rho,\alpha_{\rho^\prime})
=\int d^3p_\rho d^3p_{\rho^\prime}\psi^{*}_{n_{\rho^\prime} l_{\rho^\prime} m_{\rho^\prime}}(\alpha_{\rho^\prime};
\vec{p}_{\rho^\prime})\psi_{n_\rho l_\rho m_\rho}(\alpha_\rho;\vec{p}_\rho)\delta(p_\rho-p_{\rho^\prime}).
\end{equation}
This integral can be done analytically and is given in appendix \ref{SpecOver}. 

The interaction overlap 
$D_{\Gamma;n_\lambda l_\lambda m_\lambda s_q}^{n_{\lambda^\prime} l_{\lambda^\prime}m_{\lambda^\prime} s_{q^\prime}}$ is
\begin{align*}
D_{\Gamma;n_\lambda l_\lambda m_\lambda s_q}^{n_{\lambda^\prime} l_{\lambda^\prime} 
m_{\lambda^\prime} s_{q^\prime}}(\beta,\beta^\prime)
= \int d^3p \psi^{*}_{n_{\lambda^\prime}l_{\lambda^\prime}m_{\lambda^\prime}}
\left(\beta^\prime;\vec{p}^\prime\right)
\langle s(\vec{p}+\vec{p}_\Lambda,s_{q^\prime})|\bar{s}\Gamma c|c(\vec{p},s_q)\rangle
\psi_{n_{\lambda}l_{\lambda}m_{\lambda}}(\beta;\vec{p}),
\end{align*}
where $\beta^{(\prime)} = \sqrt{2/3}\alpha_\lambda^{(\prime)}$ is the reduced length parameter for the parent (daughter) baryon,
$p^\prime = (2m_\sigma / \tilde{m}_\Lambda) \vec{p}_\Lambda + \vec{p}$, $\tilde{m}_\Lambda = m_s+2m_\sigma$, and $m_s$ and $m_\sigma$ 
are the masses of the strange quark and each light quark, respectively. In terms of the generalized Laguerre polynomials, 
\begin{align}
D_{\Gamma;n_\lambda l_\lambda m_\lambda s_q}^{n_{\lambda^\prime} l_{\lambda^\prime} 
m_{\lambda^\prime} s_{q^\prime}}(\beta,\beta^\prime)
=&\int d^3p \text{ exp}\biggl(-\frac{p^{\prime 2}}{2\beta^{\prime 2}} - \frac{p^2}{2\beta^2}\biggr) \mathcal{L}^{l_{\lambda^\prime}
+\frac{1}{2}*}_{n_{\lambda^\prime}}\Bigg(\frac{{p^\prime}^2}{{\beta^\prime}^2}\Bigg)
\mathcal{Y}_{l_{\lambda^\prime}m_{\lambda^\prime}}^{*}(\vec{p}^{\prime})\nonumber \\
&\times
\langle s(\vec{p}_{\Lambda}+\vec{p},s_{q^\prime})|\bar{s}\Gamma c|c(\vec{p},s_q)\rangle
\mathcal{L}^{l_{\lambda}+\frac{1}{2}}_{n_{\lambda}}\Bigg(\frac{p^2}{\beta^2}\Bigg)\mathcal{Y}_{l_\lambda m_\lambda}(\vec{p}).\nonumber \\
\end{align}
The angular dependence in the exponential is eliminated by using the substitutions
\begin{equation}
\vec{p} = \vec{k} + a\vec{p}_\Lambda, \,\,\,\,\vec{p}^\prime = \vec{k} + a^\prime \vec{p}_\Lambda,
\end{equation}
where
\begin{equation}
a = - \frac{m_\sigma \alpha_\lambda^2}{\tilde{m}_\Lambda\alpha^2_{\lambda\lambda^\prime}},\,\,\,\,
a^\prime =  \frac{m_\sigma \alpha_{\lambda^\prime}^2}{\tilde{m}_\Lambda\alpha^2_{\lambda\lambda^\prime}}.
\end{equation}
and $\alpha_{\lambda\lambda^\prime} = \sqrt{(\alpha_\lambda^2+\alpha_{\lambda^\prime}^2)/2}$. 
The interaction overlap then takes the form
\begin{align}
D_{\Gamma;n_\lambda l_\lambda m_\lambda s_q}^{n_{\lambda^\prime} l_{\lambda^\prime} 
m_{\lambda^\prime} s_{q^\prime}}(\beta,\beta^\prime)
=&\text{ exp}\Bigg(\frac{-3m_\sigma^2}{2m_{\Lambda}^2}\frac{p_\Lambda^2}{\alpha_{\lambda\lambda^\prime}^2}\Bigg)
\int d^3k e^{-\alpha^2k^2}\mathcal{L}^{l_{\lambda^\prime}
+\frac{1}{2}*}_{n_{\lambda^\prime}}\Bigg(\frac{{p^\prime}^2}{{\beta^\prime}^2}\Bigg)
\mathcal{Y}^{*}_{l_{\lambda^\prime}m_{\lambda^\prime}}(\vec{p}^{\prime})\nonumber \\
&\times
\langle s(\vec{p}_{\Lambda}+\vec{p},s_{q^\prime})|\bar{s}\Gamma c|c(\vec{p},s_q)\rangle
\mathcal{L}^{l_{\lambda}+\frac{1}{2}}_{n_{\lambda}}\Bigg(\frac{p^2}{\beta^2}\Bigg)\mathcal{Y}_{l_\lambda m_\lambda}(\vec{p}),
\end{align}
where $\alpha^2 = \frac{\alpha_{\lambda}^2+\alpha^2_{\lambda^\prime}}{2\alpha^2_\lambda\alpha^2_{\lambda^\prime}}$. The Laguerre polynomials and the solid harmonics are functions of $k$ and $p_\Lambda$. The details of the semi-analytic calculations are given in appendix \ref{semianalytic}.

\section{Numerical Results}
\subsection{Form Factors}
The form factors in this work are calculated using the parameters 
for the quark model wave functions taken from \citep{RP}. The quark masses relevant for this calculation are shown in Tables \ref{hamiltonian:parameters}, while the wave function size parameters are shown in table \ref{size parameters}. The calculated form factors are parametrized to have the simple form 
\begin{equation}\label{ffpara}
F  = (a_0 + a_2 q^2+ a_4 q^4)
\exp\Bigg(\frac{-3m_\sigma^2}{2m_{\Lambda}^2}\frac{p_\Lambda^2}{\alpha_{\lambda\lambda^\prime}^2}\Bigg),
\end{equation} 
where $q^2$ is the momentum transfer $(p_{\Lambda_c}-p_\Lambda)^2$. $p_\Lambda$ is calculated in the rest frame of the parent $\Lambda_c$, and takes the form
\begin{equation}
p_\Lambda=\frac{1}{2m_{\Lambda_c}}\lambda^{1/2}(m_{\Lambda_c}^2,m_\Lambda^2,q^2).
\end{equation}
The parameters for the form factors we obtain 
are given in table \ref{form factors}.

\begin{table}[H]
\caption{Quark masses used in \citep{RP}.}
\label{hamiltonian:parameters}
\begin{center}
\begin{tabular}{|c|c|c|}
\hline
$m_q$ & $m_s$ & $m_c$ \\
GeV & GeV & GeV  \\
\hline
 0.2848 & 0.5553 & 1.8182 \\
\hline
\end{tabular}
\end{center}
\end{table}

\begin{table}[H]
\caption{Baryon masses and wave function size parameters, $\alpha_\lambda$ and $\alpha_\rho$ 
obtained from \citep{RP}. All values are in GeV.}
\label{size parameters}
\begin{center}
\begin{tabular}{|c|c|c|c|c|}
\hline
&\multicolumn{2}{c|}{Mass (GeV)}&\multicolumn{2}{c|}{Size parameters (GeV)}\\\cline{2-5}
State, $J^P$ & Experiment & Model & $\hphantom{abcd}\alpha_\lambda$ $\hphantom{abcd}$& $\alpha_\rho$\\
\hline
$\Lambda_c(2286)\frac{1}{2}^{+}$  & 2.29 & 2.27 & 0.424  & 0.393 \\
\hline
$\Lambda(1115)\frac{1}{2}^{+}$  & 1.12 & 1.10 & 0.387  & 0.372 \\
\hline
$\Lambda(1600)\frac{1}{2}_1^{+}$& 1.60 & 1.71 & 0.387 & 0.372 \\
\hline
$\Lambda(1405)\frac{1}{2}^{-}$  & 1.41 & 1.48 & 0.333 & 0.320 \\
\hline
$\Lambda(1520)\frac{3}{2}^{-}$  & 1.52 & 1.53 & 0.333 & 0.308 \\
\hline
$\Lambda(1890)\frac{3}{2}^{+}$  & 1.89 & 1.81 & 0.325 & 0.303 \\
\hline
$\Lambda(1820)\frac{5}{2}^{+}$  & 1.82 & 1.81 & 0.325 & 0.303 \\
\hline
\end{tabular}
\end{center}
\end{table}

\begin{figure}[t!]
\caption{Form factors plotted as functions of $q^2$ for transitions to (a) $\Lambda(1115)\frac{1}{2}^{+}$; (b) $\Lambda(1600)\frac{1}{2}^+$; (c) $\Lambda(1405)\frac{1}{2}^{-}$; (d) $\Lambda(1520)\frac{3}{2}^{-}$; (e) $\Lambda(1890)\frac{3}{2}^{+}$; (f) $\Lambda(1820)\frac{5}{2}^{+}$. }
\label{FF:1}
\begin{minipage}{0.5\textwidth}
\centering
\includegraphics[width=0.9\textwidth]{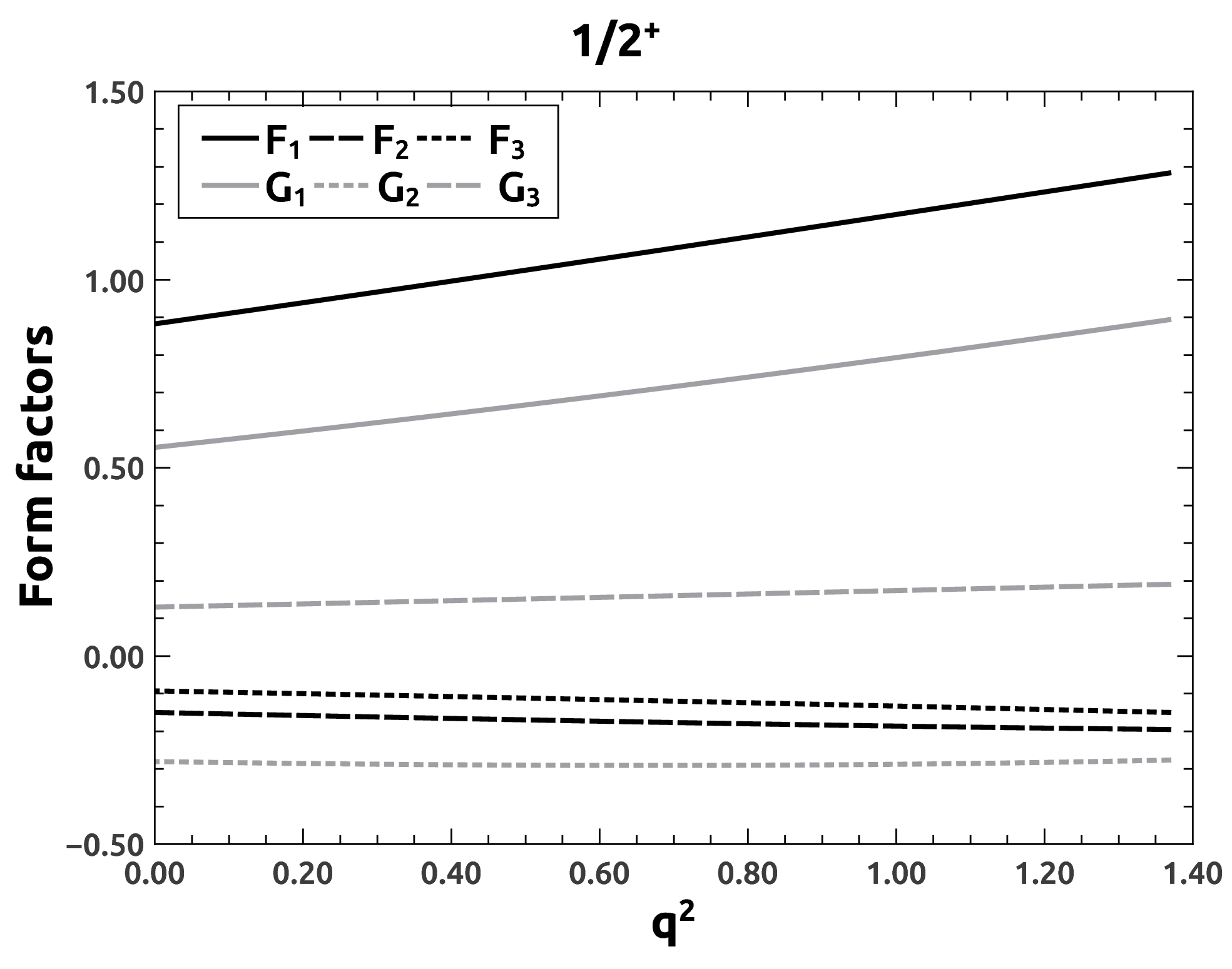}
\caption*{(a)}
\end{minipage}%
\begin{minipage}{0.5\textwidth}
\centering
\includegraphics[width=0.9\textwidth]{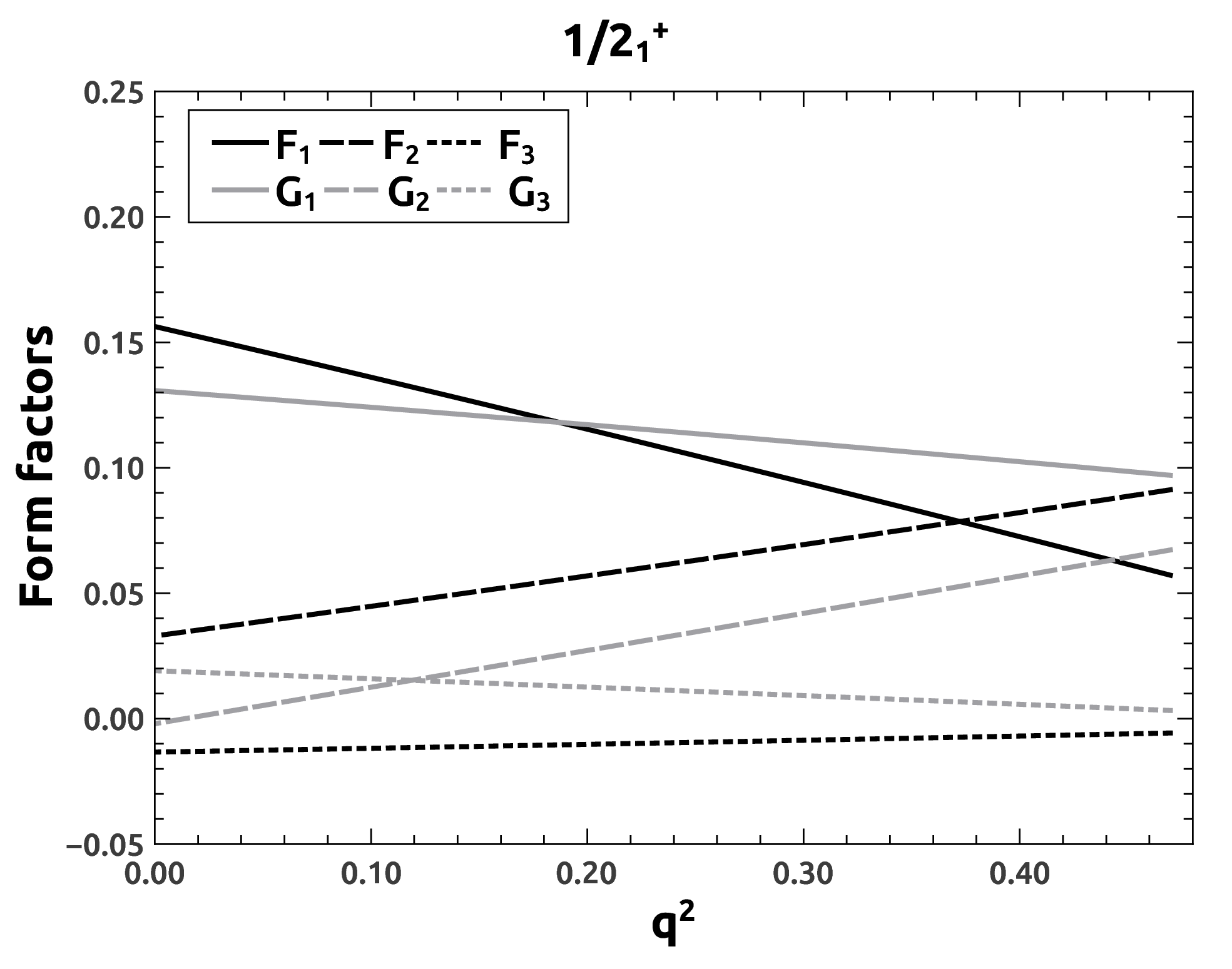}
\caption*{(b)}
\end{minipage}

\begin{minipage}{0.5\textwidth}
\centering
\includegraphics[width=0.9\textwidth]{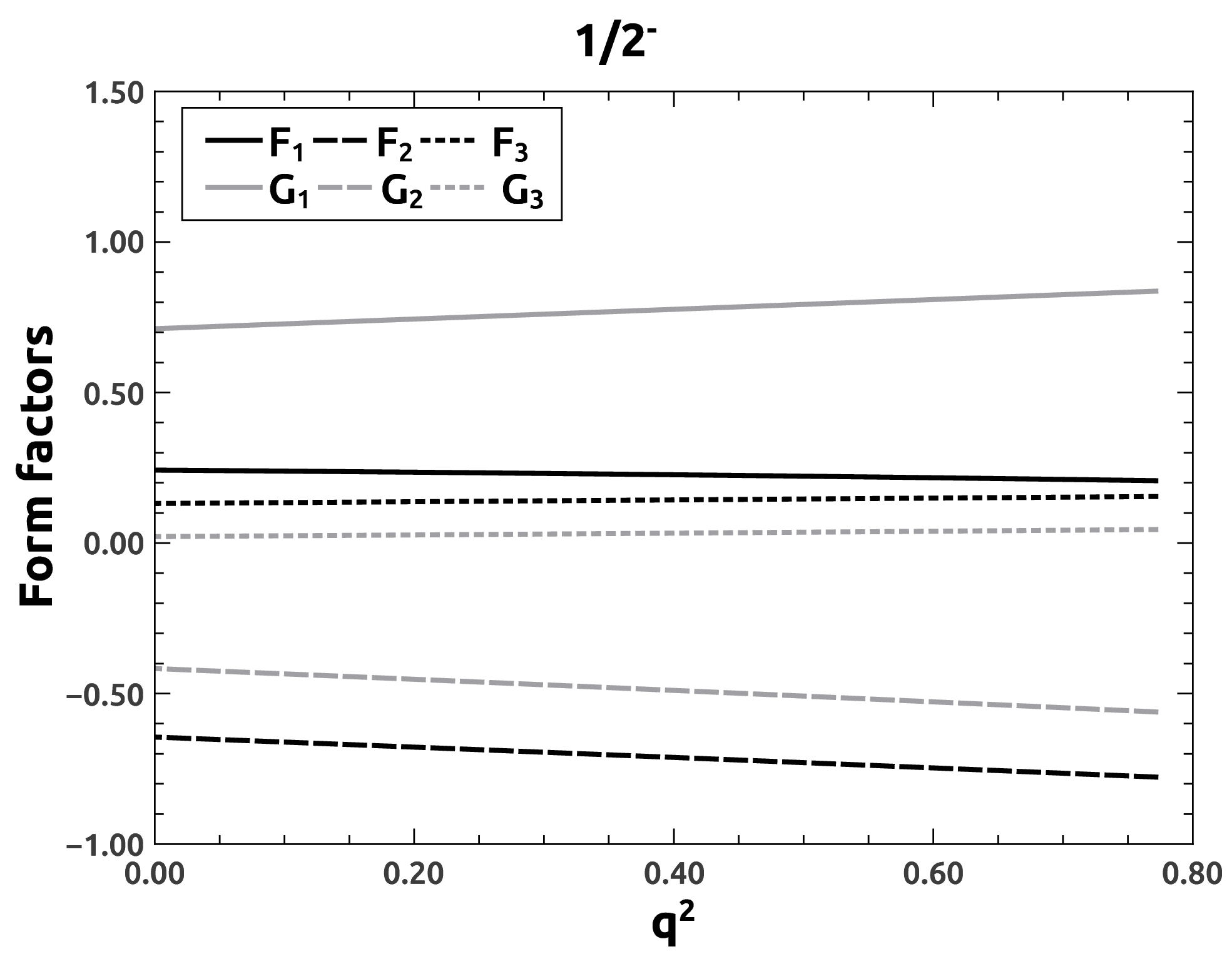}
\caption*{(c)}
\end{minipage}%
\begin{minipage}{0.5\textwidth}
\centering
\includegraphics[width=0.9\textwidth]{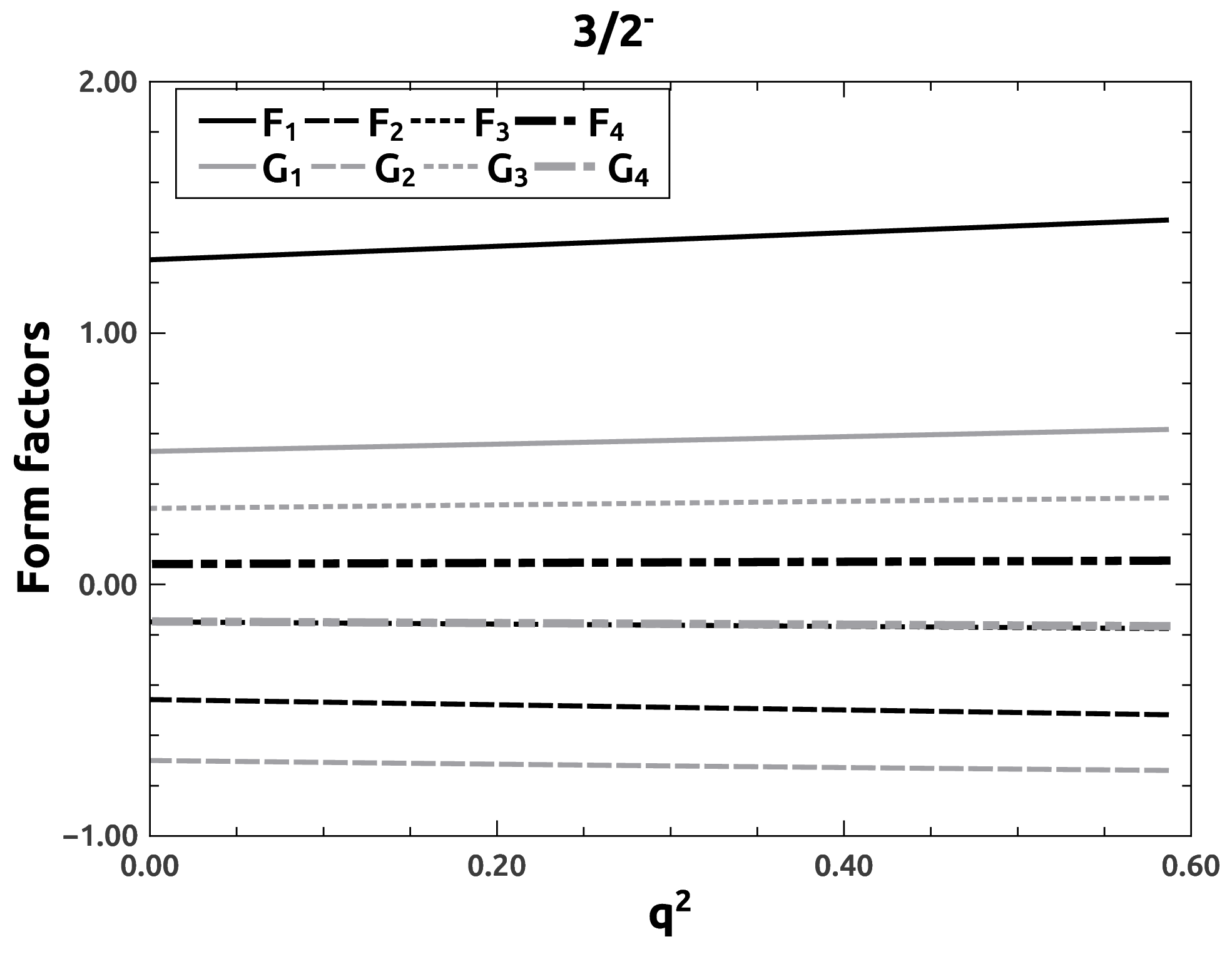}
\caption*{(d)}
\end{minipage}

\begin{minipage}{0.5\textwidth}
\centering
\includegraphics[width=0.9\textwidth]{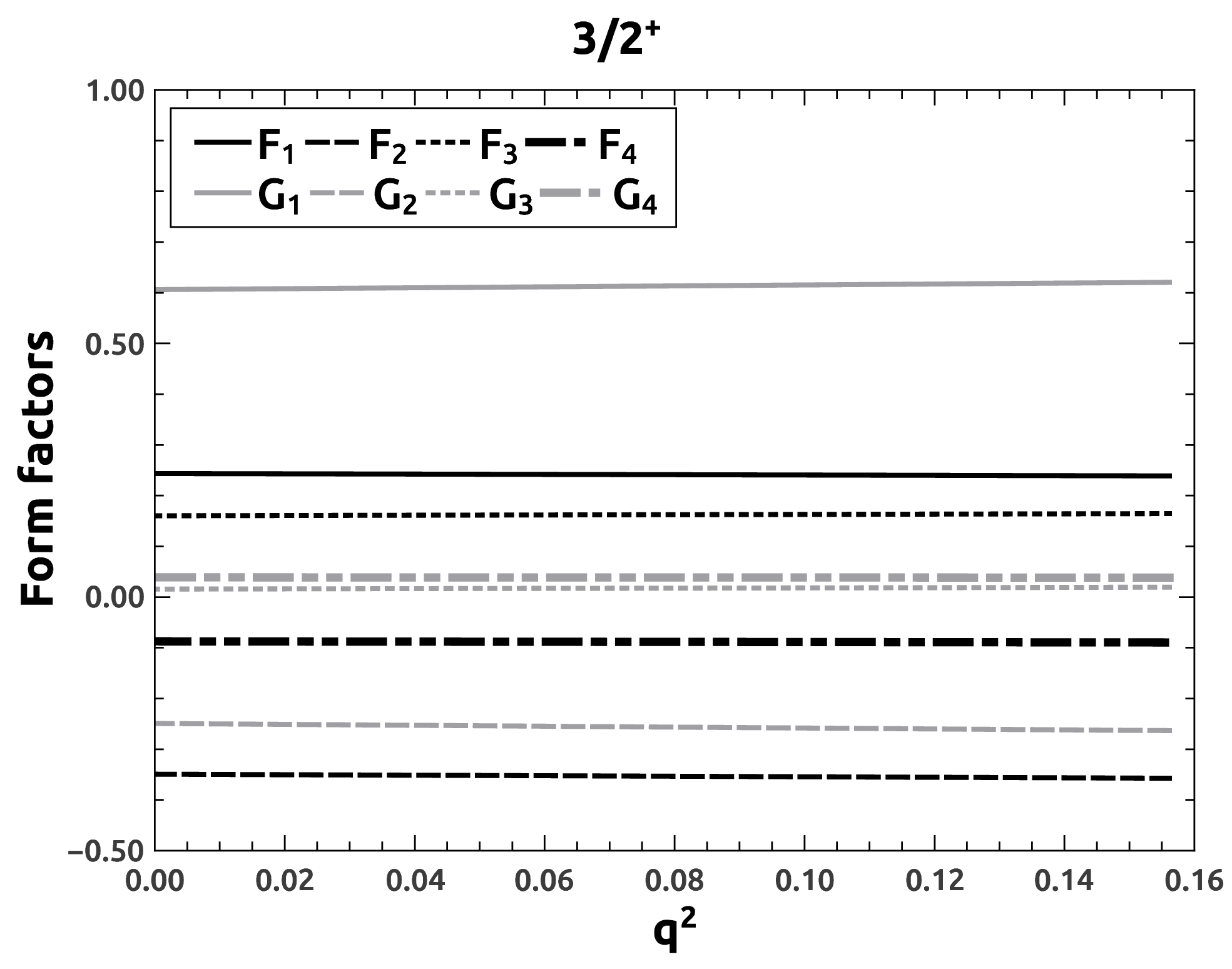}
\caption*{(e)}
\end{minipage}%
\begin{minipage}{0.5\textwidth}
\centering
\includegraphics[width=0.9\textwidth]{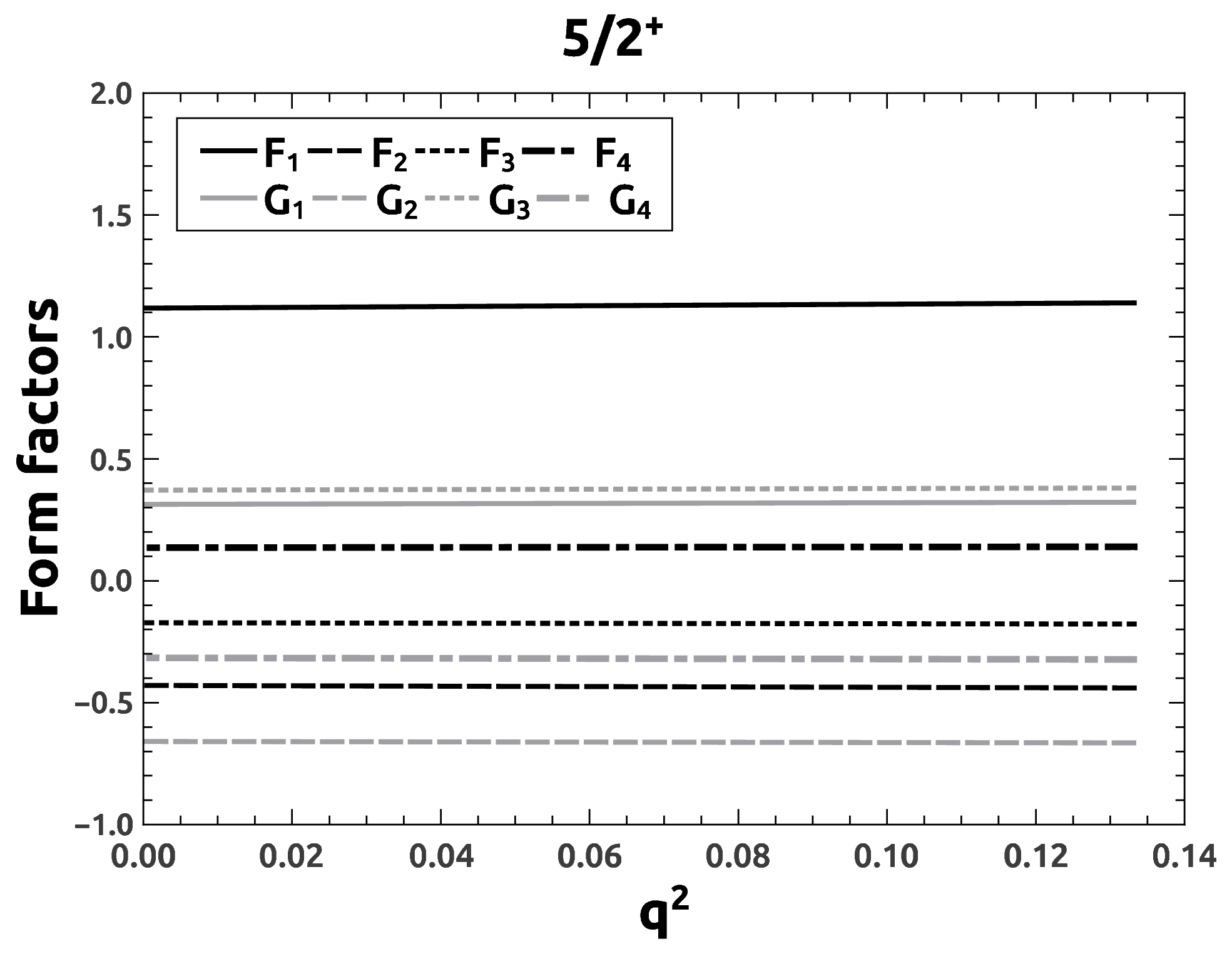}
\caption*{(f)}
\end{minipage}

\end{figure}

Figure \ref{FF:1} shows the form factors for the transitions to the ground state and the excited 
states that we consider. In the language of HQET, the form factors ($F_1$, $G_1$) associated with leading order in the $1/m_c$ expansion are dominant, while all of the others are smaller. With the exception of transitions to the $\Lambda(1600){1/2}^{+}$, all of the form factors have their largest absolute values at their respective non-recoil points.

\begin{table}[h!]
\caption{Coefficients in parametrization of the form factors, from eqn. \ref{ffpara}.}
\label{form factors}
\begin{center}
\begin{tabular}{|c|c|c|c|c|c|c|c|c|c|}
\hline
Transition & $a_n$($\text{GeV}^{-n}$) &$F_1$&$F_2$ &$F_3$ &$F_4$ &$G_1$ &$G_2$ &$G_3$ &$G_4$\\
\hline
\multirow{3}{*}{$\Lambda_c\to\Lambda(1115)$}
& $a_0$ & 1.382 & -0.235 & -0.146 & $-$ & 0.868 & -0.440 & 0.203 & $-$ \\
& $a_2$ & -0.073 & 0.022 & -0.003 & $-$ & 0.013 & -0.116 & -0.009 & $-$ \\
& $a_4$ & 0.000 &  0.006 & -0.001 & $-$ & 0.004 &  0.003 & 0.000 & $-$ \\
\hline
\multirow{3}{*}{$\Lambda_c\to\Lambda(1600)$}
& $a_0$ &  0.172 & 0.036  & -0.015 & $-$ &  0.144 & -0.002 &  0.021 & $-$\\
& $a_2$ & -0.257 & 0.121  &  0.020 & $-$ & -0.102 &  0.160 & -0.040 & $-$ \\
& $a_4$ & 0.025 & -0.008  &  -0.001 & $-$ & 0.005 &  -0.026 & 0.004 & $-$ \\
\hline
\multirow{3}{*}{$\Lambda_c\to\Lambda(1405)$}
& $a_0$ & 0.300 &  -0.797 & 0.162 & $-$ & 0.881 &  -0.516 & 0.027 & $-$\\
& $a_2$ &  -0.126 & 0.028 &  -0.010 & $-$ &  -0.058 &  -0.066 & 0.025 & $-$ \\
& $a_4$ & 0.008 &  -0.003 &  -0.000 & $-$ & 0.002 & 0.009 &  -0.001 & $-$ \\
\hline
\multirow{3}{*}{$\Lambda_c\to\Lambda(1520)$}
& $a_0$ &  1.496 & -0.530 & -0.172 & 0.094 & 0.613 & -0.810 &  0.351 & -0.170 \\
& $a_2$ & -0.080 &  0.019 & -0.005 & 0.001 & 0.005 &  0.122 & -0.010 &  0.008 \\
& $a_4$ &  0.002 &  0.003 & -0.001 & 0.000 & 0.002 & -0.001 &  0.000 &  0.000 \\
\hline
\multirow{3}{*}{$\Lambda_c\to\Lambda(1890)$}
& $a_0$ &  0.251 & -0.358 &  0.165 & -0.090 &  0.625 & -0.257 &  0.016 &  0.040\\
& $a_2$ & -0.079 & -0.107 & -0.006 &  0.004 & -0.030 & -0.041 &  0.021 & -0.011 \\
& $a_4$ &  0.005 &  0.950 &  0.000 & -0.000 &  0.001 &  0.006 & -0.001 &  0.001 \\
\hline
\multirow{3}{*}{$\Lambda_c\to\Lambda(1820)$}
& $a_0$ & 1.148 &  -0.441 &  -0.177 & 0.139 & 0.322 &  -0.677 &   0.381 &  -0.325\\
& $a_2$ &-0.059 &   0.008 &  -0.005 & 0.001 & 0.002 &   0.089 &  -0.008 &  0.013 \\
& $a_4$ & 0.002 &   0.001 &  -0.001 & 0.000 & 0.000 &  -0.002 &  0.000 &  0.000 \\
\hline
\end{tabular}
\end{center}
\end{table}

\subsection{Comparison with HQET}

\begin{figure}[t!]
\caption{Vector and axial vector form factors obtained using HQET, for the states that we treat in this work. The graphs shown are for (a) $\Lambda(1115){1/2}^{+}$; (b) $\Lambda(1600){1/2}^{+}$; (c) $\Lambda(1405){1/2}^{-}$; (d) $\Lambda(1520){3/2}^{-}$; (e) $\Lambda(1890){3/2}^{+}$; (f) $\Lambda(1820){5/2}^{+}$.}
\label{HQET1}
\begin{minipage}{0.5\textwidth}
\centering
\includegraphics[width=0.9\textwidth]{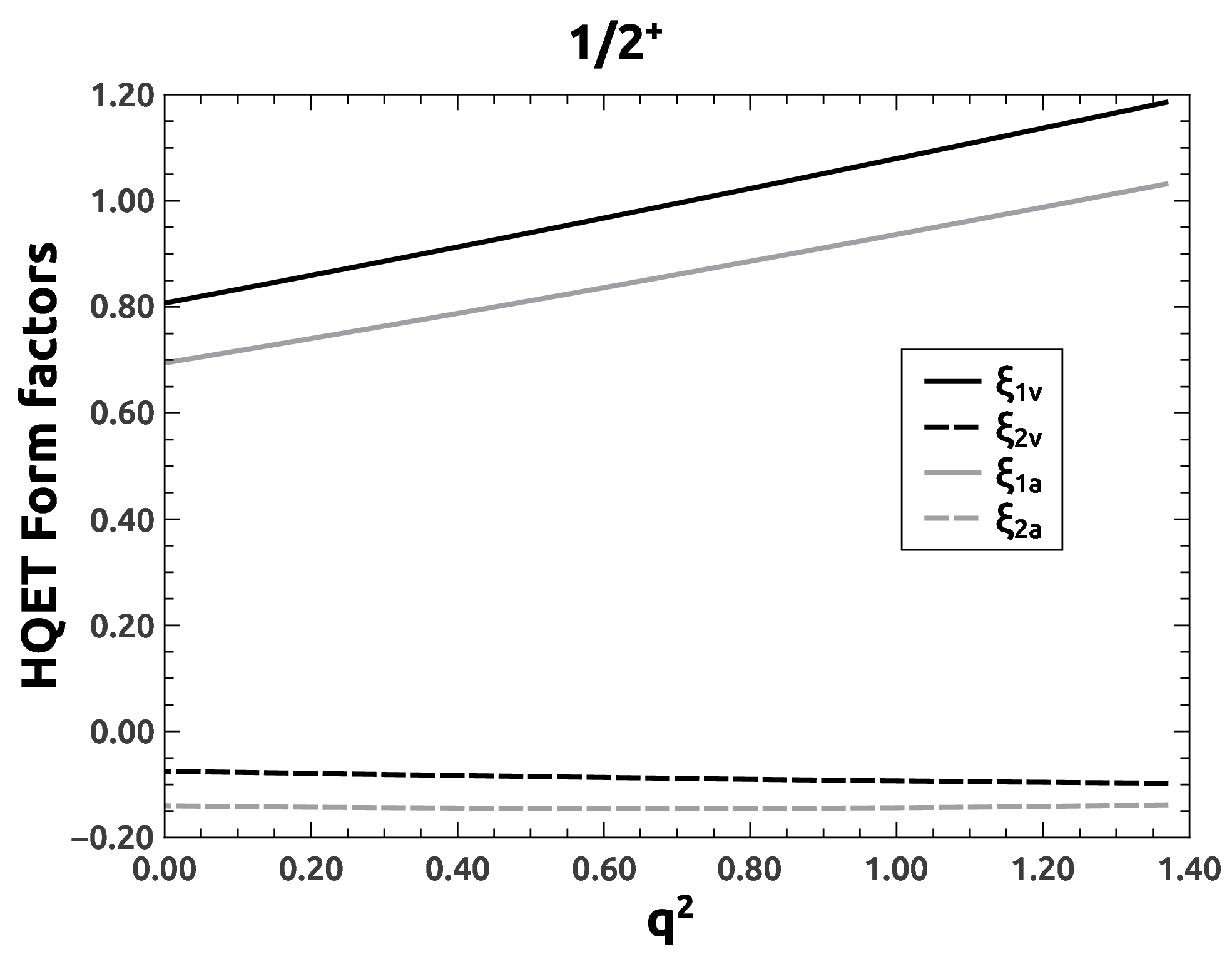}
\caption*{(a)}
\end{minipage}%
\begin{minipage}{0.5\textwidth}
\centering
\includegraphics[width=0.9\textwidth]{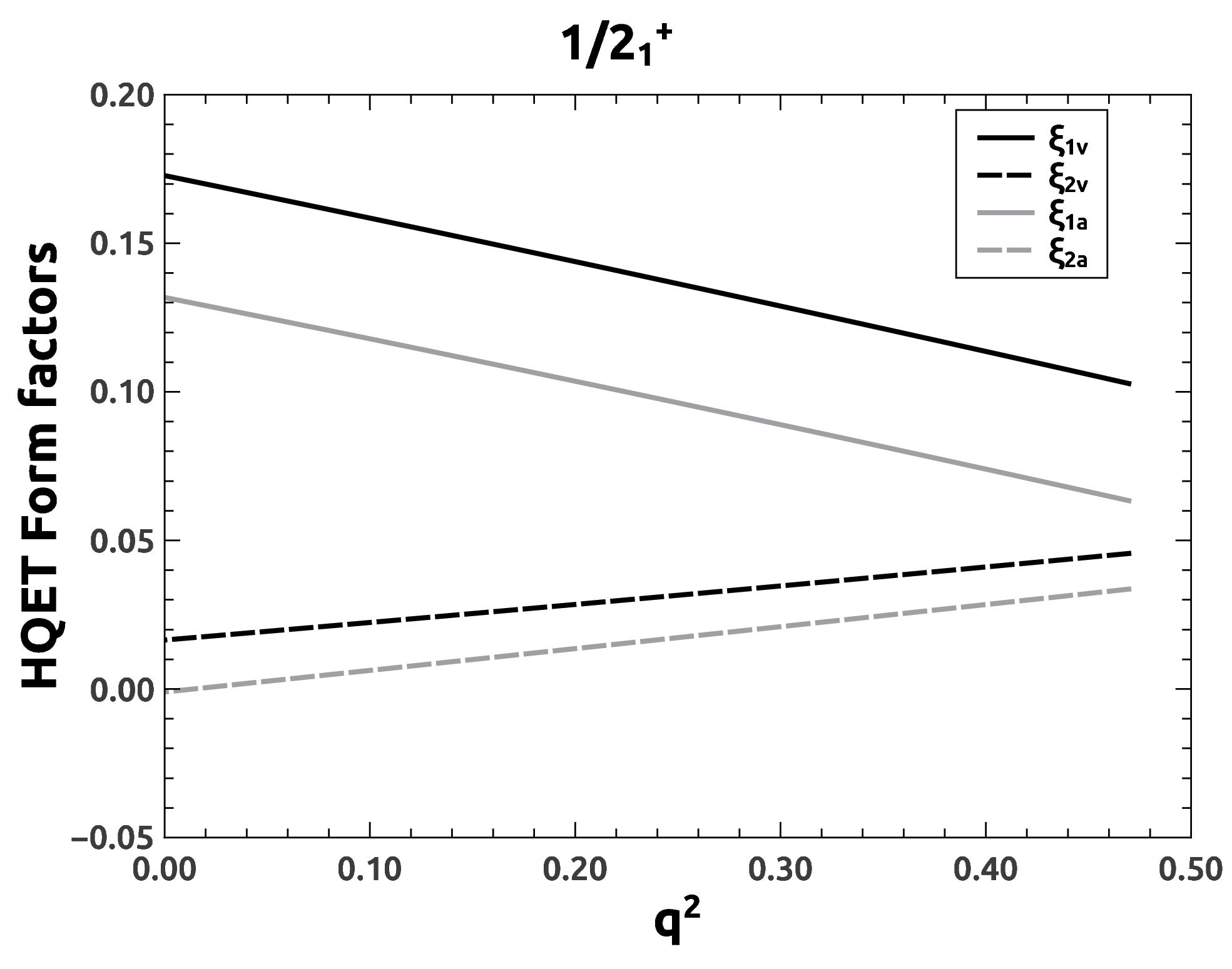}
\caption*{(b)}
\end{minipage}

\begin{minipage}{0.5\textwidth}
\centering
\includegraphics[width=0.9\textwidth]{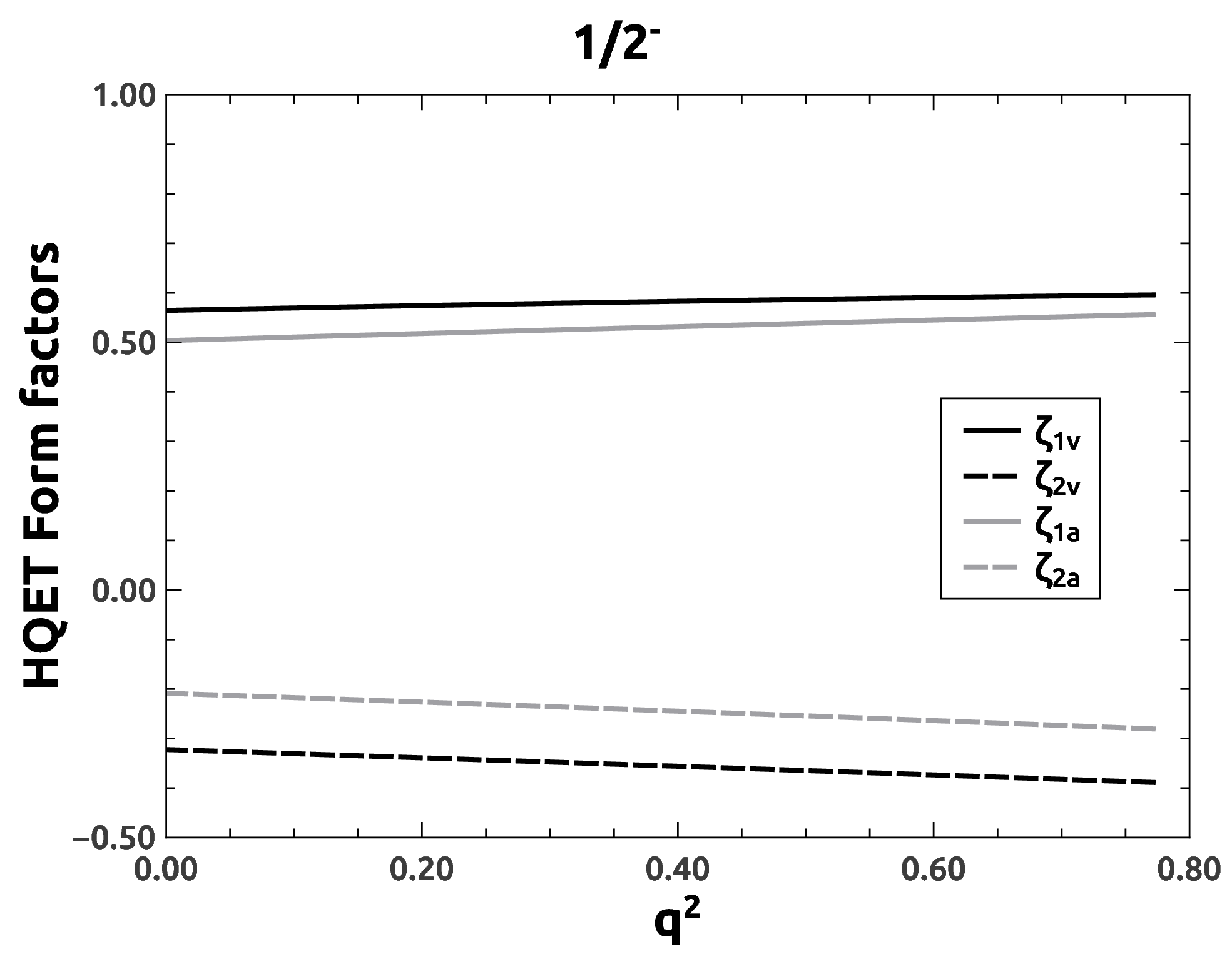}
\caption*{(c)}
\end{minipage}%
\begin{minipage}{0.5\textwidth}
\centering
\includegraphics[width=0.9\textwidth]{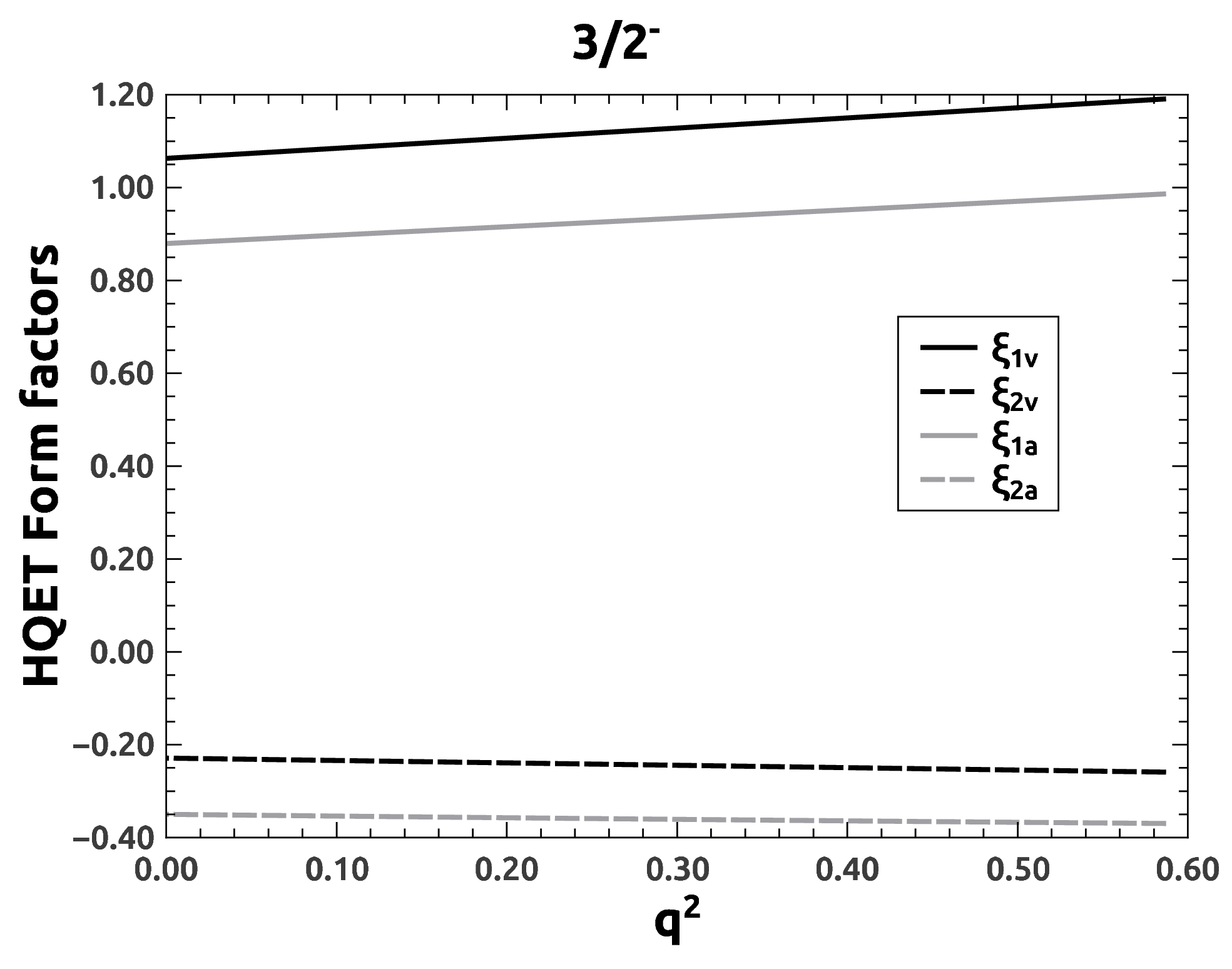}
\caption*{(d)}
\end{minipage}

\begin{minipage}{0.5\textwidth}
\centering
\includegraphics[width=0.9\textwidth]{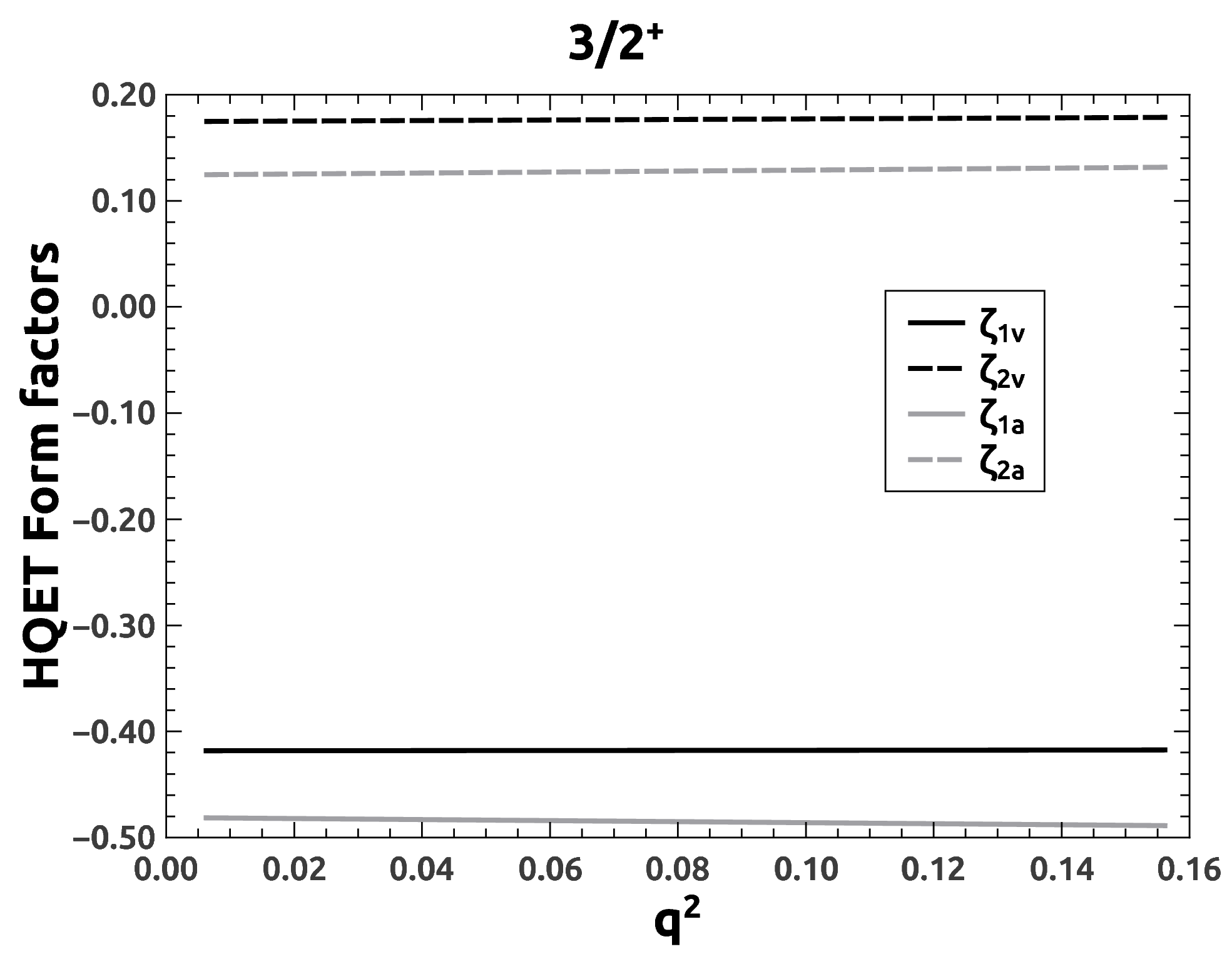}
\caption*{(e)}
\end{minipage}%
\begin{minipage}{0.5\textwidth}
\centering
\includegraphics[width=0.9\textwidth]{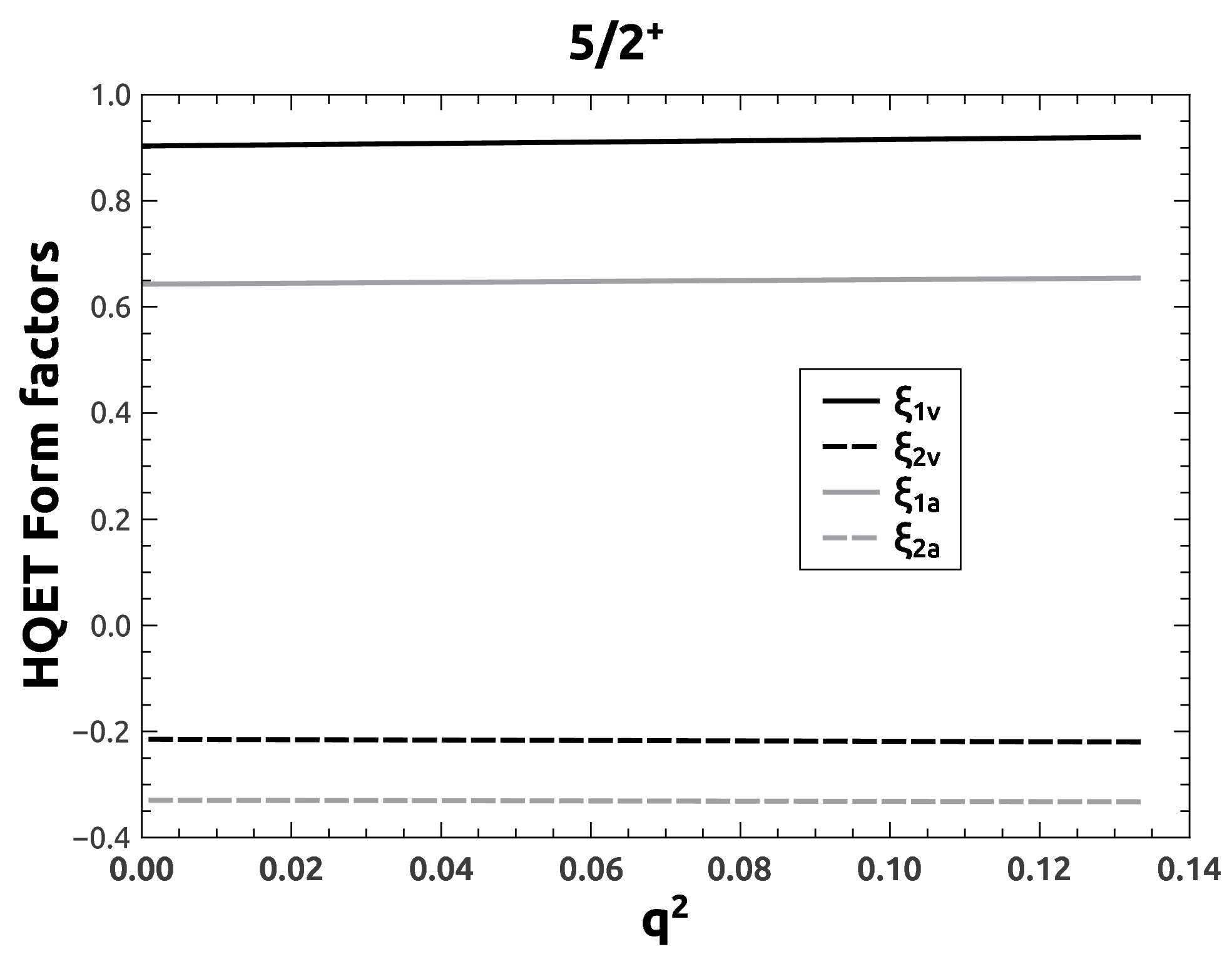}
\caption*{(f)}
\end{minipage}

\end{figure}

In Section \ref{HQETformfactors}, we obtained expressions for the general transition form factors in terms of the leading order HQET form factors. Those expressions can be inverted to write the HQET form factors in terms of the general ones. Since the pair of leading order HQET form factors are valid for both the vector and axial-vector hadronic matrix elements, we can extract them from both sets of general form factors. The expressions for $\xi_1$ and $\xi_2$ are shown in table \ref{hqetexpr}, and the curves are shown in Fig. \ref{HQET1}.

\begin{table}[h!]
\caption{The leading order HQET form factors in terms of the general form factors. The second and third columns show the expressions in terms of the vector form factors, while the fifth and sixth columns show them in terms of the axial-vector form factors. The fourth and seventh columns show the ratios $\xi_2/\xi_1 \,\,(\zeta_2/\zeta_1)$ calculated at the non-recoil point.}
\label{hqetexpr}
\begin{center}
\begin{tabular}{|r|c|c|c|c|c|c|}\hline
State, $J^P$  &\multicolumn{3}{c|}{Vector} &\multicolumn{3}{c|}{Axial Vector} \\\cline{2-7}
& $\xi_1$ ($\zeta_1$)&$\xi_2$ ($\zeta_2$)&$\xi_2/\xi_1$ ($\zeta_2/\zeta_1$)&
$\xi_1$ ($\zeta_1$)&$\xi_2$ ($\zeta_2$)&$\xi_2/\xi_1$ ($\zeta_2/\zeta_1$)\\\hline
$\Lambda(1115),\,\,\frac{1}{2}^+$&$F_1+F_2/2$&$F_2/2$ &-0.093& $G_1-G_2/2$& $G_2/2$ & -0.202\\\hline
$\Lambda(1600),\,\,\frac{1}{2}^+$&& &0.095& &  & -0.007\\\hline
$\Lambda(1405),\,\,\frac{1}{2}^-$& $-F_1+F_2/2$&$-F_2/2$ &-0.571&$-G_1-G_2/2$& $-G_2/2$ & -0.414\\\hline
$\Lambda(1520),\,\,\frac{3}{2}^-$&$F_1+F_2/2$&$F_2/2$ &-0.215& $G_1-G_2/2$& $-G_2/2$ & -0.398\\\hline
$\Lambda(1890),\,\,\frac{3}{2}^+$ & $-F_1+F_2/2$&$-F_2/2$&-0.416&$-G_1-G_2/2$& $-G_2/2$ & -0.259\\\hline
$\Lambda(1820),\,\,\frac{5}{2}^+$&$F_1+F_2/2$&$F_2/2$ &-0.238& $G_1-G_2/2$& $G_2/2$ & -0.512\\\hline
\end{tabular}
\end{center}
\end{table}
 
The leading order HQET expectation is that the extraction of $\xi_1$ and $\xi_2$ should be independent of whether they are extracted from axial or vector form factors. However, the curves we obtain indicate that there is some sensitivity to which set of form factors is used. This sensitivity can be attributed to the fact that our form factors include effects that arise in all orders of $1/m_c$, while the relationships between the $\xi_i \,\,(\zeta_i)$ and the $F_i$ and $G_i$ are obtained at leading order. Higher order terms in the $1/m_c$ expansion will modify the expressions shown in eqns. \ref{xixi:1} - \ref{xixi:5}, and hence the inverted relationships.

\subsection{Decay Widths}

\subsubsection{$\Lambda_c^+\to\Lambda^{(*)}l^+\nu_l$}

\begin{figure}[t!]
\caption{Differential decay rates $d\Gamma/dq^2$ (in units of ${\rm s}^{-1}{\rm GeV}^{-2}$) for the semileptonic decays $\Lambda_c^{+}\to\Lambda^{(*)}l^{+}\nu_l$. (a) shows the decay rate for all states considered, while (b) shows the decay rates for the excited states only.}
\label{Semidecay}
\includegraphics[width=0.6\textwidth]{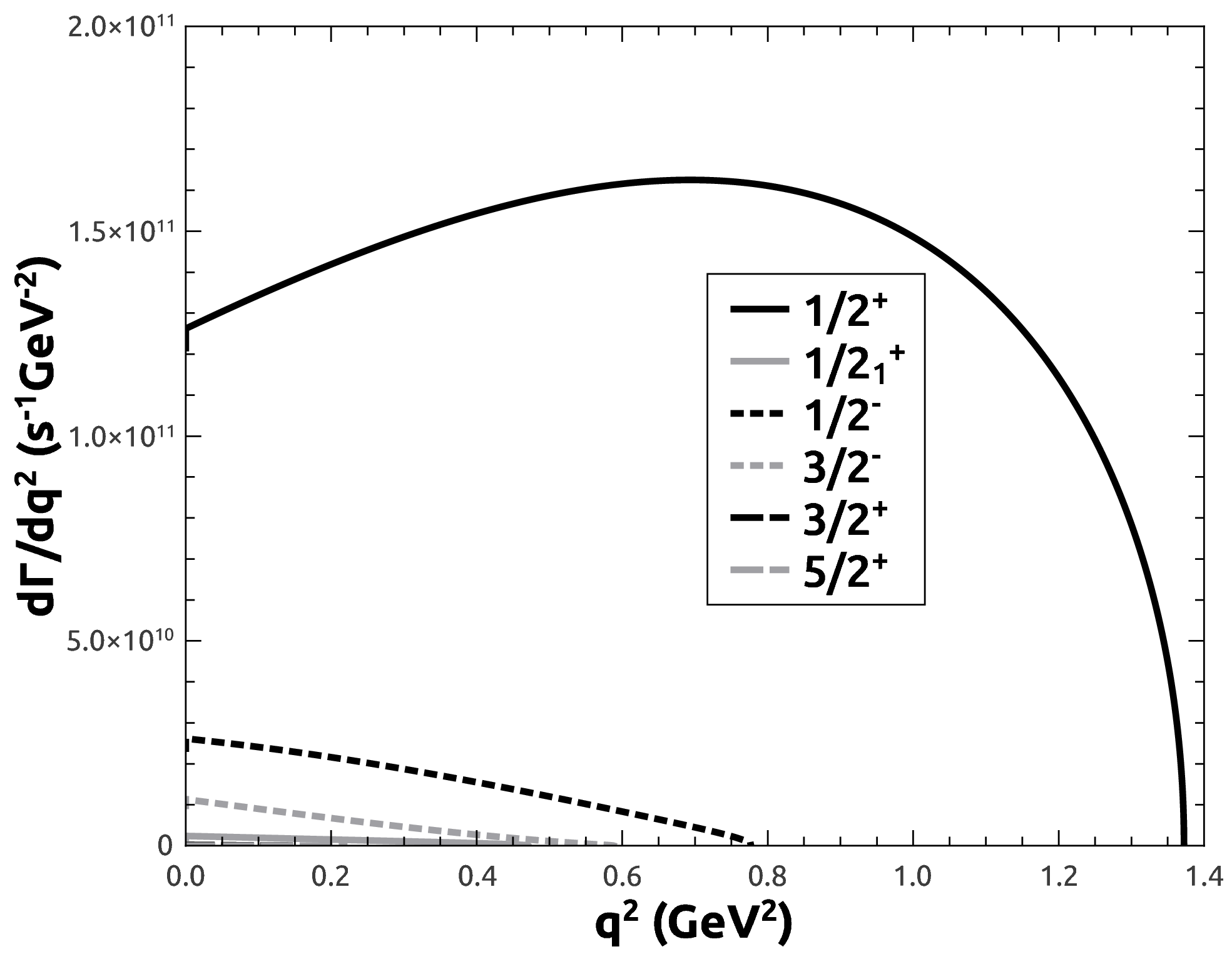}
\caption*{(a)}
\includegraphics[width=0.6\textwidth]{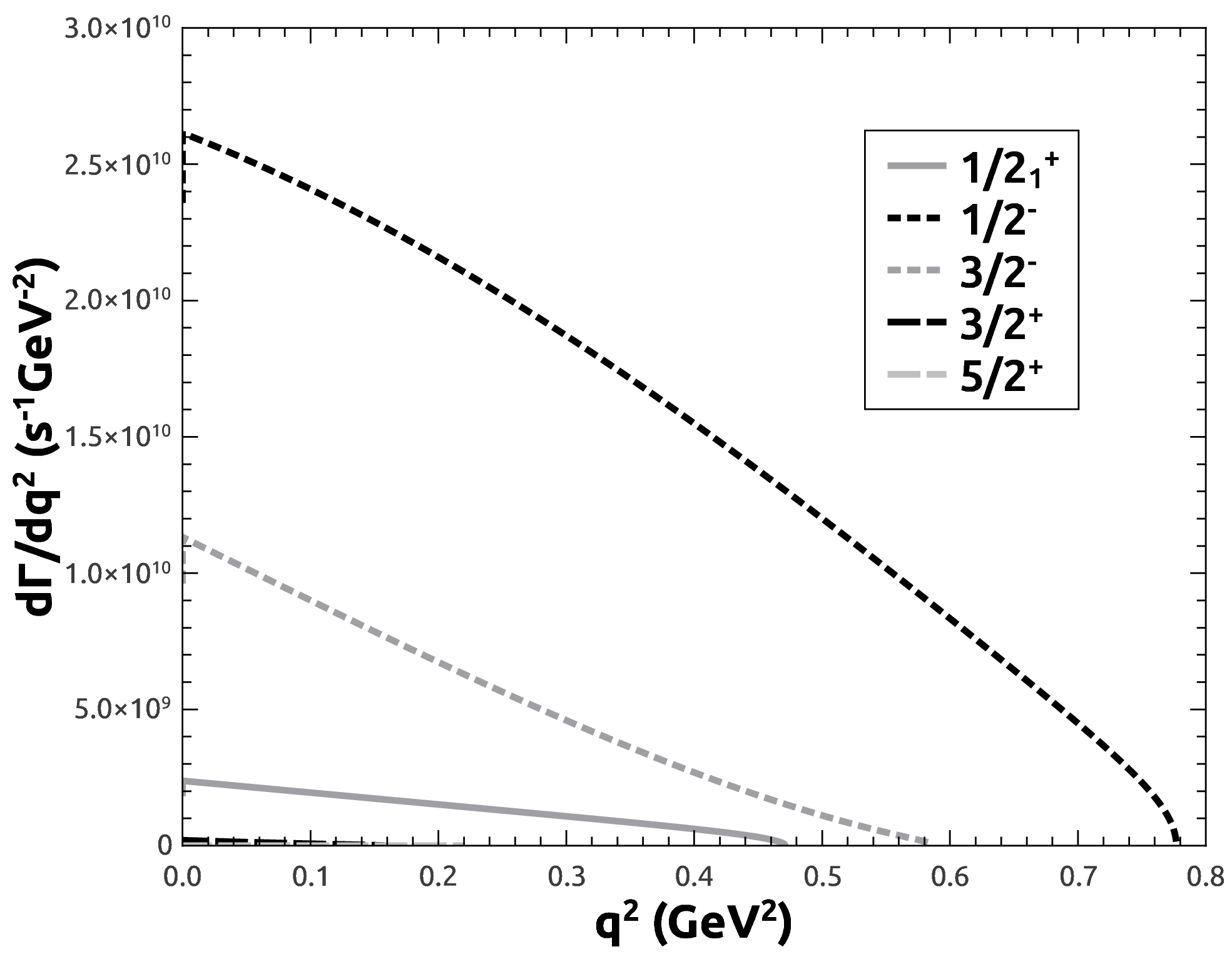}
\caption*{(b)}
\end{figure}

The differential decay rates, $d\Gamma/dq^2$ (in ${\rm s}^{-1}{\rm GeV}^{-2}$), for the semileptonic decays $\Lambda_c^{+}\to\Lambda^{(*)}l\nu_l$ are shown in figure 
\ref{Semidecay}. Fig \ref{Semidecay}(a) shows the decay rates for the transition to the elastic channel (the ground state) as well as to the excited states that we consider.
The elastic channel is dominant but the decay rates for the decays to $1/2^{-}$ and $3/2^{-}$ are significant. Fig \ref{Semidecay}(b) shows an enlarged version of the decay rates to the excited states. This figure shows that the rates for decays to the radially excited ${1/2}^{+}$, the 
$3/2^{+}$ and the $5/2^{+}$ states are small compared to the rates for the $1/2^{-}$ and $3/2^{-}$ states.

The integrated total decay widths that we obtain for $\Lambda_c^{+}\to\Lambda^{(*)}l^{+}\nu_l$ are shown in table \ref{SemidecayResult}. Also shown are the results presented in PRCI. The calculated total decay widths to 
the elastic channel are $1.92\times 10^{11}s^{-1}$ for $l=e$, and $1.86\times 10^{11}s^{-1}$ for $l=\mu$.
The branching fractions calculated are 
$\frac{\Gamma(\Lambda_c^{+}\to\Lambda l^{+}\nu_l)}{\Gamma_{\Lambda_c}} = 3.84\%$ for the electron channel, and
3.72\% for the muon channel. $\Gamma_{\Lambda_c}$ is the total decay width of the $\Lambda_c$. Table \ref{ExpResult} compares our results with other 
theoretical estimates \citep{CQM,LCSR,Oset} and the experimental results from the Belle \citep{ZUPANC} 
and BESIII \citep{BESIII,BESIII3} collaborations. 
Our results are in very good agreement with the most recent experimental result from BESIII. 

From table \ref{SemidecayResult}, it is evident that the elastic channel dominates the semileptonic decay rate of the $\Lambda_c$ but does not saturate it. We find that the branching fraction to the $\Lambda^{*}(1405)$ state with $J^P=1/2^{-}$ is $6\%$ of the total semileptonic decay, while the branching fraction to $\Lambda^{*}(1520)$ is $1\%$ of the total. Decays to the other states we consider are significantly smaller.

\begin{table}[h!]
\caption{Integrated total decay widths for $\Lambda_c^{+}\to\Lambda^{(*)}l^{+}\nu_l$ in units of $10^{11} s^{-1}$, for the states we consider in this work. Also shown are the results obtained in PRCI. The last row shows the branching fraction of the elastic decay channel, where $\Gamma_{\text{Total}}$ is the total semileptonic decay width assuming the decays shown in the table saturate the semileptonic decays.}
\label{SemidecayResult}
\begin{tabular}{|P{1cm}|P{2cm}|P{2.5cm}|P{2.5cm}|P{2.5cm}|}
\hline
\multirow{3}{*}{Spin} & \multirow{3}{*}{Mass (GeV)} & \multicolumn{3}{c|}{Model estimates}  \\
\cline{3-5}
& & \multicolumn{2}{c|}{This Work} & PRCI \citep{PRC}\\
\cline{3-4}
& &$\Lambda^{+}_c\to\Lambda^{*}e^{+}\nu_e$ & $\Lambda^{+}_c\to\Lambda^{*}\mu^{+}\nu_\mu$ & \\
\hline
$\frac{1}{2}^{+}$& 1.115  	&  $ 1.92$ 						& 1.86 	&  $2.10$ \\
\hline
$\frac{1}{2_1}^{+}$& 1.600  &  $ 0.63\times 10^{-2}$   & $0.55	\times 10^{-2}$ 	&  $2.00\times 10^{-2}$ \\
\hline
$\frac{1}{2}^{-}$& 1.405 	&  $ 0.12$ 						& 0.11 	&  $0.19$ \\
\hline
$\frac{3}{2}^{-}$& 1.519 	&  $ 2.97\times 10^{-2}$ 	& $2.6\times 10^{-2}$	&  $5.00\times 10^{-2}$\\
\hline
$\frac{3}{2}^{+}$& 1.890 	&  $ 1.58\times 10^{-4}$ 	& $1.01\times 10^{-4}$& $--$ \\
\hline
$\frac{5}{2}^{+}$& 1.820 	&  $ 0.66\times 10^{-4}$ 	& $0.42\times 10^{-4}$& $--$ \\
\hline
\multicolumn{2}{|c|}{ $\text{Total}$}			   		&	$2.08$ 						& 2.00 	& $2.36$ \\
\hline
\multicolumn{2}{|c|}{ $\Gamma_{\Lambda_c^{+}\to\Lambda l^{+}\nu_l} / \Gamma_{\text{total}}$}  & 0.92 		& 0.93 & 0.89 \\
\hline
\end{tabular}
\end{table}

\begin{table}[h!]
\caption{Branching fractions of the semileptonic decay $\Lambda_c^{+}\to\Lambda(1115)l^+\nu_l$,
compared with other theoretical estimates and experimental results. In the table, CQM refers to the covariant quark model of reference \cite{CQM}, while LCSR refers to the light cone sum rules of reference \cite{LCSR}.}
\label{ExpResult}
\begin{tabular}{|P{2.5cm}|P{1.5cm}|P{1.5cm}|P{2.5cm}|P{2.5cm}|P{2cm}|P{2cm}|}
\hline
Branching & \multicolumn{4}{c|}{Model estimates ($\%$)}&\multicolumn{2}{c|}{Experimental results($\%$)}\\
\cline{2-7}
 fraction & This work & PRCI \citep{PRC} & CQM \cite{CQM} & LCSR \cite{LCSR} & Belle \cite{ZUPANC} & BESIII \cite{BESIII,BESIII3} \\
\hline
$\Gamma_{\Lambda^{+}_c\to\Lambda e^{+}\nu_e}/\Gamma_{\Lambda_c^{+}}$ & 3.84 & \multirow{2}{*}{4.2} & 2.78 
& $3.05\pm 0.27$ &$2.9\pm 0.5$ & $3.63\pm 0.38\pm 0.20$\\
\cline{1-2}\cline{4-7}
$\Gamma_{\Lambda^{+}_c\to\Lambda \mu^{+}\nu_\mu}/ \Gamma_{\Lambda_c^{+}}$ & 3.72 &  & 2.69  & $1.96\pm 0.32$
& $2.7\pm 0.6$ & $3.49\pm 0.46\pm 0.27$ \\
\hline
\end{tabular}
\end{table}

\subsubsection{$\Lambda_c^+\to\Sigma\pi l^+\nu_l$ and $\Lambda_c^+\to N\bar{K} l^+\nu_l$}

Table \ref{Strongdecay} lists the states that we have included in this study, their total and partial widths in the $\Sigma\pi$ and $N\bar{K}$ channels, and the corresponding strong coupling constants,
($g_{\Lambda\Sigma\pi}$, $g_{\Lambda N\bar{K}}$). The $\Lambda(1405)$ lies just below the $N\bar{K}$ threshold, so its coupling to this channel must be estimated by other means. We use the value estimated by Schat, Scoccola and Gobbi \citep{Gobbi}, but also explore the effects on the decay rate of allowing departures from their value.

\begin{table}[t!]
\caption{Parameters of the excited $\Lambda$ states that we use in our study. Shown are their total decay widths, and their partial decay widths and the strong couplings for the decays $\Lambda^{*}\to \Sigma\pi$ and $\Lambda^{*}\to N\bar{K}$.}
\label{Strongdecay}
\begin{tabular}{|P{2cm}|P{2cm}|P{2cm}|P{2cm}|P{2cm}|P{2cm}|P{2cm}|}
\hline
\multirow{2}{*}{Spin of $\Lambda^{(*)}$ } & \multirow{2}{*}{Mass(GeV)} &
Total width
& \multicolumn{2}{c|}{Partial width (MeV)} & \multicolumn{2}{c|}{Strong coupling constant}\\
\cline{4-7}
& &(MeV) & $\Gamma_{\Lambda^{*}\to\Sigma\pi}$ &$\Gamma_{\Lambda^{*}\to N\bar{K}}$ & $g_{\Lambda\Sigma\pi}$ & $g_{\Lambda N\bar{K}}$\\
\hline
$\frac{1}{2}^{+}$  & 1.115 & - & - & - &  15.73 &  14.03\\
\hline
$\frac{1}{2}^{+}_1$& 1.600 & 150 & 52.5 & 33.8 & 8.21 & 5.76 \\
\hline
$\frac{1}{2}^{-}$  & 1.405 & 50.5 & 50.5 & - & 1.57  & 1.90 \\
\hline
$\frac{3}{2}^{-}$  & 1.519 & 15.7 & 6.6 & 7.1 & 3.64 & 15.38 \\
\hline
$\frac{3}{2}^{+}$  & 1.890 & 100 & 6.5 & 27.5 & 0.14 & 1.05 \\
\hline
$\frac{5}{2}^{+}$  & 1.820 & 80.0 & 8.8 & 48.0 & 0.40 & 8.45\\
\hline
\end{tabular}
\end{table}

\begin{figure}[t!]
\caption{(a) The differential decay rates $d\Gamma/dq^2$ for the four-body decay 
$\Lambda^{+}_c\to \Lambda^{(*)}l^{+}\nu_l\to \Sigma\pi l^{+}\nu_l$.
(b) Comparison of the decay rates of the four-body semileptonic decay to the elastic channel $\Lambda_c^{+}\to\Lambda l^{+}\nu_l$.}
\label{DWSigmaPi}
\includegraphics[width=0.6\textwidth]{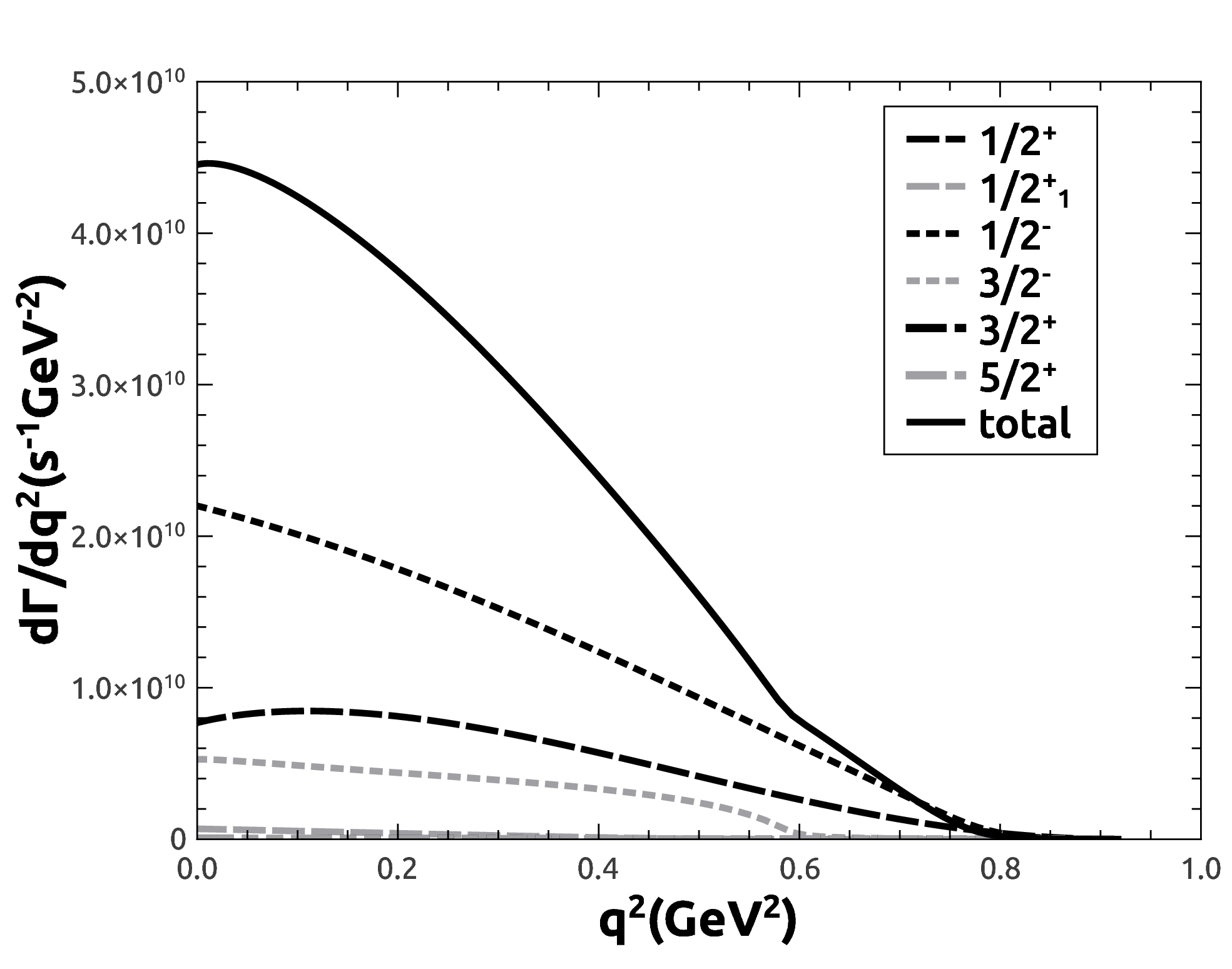}
\caption*{(a)}

\includegraphics[width=0.6\textwidth]{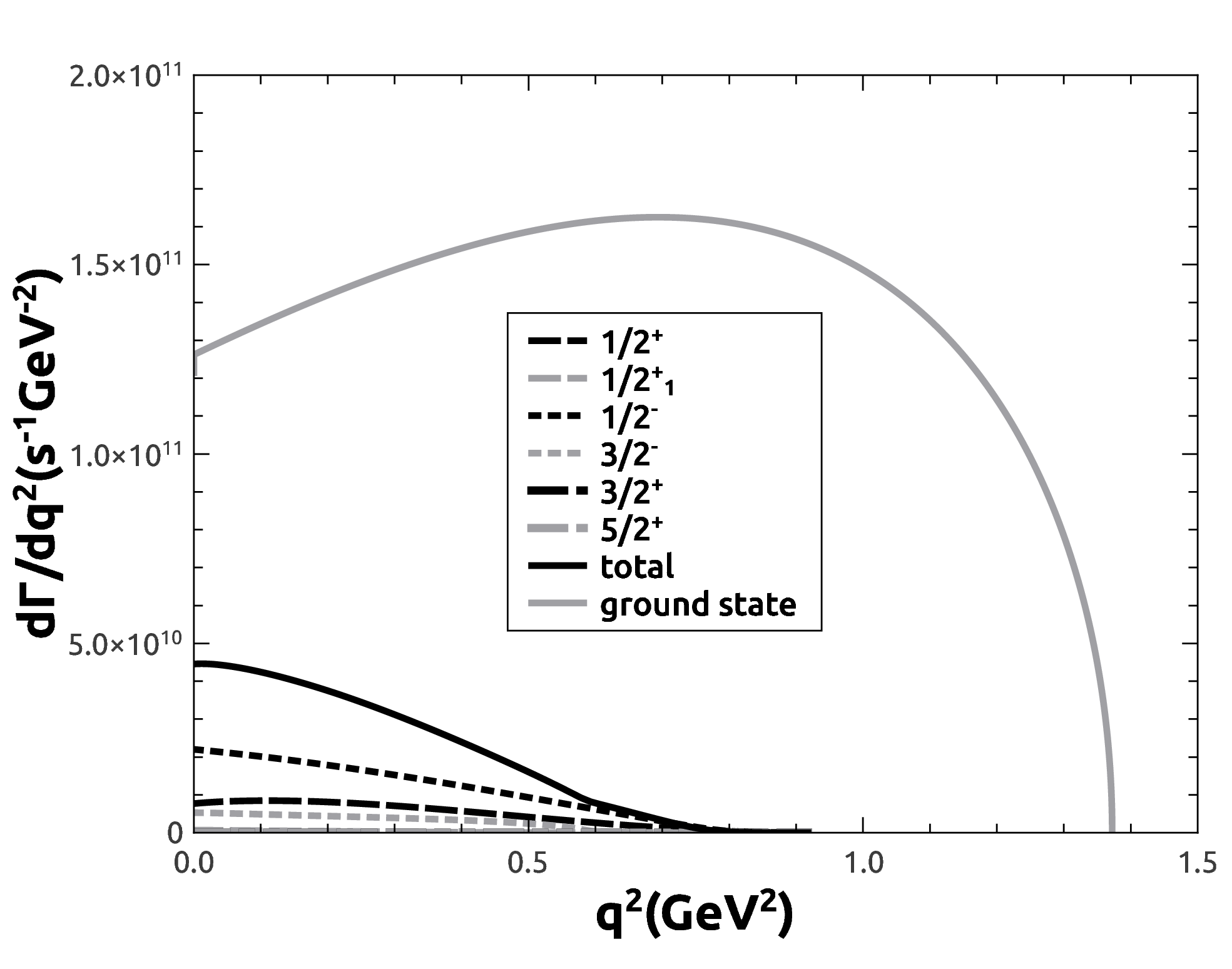}
\caption*{(b)}
\end{figure}

\begin{figure}[t!]
\caption{The differential decay rates $d\Gamma/dS_{\Sigma\pi}$ for decays via the states we consider, along with the coherent total. The black solid curve shows the coherent total decay rate for the $\Sigma\pi l\nu$ final state.}
\label{DWSigmaPi2}
\begin{center}
\includegraphics[width=0.6\textwidth]{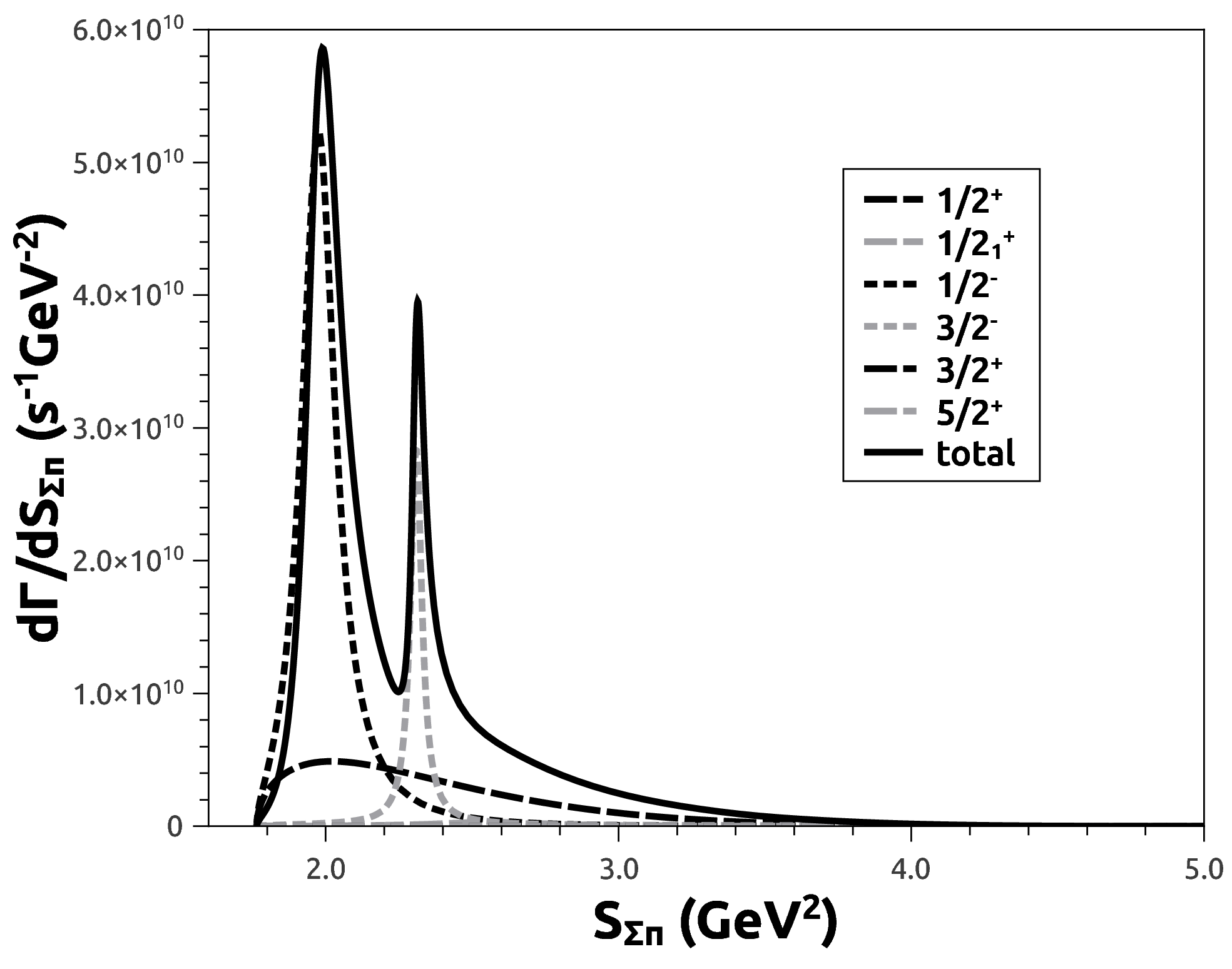}
\end{center}
\end{figure}

Figures \ref{DWSigmaPi} and \ref{DWSigmaPi2} show the differential decay rates $d\Gamma/dq^2$ and $d\Gamma/dS_{\Sigma\pi}$, respectively,  for the decays $\Lambda^{+}_c\to \Lambda^{(*)}l^{+}\nu_l\to \Sigma\pi l^{+}\nu_l$. 
The dominant contribution to this total decay width is through the $\Lambda(1405)$ resonance. Transitions through the 
$\Lambda(1115)$ and $\Lambda(1520)$ also provide a significant contribution to the total decay rate for $\Lambda_c^+\to\Sigma\pi l^+\nu_l$. The contributions from the transitions through the $\Lambda(1600)$, $\Lambda(1890)$ and $\Lambda(1820)$ are small. 

\begin{figure}[t!]
\caption{(a) The differential decay rates $d\Gamma/dq^2$ for the decay 
$\Lambda^{+}_c\to \Lambda^{(*)}l^{+}\nu_l\to N\bar{K} l^{+}\nu_l$. 
(b) Comparison the decay rates to the excited states with the semileptonic decay 
to the elastic channel $\Lambda_c^{+}\to\Lambda(1115) l^{+}\nu_l$.}
\label{DWNK}
\includegraphics[width=0.6\textwidth]{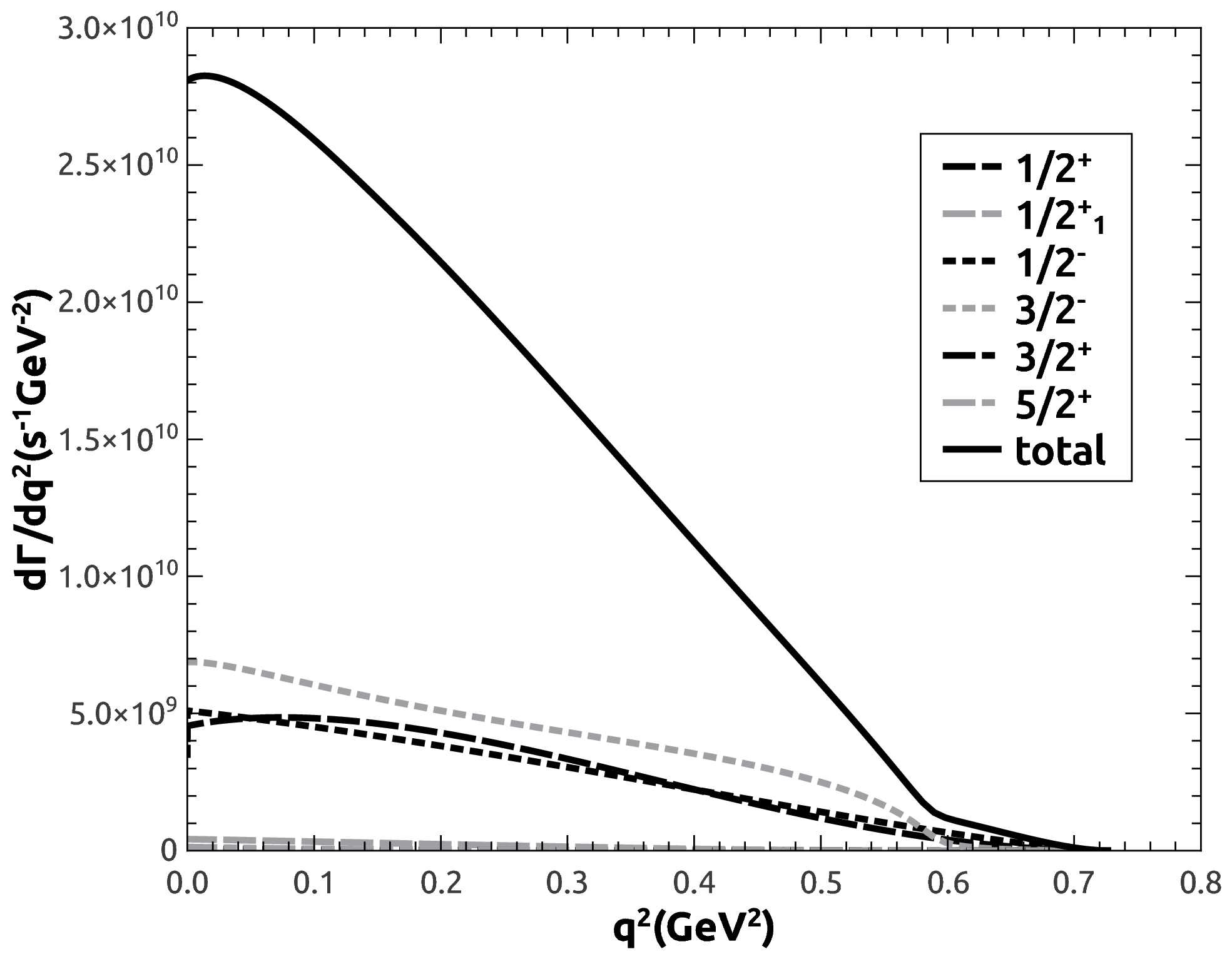}
\caption*{(a)}

\includegraphics[width=0.6\textwidth]{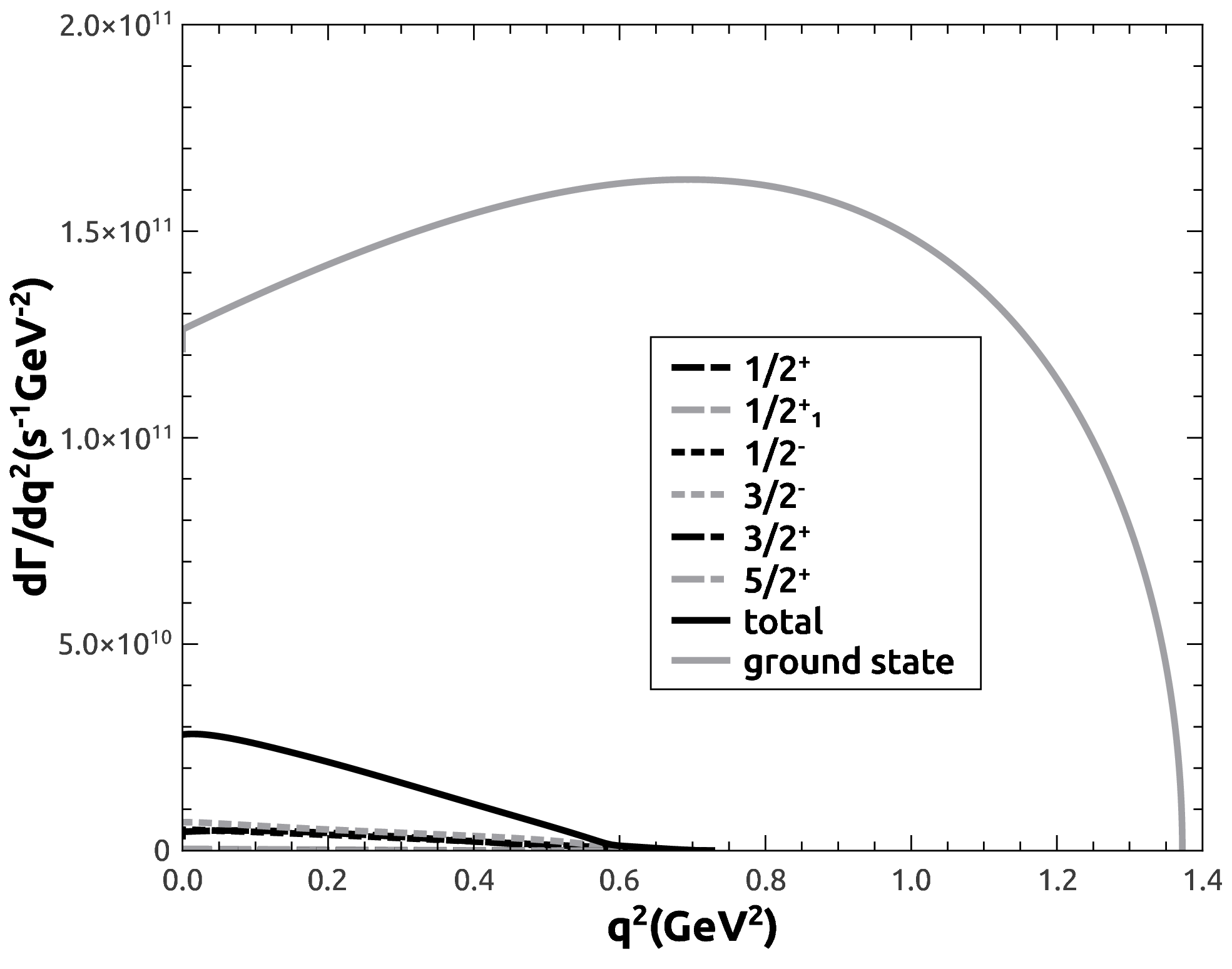}
\caption*{(b)}

\end{figure}

\begin{figure}
\caption{(a) The differential decay rate 
through each of the $\Lambda^{*}$ states considered in this calculation along with the coherent total.
The black solid curve shows the differential decay rate for $\Lambda_c\to NKl\nu_l$.
(b) The differential decay rate $d\Gamma/dS_{N\bar{K}}$ for different values of $g_{\Lambda N\bar{K}}$ for $\Lambda (1405)$ state.}
\label{DWNK2}
\includegraphics[width=0.6\textwidth]{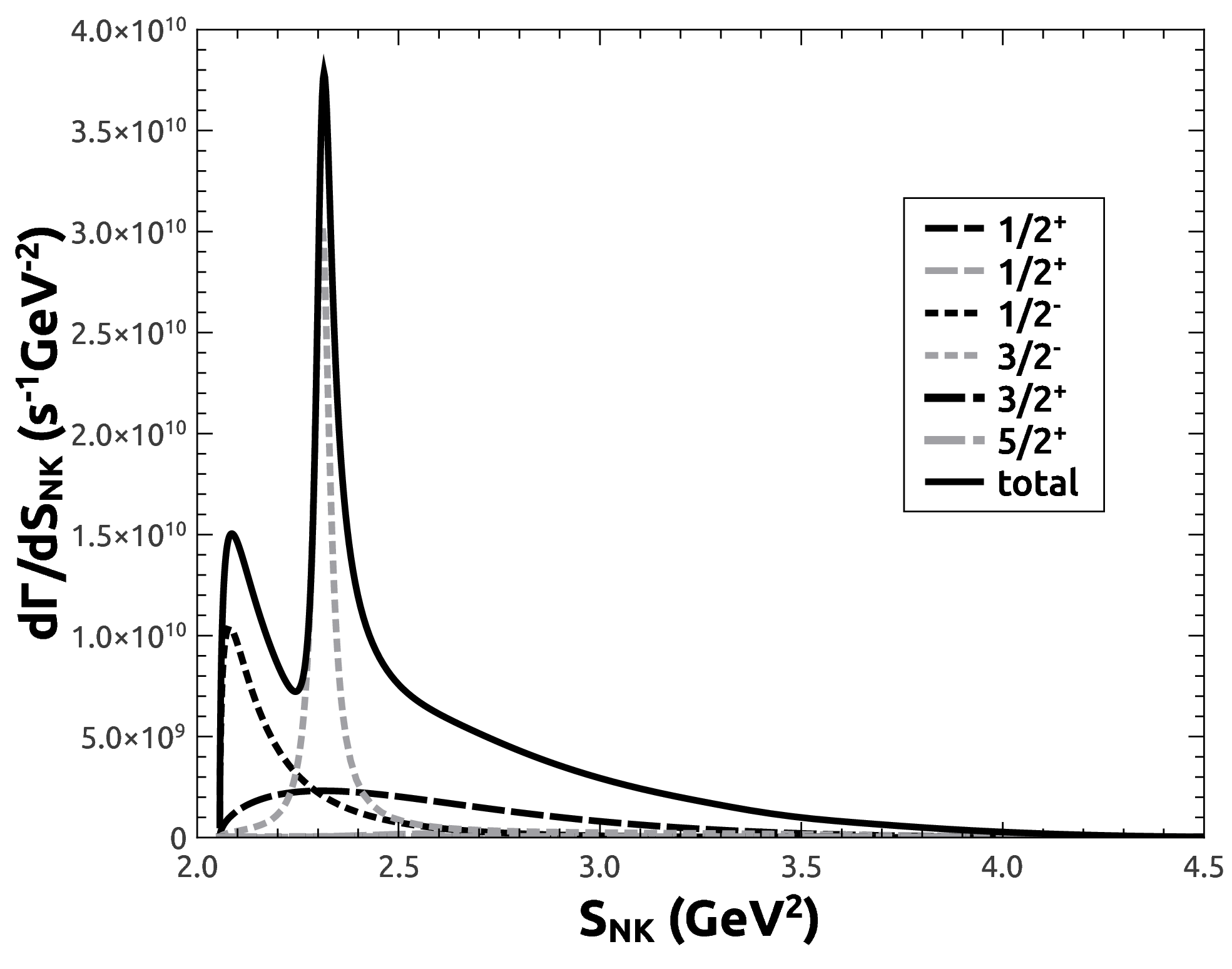}
\caption*{(a)}

\includegraphics[width=0.6\textwidth]{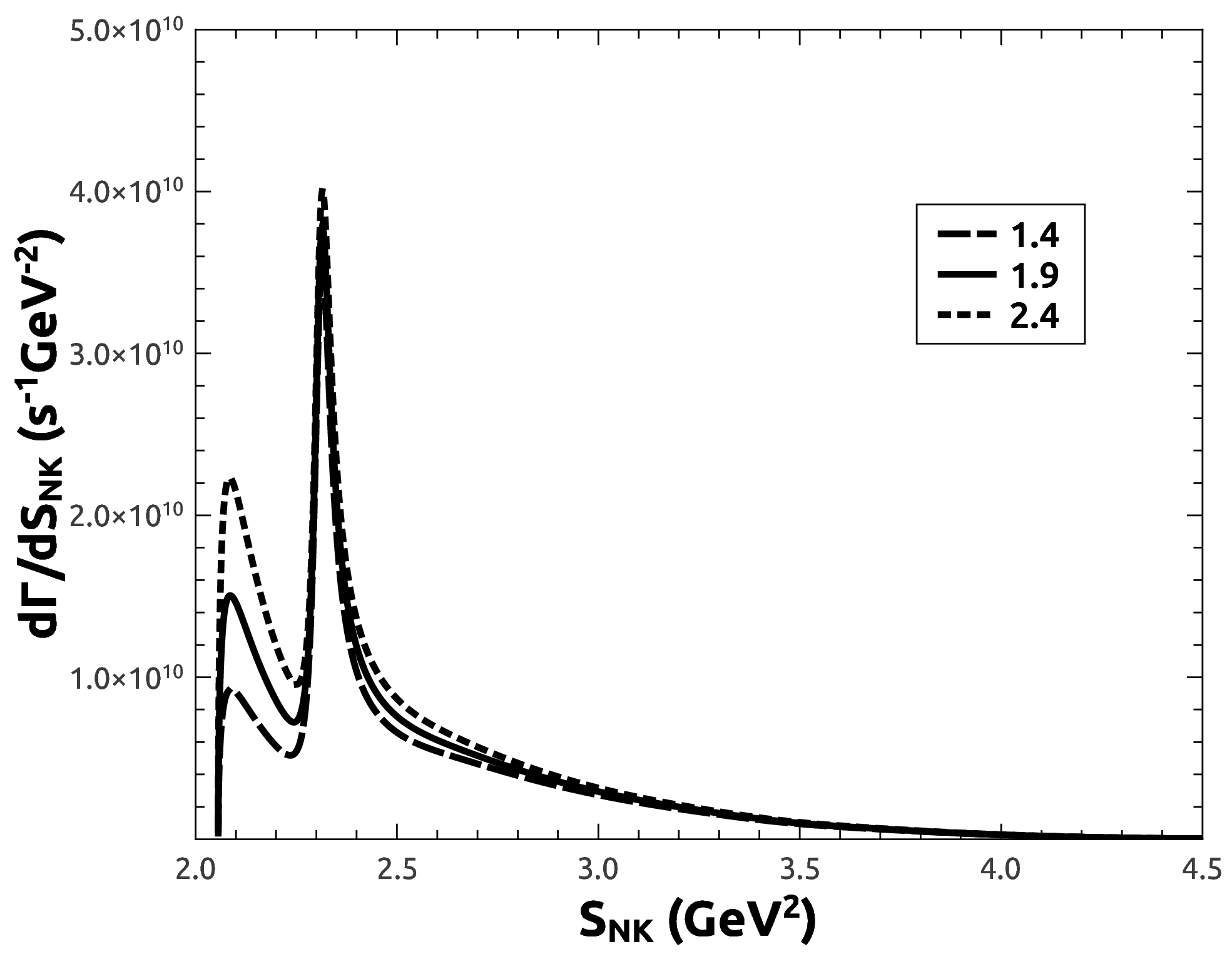}
\caption*{(b)}

\end{figure}

The differential decay rates $d\Gamma/dq^2$ and $d\Gamma/dS_{N\bar{K}}$ for the decay 
$\Lambda^{+}_c\to \Lambda^{(*)}l^{+}\nu_l\to N\bar{K} l^{+}\nu_l$ are shown in figs \ref{DWNK} and \ref{DWNK2}, respectively. 
In the total decay width, the transition through the $\Lambda(1520)$ is dominant, and the transition through
the sub-threshold $\Lambda(1405)$ is still large. The contributions to the total rate, from transitions through the 
$\Lambda(1600)$, $\Lambda(1890)$ and $\Lambda(1820)$, are small.

The integrated total decay widths are shown in table \ref{AllDecResult}. In this calculation, we assume 
$\Lambda_c^{+}\to\Lambda^{*}\to\Sigma\pi$ and 
$\Lambda_c^{+}\to\Lambda^{*}\to N\bar{K}$ are the dominant decay modes, and that these two decay modes are saturated by contributions from the states we consider. We also assume that other semileptonic decay modes of the $\Lambda_c$ are suppressed. The total decay width $\Gamma_{\Lambda_c^{+}\to\Lambda^{(*)}l^{+}\nu_l\to\Sigma\pi l^{+}\nu_l}$ and 
$\Gamma_{\Lambda_c^{+}\to\Lambda^{(*)}l^{+}\nu_l\to N\bar{K} l^{+}\nu_l}$ 
are calculated to be $\sim 0.18\times 10^{11}s^{-1}$, and $\sim 0.1\times 10^{11}s^{-1}$ respectively. 
We calculate the semileptonic branching fraction 
$\frac{\Gamma_{\Lambda_c^{+}\to\Lambda^{(*)}l^{+}\nu_l\to\Sigma\pi l^{+}\nu_l}}{\Gamma_{\Lambda_c}}$ to be $0.08$, while the branching 
fraction $\frac{\Gamma_{\Lambda_c^{+}\to\Lambda^{(*)}l^{+}\nu_l\to N\bar{K} l^{+}\nu_l}}{\Gamma_{\Lambda_c}}$ is $0.04$.
Our calculation contradicts the assumption of the CLEO collaboration \citep{PDG2} that the elastic channel saturates 
the semileptonic decay of $\Lambda_c$. In our model, we find the branching fractions for the multi-particle final states 
are $12\%$ of the total semileptonic decay. This suggests that the semileptonic decay of $\Lambda_c^{+}$ is not saturated to decay to 
elastic channel and further investigation is needed to see evidence of the channels we have discussed here. 

We have treated the $\Lambda(1405)$ as a three-quark state that is the lightest excitation of the $\Lambda$, with  $J^P = 1/2^{-}$. In fig. \ref{DWSigmaPi}, this state contributes a clear resonant structure at $\sqrt{S_{\Sigma\pi}} \approx 1405 $. This would suggest that examination of the decay channel $\Lambda_c^{+}\to \Sigma\pi l^{+}\nu_l$ would provide confirmation of this state as a three-quark state. If no evidence is found for this resonance then it may well be that this state is not a simple three-quark state. 

There are a number of other conjectures regarding the structure of the $\Lambda(1405)$. It has been suggested that it could be a dynamically generated molecular state of $\bar{K}N$ and $\Sigma\pi$ \citep{Jennings,Dalitz,Dalitz2,Ramos,Oset}, or a multi-quark state \citep{Choe}. Recently, Roca and Oset \citep{Roca} explained it as a molecular state of $\bar{K}N$. Hall {\it et al.} \citep{Hall} drew the same conclusion based on a lattice simulation. Ikeno and Oset \citep{Oset} have estimated the semileptonic decay rate of the $\Lambda_c$ to this state, assuming that it is a dynamically-generated molecular state. They obtained a value of $2\times 10^{-5}$ for the branching fraction $\mathcal{B}(\Lambda_c^{+}\to \Lambda(1405) l^{+}\nu_l)=2\times 10^{-5}$. For $\Lambda_c^{+}\to \Lambda(1405) l^{+}\nu_l\to \Sigma\pi l^{+}\nu_l$ our branching fraction/ratio is $\sim 2.0\times 10^{-3}$, while for $\Lambda_c^{+}\to \Lambda(1405) l^{+}\nu_l\to N\bar{K} l^{+}\nu_l$ it is $\sim 0.4\times 10^{-3}$. Our values are therefore about 20 times larger than the prediction by Ikeno and Oset.

\begin{table}[t!]
\caption{Integrated decay widths for $\Lambda_c^{+}\to\Lambda^{(*)}l^{+}\nu_l\to\Sigma\pi l^{+}\nu_l$ and 
$\Lambda_c^{+}\to\Lambda^{(*)}l^{+}\nu_l\to N\bar{K} l^{+}\nu_l$ in units of $10^{11} s^{-1}$, for individual
$\Lambda^{(*)}$ states for both $l^{+}=e^{+}$ and $l^{+}=\mu^{+}$. The last row shows the coherent totals for the four-body decays $\Lambda_c^+\to\Sigma\pi l\nu_l$ and $\Lambda_c^+\to NKl\nu_l$.}
\label{AllDecResult}
\begin{tabular}{|P{2cm}|P{2cm}|P{3cm}|P{3cm}|P{3cm}|P{3.1cm}|}
\hline
Spin of $\Lambda^{(*)}$ & Mass(GeV) 
& $\Gamma_{\Lambda_c\to\Lambda^{*} e^{+} \nu_e \to \Sigma\pi e^{+} \nu_e}$
& $\Gamma_{\Lambda_c\to\Lambda^{*} \mu^{+} \nu_\mu \to \Sigma\pi \mu^{+} \nu_\mu}$
& $\Gamma_{\Lambda_c\to\Lambda^{*} e^{+} \nu_e \to N\bar{K} e^{+} \nu_e}$
& $\Gamma_{\Lambda_c\to\Lambda^{*} \mu^{+} \nu_\mu \to N\bar{K} \mu^{+} \nu_\mu}$ \\
\hline
$\frac{1}{2}^{+}$  & 1.115 &  $ 4.18\times 10^{-2}$ &$ 3.81\times 10^{-2}$  & $ 1.86\times 10^{-2}$ & $1.65\times 10^{-2}$\\
\hline
$\frac{1}{2}^{+}_1$& 1.600 &  $ 1.66\times 10^{-3}$ &$ 1.44 \times 10^{-4}$ & $ 9.79\times 10^{-4}$ & $8.42\times 10^{-4}$\\
\hline
$\frac{1}{2}^{-}$  & 1.405 &  $ 9.54\times 10^{-2}$ &$ 8.78 \times 10^{-2}$ & $ 1.82\times 10^{-2}$ & $1.65\times 10^{-2}$\\
\hline
$\frac{3}{2}^{-}$  & 1.519 &  $ 2.21\times 10^{-2}$ &$ 1.98 \times 10^{-2}$ & $ 2.54\times 10^{-2}$ & $2.25\times 10^{-2}$\\
\hline
$\frac{3}{2}^{+}$  & 1.890 &  $ 2.11\times 10^{-5}$ &$ 1.66 \times 10^{-5}$ & $ 1.12\times 10^{-4}$ & $9.00\times 10^{-5}$\\
\hline
$\frac{5}{2}^{+}$  & 1.820 &  $ 2.23\times 10^{-5}$ &$ 1.74 \times 10^{-5}$ & $ 1.81\times 10^{-4}$ & $1.43\times 10^{-4}$\\
\hline
$\text{Total}$     &	     & $18.31\times 10^{-2}$  &$ 16.59\times 10^{-2}$ & $ 9.33\times 10^{-2}$ & $8.23\times 10^{-2}$\\
\hline
\end{tabular}
\end{table}

\section{Conclusions and Outlook}

In this work, semileptonic decays of the $\Lambda_c^{+}$ have been studied using a constituent quark model to calculate the required form factors. These form factors for the $\Lambda_c\to\Lambda^{(*)}$ transitions have been obtained both 
analytically and numerically, using the harmonic oscillator basis to describe the baryon wave functions.
The form factors obtained in this model are compared with the HQET expectations at leading order, and are seen to be largely consistent with those expectations.
The decay rates of $\Lambda_c^{+}$ to the ground state and a number of excited $\Lambda$ states have been evaluated.

The original motivation for this work was that there was no model independent calculation for 
$\mathcal{B}(\Lambda_c^{+}\to pK^{-}\pi^{+})$ reported in the previous edition of PDG \citep{PDGold}. PDG estimated
$B(\Lambda_{c}^{+}\to pK^{-}\pi^{+})=RfF \frac{B(D\to Xl^{+}\nu_l)}{1+|\frac{V_{cd}}{V_{cs}}|^2}\tau(\Lambda_c^{+})$, based on 
the measurements by the ARGUS \citep{ARGUS2} and CLEO \citep{CLEO2} Collaborations, using the semileptonic decays of the $\Lambda_c$. They assumed that $f=\frac{B(\Lambda_c^{+}\to \Lambda l^{+} \nu_l)}{B(\Lambda_c^{+}\to X_s l^{+}\nu_l)} = 1.0$ 
and $F=\frac{B(\Lambda_c^{+}\to X_s l^{+}\nu_l)}{B(D\to X_sl^{+}\nu_l)}=1.0$. 
The latest edition reports a model independent measurement that makes the old estimate obsolete. A. Zupanc {\it et al.}
(Belle Collaboration) \citep{ZUPANC} and M. Ablikim (BESIII Collaboration) \citep{BESIII2} measured $B(\Lambda_{c}^{+}\to pK^{-}\pi^{+})$ to be $6.84^{+0.32}_{-0.40}$ and $5.84\pm 0.27\pm 0.23\%$ respectively.
PDG reports their fit for $B(\Lambda_{c}^{+}\to pK^{-}\pi^{+})$ to be $6.35\pm 0.33\%$.
This result lets us estimate the branching fraction $f= 0.87^{+0.13}_{-0.17}$, still assuming that $F=1.0$.

We have calculated branching fractions of the semileptonic decays and they are in a good agreement with the 
calculations done by Pervin {\it et al.} in PRCI \citep{PRC}. The branching fraction of 
the decay to the elastic channel has been calculated to be $3.84\%$ (for $l=e$)
and $3.72\%$ (for $l=\mu$). Our prediction is in agreement with the recent
results from BESIII \citep{BESIII, BESIII3} that measured it to be $3.63\pm 0.38\pm 0.20\%$ (for $l=e$) and $3.49\pm 0.46\pm 0.27$ (for $l=\mu$). 

We have used the form factors obtained to examine the semileptonic decays to two four-particle final states, namely $\Sigma\pi l^+\nu_l$ and $NKl^+\nu_l$. We find that the branching fraction for these two channels totals $12\%$ of the inclusive semileptonic decay $\Lambda_c\to X_s l^{+}\nu_l$. We estimate $f=0.88$, in disagreement with the CLEO \citep{CLEO} assumption that the decay to the ground state $\Lambda(1115)$ saturates the semileptonic decays of the $\Lambda_c$. 

The two lowest-lying $\Lambda$ resonances, the $\Lambda(1405) 1/2^{-}$ and $\Lambda(1520) 3/2^{-}$ are seen to be important in
both the rate and the shape of the spectrum.
The $\Lambda(1405)$ produces a sharp resonant structure in the $S_{\Sigma\pi}$ spectrum, suggesting that this state may be detectable in the $\Lambda_c^{+}\to\Sigma\pi l^{+}\nu_l$ transition. The $\Lambda(1520)$ also generates sharp resonant structures in both the $\Sigma\pi$ and $N\bar{K}$ decay spectra. This state may therefore also be detectable in these channels. This can have some impact on baryon phenomenology, as it would confirm these states as orbital excitations of the ground states $\Lambda$. The broader resonances that were included in the study are less likely to be identifiable in the decay spectra. 

In this calculation we have assumed that the states we include saturate the resonant decays of the $\Lambda_c$. The available phase space limits the number of excited states that can contribute significantly to the semileptonic decay rate. There is ample phase space to produce some of the lighter excitations, such as the $\Lambda(1670)1/2^-$ and $\Lambda(1690)3/2^-$, but the very small wave function overlap with that of the $\Lambda_c$ means that the form factors are tiny, so that the decays are very effectively suppressed. 

The work presented in this manuscript can be extended in a number of directions. The form factors calculated here may be used to study any of the polarization observables that can arise in 
these semileptonic decays. With a suitable parametrization of the factorization assumption, they can also be used to examine a number of nonleptonic decays of the $\Lambda_c$. The form factors were evaluated using the harmonic oscillator basis, and this leads to form factors that have exponential dependence on $q^2$. One possible extension of the project would be to use a different basis, such as the sturmian basis, to extract the form factors. This basis leads to form factors  with multipole dependence on $q^2$, closer to popular expectations. The semianalytic method we have developed for use with the harmonic oscillator basis can easily be adapted for the sturmian basis. The semileptonic decays of the $\Lambda_b$ to both charmed and charmless final states may also be re-examined.

\section*{Acknowledgements}
We gratefully acknowledge the support of the Department of Physics, the College of Arts and Sciences, and the
Office of Research at Florida State University. This research is supported by the U.S. Department of Energy under
contracts DE-SC0002615.

\appendix
\section{Semi Analytic Treatment of Hadronic Matrix Elements}\label{semianalytic}
The hadronic matrix element can be written in the form,
\begin{equation}
\langle \Lambda|\bar{s}\Gamma c|\Lambda_c \rangle =  \sum_{b^\prime,b} h_{b^\prime}^{\Lambda^{*}}
h_b^{\Lambda_c}\delta_{s_1^\prime s_1}\delta_{s_2^\prime s_2}(-1)^{l_{\lambda^\prime}+l_\lambda}\times B_{n_\rho l_\rho m_\rho}^{n_{\rho^\prime}l_{\rho^\prime}m_{\rho^\prime}}(\alpha_\rho,\alpha_{\rho^\prime})D_{\Gamma;n_\lambda l_\lambda m_\lambda s_q}^{n_{\lambda^\prime} l_{\lambda^\prime} m_{\lambda^\prime} s_{q^\prime}}(\alpha_\lambda,\alpha_{\lambda^\prime}),
\end{equation}
where the coefficients $h^A_{b(b^\prime)}$ are the products of the normalization of the baryon state $A$, the expansion coefficients 
$\eta_i$, and the various Clebsch-Gordan coefficients that appear in the parent (daughter) baryon wave function.
The indices $b(b^\prime)$ contain all the relevant quantum numbers being summed over for the parent (daughter) baryon state. 
$B_{n_\rho l_\rho m_\rho}^{n_{\rho^\prime}l_{\rho^\prime}m_{\rho^\prime}}$ is the spectator overlap,
\begin{align*}
B_{n_\rho l_\rho m_\rho}^{n_{\rho^\prime}l_{\rho^\prime}m_{\rho^\prime}}(\alpha_\rho,\alpha_{\rho^\prime})
=& \int d^3p_\rho d^3p_{\rho^\prime}\psi^{*}_{n_{\rho^\prime} l_{\rho^\prime} m_{\rho^\prime}}(\alpha_{\rho^\prime};
\vec{p}_{\rho^\prime})\psi_{n_\rho l_\rho m_\rho}(\alpha_\rho;\vec{p}_\rho)\delta(p_\rho-p_{\rho^\prime})\\
= &\delta_{l_\rho l_{\rho^\prime}}\delta_{m_\rho m_{\rho^\prime}}
N^{*}_{n_{\rho^\prime}l_{\rho^\prime}}(\alpha_{\rho^\prime})N_{n_\rho l_\rho}(\alpha_\rho)
\int d^3p_\rho \text{ exp}( -p^2_\rho/2\alpha_\rho^2)\text{ exp}( -p^2_\rho/2\alpha_{\rho^\prime}^2)\\
& \times 
\mathcal{Y}^{*}_{l_{\rho^\prime} m_{\rho^\prime}}(\vec{p}_\rho)\mathcal{Y}_{l_\rho m_\rho}(\vec{p}_\rho)
\mathcal{L}_{n_{\rho^\prime}}^{l_{\rho^\prime}+1/2 *}(p_\rho^2/\alpha_{\rho^\prime}^2)
\mathcal{L}_{n_{\rho}}^{l_{\rho}+1/2}(p_\rho^2/\alpha_\rho^2)\\
 = & N^{*}_{n_{\rho^\prime}l_\rho}(\alpha_{\rho^\prime})N_{n_\rho l_\rho}(\alpha_\rho)
\int d^3p_\rho \text{ exp}( -b^2 p^2_\rho)\\
& \times 
\mathcal{Y}^{*}_{l_{\rho} m_{\rho}}(\vec{p}_\rho)\mathcal{Y}_{l_\rho m_\rho}(\vec{p}_\rho)
\mathcal{L}_{n_{\rho^\prime}}^{l_{\rho}+1/2 *}(p_\rho^2/\alpha_{\rho^\prime}^2)
\mathcal{L}_{n_{\rho}}^{l_{\rho}+1/2}(p_\rho^2/\alpha_\rho^2),\\
\end{align*}
where $b^2 = \frac{\alpha_\rho^2 + \alpha_{\rho^\prime}^2}{2\alpha_\rho^2\alpha_{\rho^\prime}^2}$. The results for the overlap integrals that appear in this calcuation are shown in appendix \ref{SpecOver}.

The interaction overlap 
$D_{\Gamma;n_\lambda l_\lambda m_\lambda s_q}^{n_{\lambda^\prime} l_{\lambda^\prime}m_{\lambda^\prime} s_{q^\prime}}$ is
\begin{align}\label{SA1}
D_{\Gamma;n_\lambda l_\lambda m_\lambda s_q}^{n_{\lambda^\prime} l_{\lambda^\prime} 
m_{\lambda^\prime} s_{q^\prime}}(\beta,\beta^\prime)
=&\text{ exp}\Bigg(\frac{-3m_\sigma^2}{2m_{\Lambda}^2}\frac{p_\Lambda^2}{\alpha_{\lambda\lambda^\prime}^2}\Bigg)
\int d^3k e^{-\alpha^2k^2}\mathcal{L}^{l_{\lambda^\prime}
+\frac{1}{2}*}\Bigg(\frac{{p^\prime}^2}{\beta^{\prime 2}}\Bigg)
\mathcal{Y}^{*}_{l_{\lambda^\prime}m_{\lambda^\prime}}(\vec{p}^{\prime})\nonumber \\
&\times
\langle s(\vec{p}_{\Lambda}+\vec{p},s_{q^\prime})|\bar{s}\Gamma c|c(\vec{p},s_q)\rangle
\mathcal{L}^{l_{\lambda}+\frac{1}{2}}\Bigg(\frac{p^2}{\beta^2}\Bigg)\mathcal{Y}_{l_\lambda m_\lambda}(\vec{p}),
\end{align}
where, $\beta^{(\prime)} = \sqrt{2/3}\alpha_\lambda^{(\prime)}$,
$\alpha^2 = \frac{\alpha_{\lambda}^2+\alpha^2_{\lambda^\prime}}{2\alpha^2_\lambda\alpha^2_{\lambda^\prime}}$
and
$\alpha_{\lambda\lambda^\prime} = \sqrt{(\alpha_\lambda^2+\alpha_{\lambda^\prime}^2)/2}$.

Using the changes in variable $\vec{p}=\vec{k}+a\vec{p}_\Lambda$, and $\vec{p}^\prime=\vec{k}+a^\prime\vec{p}_\Lambda$,
where $a=-\frac{m_\sigma}{\tilde{m}_\Lambda}\frac{\alpha_\lambda^2}{2\alpha_{\lambda\lambda^\prime}^2}$ and
$a^\prime=\frac{m_\sigma}{\tilde{m}_\Lambda}\frac{\alpha_{\lambda^\prime}^2}{2\alpha_{\lambda\lambda^\prime}^2}$,
the solid harmonics take the form $\mathcal{Y}_{lm}(a\vec{p}_1+b\vec{p}_2)$, and can be decomposed using the addition theorem as,
\begin{align*}
\mathcal{Y}_{lm}(b_1\vec{p}_1+b_2\vec{p}_2) &= \sum_{\lambda=0}^l\sum_{\mu_\lambda={-\lambda}}^\lambda 
B_{lm}^{\lambda \mu_\lambda}\mathcal{Y}_{\lambda \mu_\lambda}(\vec{p}_1)\mathcal{Y}_{l-\lambda m-\mu_\lambda}(\vec{p}_2),\\
\end{align*}
where
\begin{align*}
B_{lm}^{\lambda \mu_\lambda} &= \sqrt{\frac{4\pi(2l+1)!}{(2\lambda+1)!(2l-2\lambda+1)!}}b_1^\lambda 
b_2^{l-\lambda}C^{lm}_{\lambda \mu_\lambda l-\lambda m-\mu_\lambda}. \\
\end{align*}
Equation \ref{SA1} then takes the form
\begin{align*}
D_{\Gamma;n_\lambda l_\lambda m_\lambda s_q}^{n_{\lambda^\prime} l_{\lambda^\prime} 
m_{\lambda^\prime} s_{q^\prime}}(\beta,\beta^\prime)
=\text{exp}\Bigg(\frac{-3m_\sigma^2}{2m_{\Lambda}^2}\frac{p_\Lambda^2}{\alpha_{\lambda\lambda^\prime}^2}\Bigg)
\sum_{\lambda^\prime\mu_{\lambda^\prime}}\sum_{\lambda\mu_\lambda}
B_{l_{\lambda^\prime}m_{\lambda^\prime}}^{\lambda^\prime\mu_{\lambda^\prime}*}
B_{l_\lambda m_\lambda}^{\lambda\mu_\lambda}\mathcal{I}_{\Gamma;n_\lambda l_\lambda s_q:\lambda\mu_\lambda}^{n_{\lambda^\prime} l_{\lambda^\prime} s_{q^\prime}:\lambda^\prime\mu_{\lambda^\prime}}(\alpha^2),
\end{align*}
where
\begin{align}\label{SAI}
\mathcal{I}_{\Gamma;n_\lambda l_\lambda s_q:\lambda\mu_\lambda}^{n_{\lambda^\prime} l_{\lambda^\prime} s_{q^\prime}:
\lambda^\prime\mu_{\lambda^\prime}}(\alpha^2) 
= &\int d^3k e^{-\alpha^2 k^2}\mathcal{L}_{n_{\lambda^\prime}}^{l_{\lambda^\prime}+\frac{1}{2}*}
(\frac{p^{\prime^2}}{\beta^{\prime 2}})\mathcal{Y}_{\lambda^\prime\mu_{\lambda^\prime}}^{*}
(\vec{k})\mathcal{Y}^{*}_{l-\lambda^\prime m-\mu_{\lambda^\prime}}(\vec{p}_\Lambda)\nonumber\\
&\times
\langle s(\vec{p}_\Lambda + \vec{p},s_{q^\prime})|\bar{s}\Gamma c| c(\vec{p},s_q)\rangle \mathcal{L}_{n_\lambda}^{l_\lambda+\frac{1}{2}}(\frac{p^2}{\beta^2})\mathcal{Y}_{\lambda\mu_\lambda}(\vec{k})\mathcal{Y}_{l-\lambda m-\mu_\lambda}(\vec{p}_\Lambda).
\end{align}
The quark current $\langle s(\vec{p}_\Lambda + \vec{p},s_{q^\prime})|\bar{s}\Gamma c| c(\vec{p},s_q)\rangle $
can be written in its most general form as
\begin{equation}\label{spinme}
\langle s(\vec{p}_\Lambda + \vec{p},s_{q^\prime})|\bar{s}\Gamma c| c(\vec{p},s_q)\rangle=\sum_{l=0}^2
\sum_{m_l = -l}^l \xi_{\Gamma;lm_l}^{s_{q^\prime}s_q}(\vec{k},\vec{p}_\Lambda)\mathcal{Y}_{lm_l}(\vec{k}),
\end{equation}
where $\xi_{\Gamma;lm_l}^{s_{q^\prime}s_q}(\vec{k},\vec{p}_\Lambda)$ can be expanded in Legendre polynomials as
\begin{equation}\label{lagexp}
\xi_{\Gamma;lm_l}^{s_{q^\prime}s_q}(\vec{k},\vec{p}_\Lambda)  = \sum_{l_0=0}^{\infty}\zeta_{\Gamma;lm_l l_0}^{s_{q^\prime}s_q}(k,p_\Lambda) P_{l_0}(x).
\end{equation}
Here, $x = \hat{k}\cdot\hat{p}_\Lambda$. The coefficients $\zeta_{\Gamma;lm_l l^\prime}^{s_{q^\prime}s_q}(k,\vec{p}_\Lambda)$ are obtained as 
\begin{equation}
\zeta_{\Gamma;lm_l l_0}^{s_{q^\prime}s_q}(k,p_\Lambda) = \bigg(\frac{2l_0+1}{2}\bigg)\int_{-1}^{1} dx \xi_{\Gamma;lm_l}^{s_{q^\prime}s_q}(\vec{k},\vec{p}_\Lambda)P_{l_0}(x),
\end{equation}
and the integral on the right hand side is evaluated numerically. In practice, the sum in eq. \ref{lagexp} includes a finite number of terms, determined by the values of $l_\lambda$, $n_\lambda$, $l_{\lambda^\prime}$, $n_{\lambda^\prime}$ and the maximum value of $l$ in eq. \ref{spinme}.
The Legendre polynomial $P_{l_0}(x)$ can be written as 
\begin{equation*}
P_{l_0}(x) = \bigg(\frac{4\pi}{2l_0+1}\bigg)\sum_{m_0=-l_0}^{l_0}Y^{*}_{l_0m_0}(\hat{p}_\Lambda)Y_{l_0m_0}(\hat{k}),
\end{equation*}
Thus,  $\xi_{\Gamma;lm_l}^{s_{q^\prime}s_q}(\vec{k},\vec{p}_\Lambda)$ takes the form
\begin{equation}
\xi_{\Gamma;lm_l}^{s_{q^\prime}s_q}(\vec{k},\vec{p}_\Lambda)  = 2\pi\sum_{l_0=0}^{\infty}\sum_{m_0=-l_0}^{l_0}Y^{*}_{l_0m_0}(\hat{p}_\Lambda)Y_{l_0m_0}(\hat{k})\bigg[\int_{-1}^{1}dx  \xi_{\Gamma;lm_l}^{s_{q^\prime}s_q}(\vec{k},\vec{p}_\Lambda)P_{l_0}(x)\bigg].
\end{equation}

\noindent
The product of the Laguerre polynomials 
$\mathcal{L}_{n_{\lambda^\prime}}^{l_{\lambda^\prime}+\frac{1}{2}*}(p^{\prime 2}/\beta^{\prime 2})\times 
\mathcal{L}_{n_\lambda}^{l_\lambda+\frac{1}{2}}(p^2/\beta^2)$ can be written as
\begin{equation}
\mathcal{L}_{n_{\lambda^\prime}}^{l_{\lambda^\prime}+\frac{1}{2}*}\bigg(\frac{{p^\prime}^2}{\beta^{\prime 2}}\bigg)
\mathcal{L}_{n_\lambda}^{l_\lambda+\frac{1}{2}}\bigg(\frac{p^2}{\beta^2}\bigg)=\sum_{\delta\rho\sigma}D_{\delta\rho\sigma}k^{2\delta}
p_\Lambda^{2\rho} (\vec{k}\cdot\vec{p}_\Lambda)^\sigma,
\end{equation}
where $(\vec{k}\cdot\vec{p}_\Lambda)$ can be expanded as
\begin{align*}
(\vec{k}\cdot\vec{p}_\Lambda)^\sigma &= \sum_{l_\sigma=0}^{\sigma}k^\sigma p_\Lambda^\sigma c_{l_\sigma}^{\prime\sigma}P_{l_\sigma}(\hat{k}\cdot\hat{p}_\Lambda)\\
&=\sum_{l_\sigma=0}^{\sigma}k^\sigma p_\Lambda^\sigma C_{l_\sigma}^{\prime\sigma}\frac{4\pi}{2l_\sigma+1}Y_{l_\sigma m_\sigma}(\hat{k})Y^{*}_{l_\sigma m_\sigma}(\hat{p}_\Lambda),
\end{align*}
where $P_{l_\sigma}(\hat{k}\cdot\hat{p}_\Lambda)$ is the Legendre polynomial and $C_{l_\sigma}^{\prime\sigma}$ is defined as
\begin{align*}
C_{2s}^{\prime 2r} =& (2r)!\frac{2^{2s}(4s+1)(r+s)!}{(2r+2s+1)!(r-s)!}, C^{\prime 2r}_{2s+1}=0,\\
C_{2s+1}^{\prime 2r+1} =& (2r+1)!\frac{2^{2s+1}(4s+3)(r+s+1)!}{(2r+2s+3)!(r-s)!}, C^{\prime 2r+1}_{2s}=0.
\end{align*}

Eqn. \ref{SAI} therefore becomes
\begin{eqnarray}\label{totme}
\mathcal{I}_{\Gamma;n_\lambda l_\lambda s_q:\lambda\mu_\lambda}^{n_{\lambda^\prime} l_{\lambda^\prime} s_{q^\prime}:\lambda^\prime\mu_{\lambda^\prime}}(\alpha^2) = &&\sum_{l=0}^2
\sum_{m_l = -l}^l\sum_{l_0=0}^{\infty}\sum_{m_0=-l_0}^{l_0} \sum_{\delta\rho\sigma}\sum_{l_\sigma=0}^{\sigma}\int d^3k e^{-\alpha^2 k^2}\xi_{\Gamma;lm_l}^{s_{q^\prime}s_q}(\vec{k},\vec{p}_\Lambda)\nonumber\\
& \times& 
D_{\delta\rho\sigma}k^{2\delta}
p_\Lambda^{2\rho}  k^\sigma p_\Lambda^\sigma C_{l_\sigma}^{\prime\sigma}\frac{4\pi}{2l_\sigma+1}
 \mathcal{Y}^{*}_{l-\lambda^\prime m-\mu_{\lambda^\prime}}(\vec{p}_\Lambda)Y^{*}_{l_0m_0}(\hat{p}_\Lambda)
 Y^{*}_{l_\sigma m_\sigma}(\hat{p}_\Lambda)\mathcal{Y}_{l-\lambda m-\mu_\lambda}(\vec{p}_\Lambda)\nonumber\\
&\times&
 Y_{l_0m_0}(\hat{k})Y_{l_\sigma m_\sigma}(\hat{k})
 \mathcal{Y}_{\lambda^\prime\mu_{\lambda^\prime}}^{*}(\vec{k})\mathcal{Y}_{lm_l}(\vec{k})\mathcal{Y}_{\lambda\mu_\lambda}(\vec{k}).
\end{eqnarray}
The angular integrations in eq. \ref{totme} are evaluated using the properties of the spherical and solid harmonics. The remaining integral is done numerically.

\section{Spectator Overlap}\label{SpecOver}
The spectator overlaps for the set of quantum numbers ($n_\rho l_\rho m_\rho$,$n_{\rho^\prime}l_{\rho^\prime}m_{\rho^\prime}$)
used in our calculation are listed in this appendix. We define $\alpha_{\rho\rho^\prime} = \sqrt{\alpha_\rho^2 + \alpha_{\rho^\prime}^2/2}$.
\begin{align*}
B_{000}^{000} & =\left[\frac{\alpha_\rho\alpha_{\rho^\prime}}{\alpha_{\rho\rho^\prime}^2}\right]^{3/2}, \,\,\,\,
B_{010}^{010} =\left[\frac{\alpha_\rho\alpha_{\rho^\prime}}{\alpha_{\rho\rho^\prime}^2}\right]^{5/2}, \\
B_{000}^{100} &=-B^{000}_{100} =\sqrt{\frac{3}{2}}\left[\frac{\alpha_\rho\alpha_{\rho^\prime}}{\alpha_{\rho\rho^\prime}^2}\right]^{3/2}
\frac{\alpha_{\rho^\prime}^2-\alpha_\rho^2}{\alpha_{\rho^\prime}^2+\alpha_\rho^2}, \\
B_{100}^{100} & =\left[\frac{\alpha_\rho\alpha_{\rho^\prime}}{\alpha_{\rho\rho^\prime}^2}\right]^{3/2}
\left[\frac{5}{2}\left(\frac{\alpha_\rho\alpha_{\rho^\prime}}{\alpha_{\rho\rho^\prime}^2}\right)^2 - \frac{3}{2}\right],\\
B_{010}^{110} &=-B^{010}_{110} = \sqrt{\frac{5}{2}}\left[\frac{\alpha_\rho\alpha_{\rho^\prime}}{\alpha_{\rho\rho^\prime}^2}\right]^{5/2}
\frac{\alpha_{\rho^\prime}^2-\alpha_\rho^2}{\alpha_{\rho^\prime}^2+\alpha_\rho^2}, \\
B_{110}^{110} & =\left[\frac{\alpha_\rho\alpha_{\rho^\prime}}{\alpha_{\rho\rho^\prime}^2}\right]^{5/2}
\left[\frac{7}{2}\left(\frac{\alpha_\rho\alpha_{\rho^\prime}}{\alpha_{\rho\rho^\prime}^2}\right)^2 - \frac{5}{2}\right].\\
\end{align*}

\section{Analytic Expressions For The Form Factors}\label{AnalyticFormFactors}

The analytical expressions for the form factors for transition to $\Lambda^{*}$ states with the $J^P$ are shown. We obtained these form factors using the single component wave-functions in the harmonic oscillator basis. 
\subsection{\texorpdfstring{$1/2^{+}$}{}}

\begin{align*}
 F_1 &=I_H\left[1+\frac{m_\sigma}{\alpha_{\lambda\lambda^\prime}^2}
 \left(\frac{\alpha_{\lambda^\prime}^2}{m_s}-\frac{\alpha_\lambda^2}{m_c}\right)\right],\\
  F_2 &=-I_H\left[\frac{m_{\sigma}}{m_s}\frac{\alpha_{\lambda^\prime}^2}{\alpha_{\lambda\lambda^\prime}^2}
 -\frac{\alpha_{\lambda}^2\alpha_{\lambda^\prime}^2}{4m_cm_s\alpha_{\lambda\lambda^\prime}^2}\right],\\
 F_3 &= -I_H\frac{m_{\sigma}}{m_c}\frac{\alpha_{\lambda}^2}{\alpha_{\lambda^{\prime}}^2},\\
G_1 & = I_H\left[1-\frac{\alpha_{\lambda}^2{\alpha_{\lambda^\prime}}^2}
{12\alpha_{\lambda\lambda^{\prime}}^2m_cm_s}\right],\\
 G_2&=-I_H\left[\frac{m_{\sigma}}{m_s}\frac{\alpha_{\lambda^\prime}^2}{\alpha_{\lambda\lambda^\prime}^2}
 +\frac{\alpha_{\lambda}^2\alpha_{\lambda^\prime}^2}{12m_cm_s\alpha_{\lambda\lambda^\prime}^2}
 \left(1+\frac{12m_\sigma^2}{\alpha_{\lambda{\lambda^\prime}}^2}\right)\right],\\
G_3&=I_H\left[\frac{m_{\sigma}}{m_c}\frac{\alpha_{\lambda^\prime}^2}{\alpha_{\lambda\lambda^\prime}^2}
 +\frac{m_{\sigma}^2\alpha_{\lambda}^2\alpha_{\lambda^\prime}^2}{m_cm_s\alpha_{\lambda\lambda^\prime}^4}\right],
\end{align*}

where
\begin{align*}
 I_H=\left(\frac{\alpha_\lambda\alpha_{\lambda^\prime}}{\alpha_{\lambda\lambda^\prime}^2}\right)^{3/2}
 \exp\left(-\frac{3m_\sigma^2}{2m_\Lambda^2}\frac{p_\Lambda^2}{\alpha_{\lambda\lambda^\prime}^2}\right).
\end{align*}

\vspace{0.5cm}

\subsection{\texorpdfstring{$1/2_{1}^{+}$}{}}

\begin{align*}
 F_1 & =I_H\frac{1}{2\alpha_{\lambda\lambda^\prime}^2}\left[(\alpha_\lambda^2-\alpha_{\lambda^\prime}^2)
 -\frac{m_\sigma}{3\alpha_{\lambda\lambda^\prime}^2}
 \left(\frac{\alpha_{\lambda^\prime}^2}{m_s}(7\alpha_\lambda^2-3\alpha_{\lambda^\prime}^2) 
 + \frac{\alpha_\lambda^2}{m_c}(7\alpha_{\lambda^\prime}^2-3\alpha_\lambda^2)\right)\right],\\
  F_2 & =-I_H\frac{\alpha_{\lambda^\prime}^2}{6m_s\alpha_{\lambda\lambda^\prime}^4}
  (7\alpha_\lambda^2-3\alpha_{\lambda^\prime}^2)\left[m_\sigma-\frac{\alpha_\lambda^2}{4m_c}\right],\\
 F_3 & = I_H\frac{\alpha_\lambda^2m_\sigma}{6m_c\alpha_{\lambda\lambda^\prime}^4}
  (7\alpha_{\lambda^\prime}^2-3\alpha_\lambda^2),\\
G_1 & = I_H\left[\frac{(\alpha_{\lambda}^2-\alpha_{\lambda^\prime}^2)}{2\alpha_{\lambda\lambda^{\prime}}^2}
- \frac{\alpha_\lambda^2\alpha_{\lambda^\prime}^2}{72\alpha_{\lambda\lambda^\prime}^4m_cm_s}(7\alpha_\lambda^2-3\alpha_{\lambda^\prime}^2)\right],\\
 G_2 & = -I_H\frac{\alpha_{\lambda^\prime}^2}{6m_s\alpha_{\lambda\lambda^\prime}^4}
 \left[(7\alpha_\lambda^2-3\alpha_{\lambda^\prime}^2)\left(m_\sigma+\frac{\alpha_\lambda^2}{6m_c}\right)
  +\frac{7m_\sigma^2\alpha_{\lambda}^2}{m_c\alpha_{\lambda\lambda^\prime}^2}(\alpha_\lambda^2-\alpha_{\lambda^\prime}^2)\right],\\
G_3 & = -I_H\frac{\alpha_\lambda^2m_\sigma}{6m_c\alpha_{\lambda\lambda^\prime}^4}
\left[(7\alpha_{\lambda^\prime}^2-3\alpha_\lambda^2)
+\frac{7m_\sigma\alpha_{\lambda^\prime}^2}{m_c\alpha_{\lambda\lambda^\prime}^2}(\alpha_\lambda^2-\alpha_{\lambda^\prime}^2)
 \right],
 \end{align*}

where
\begin{align*}
 I_H=\sqrt{\frac{3}{2}}\left(\frac{\alpha_\lambda\alpha_{\lambda^\prime}}{\alpha_{\lambda\lambda^\prime}^2}\right)^{3/2}
 \exp\left(-\frac{3m_\sigma^2}{2m_\Lambda^2}\frac{p_\Lambda^2}{\alpha_{\lambda\lambda^\prime}^2}\right).
\end{align*}

\vspace{0.5cm}

\subsection{\texorpdfstring{$1/2^{-}$}{}}

\begin{align*}
F_1&=I_H\frac{\alpha_{\lambda}}{6}\left[\frac{3}{m_s}-\frac{1}{m_c}\right],\\
F_2&=-I_H\left[\frac{2m_{\sigma}}{\alpha_{\lambda}}-\frac{\alpha_{\lambda}}{2m_s}
 +\frac{2m_{\sigma}^2\alpha_{\lambda}}{m_c\alpha_{\lambda\lambda^\prime}^2}
 -\frac{m_{\sigma}\alpha_\lambda}{6m_cm_s\alpha_{\lambda\lambda^\prime}^2}
 (3\alpha_{\lambda}^2-2\alpha_{\lambda^\prime}^2)\right],\\
F_3&= I_H\frac{2m_{\sigma}^2\alpha_{\lambda}}{m_c\alpha_{\lambda\lambda^{\prime}}^2},\\
G_1&=I_H\left[\frac{2m_{\sigma}}{\alpha_{\lambda}}-\frac{\alpha_{\lambda}}{6m_c}
  +\frac{m_{\sigma}\alpha_\lambda}{6m_cm_s\alpha_{\lambda\lambda^\prime}^2}
 (3\alpha_{\lambda}^2-2\alpha_{\lambda^\prime}^2)\right],\\
G_2&=I_H\left[-\frac{2m_{\sigma}}{\alpha_{\lambda}}+\frac{\alpha_{\lambda}}{2m_s}
 +\frac{\alpha_\lambda}{3m_c}\right],\\
G_3 &= I_H\frac{\alpha_\lambda}{3m_c}\left[1-\frac{m_\sigma} 
{2m_s\alpha_{\lambda\lambda^\prime}^2}(3\alpha_{\lambda}^2-2\alpha_{\lambda^\prime}^2)\right],
\end{align*}

where
\begin{align*}
 I_H=\left(\frac{\alpha_\lambda\alpha_{\lambda^\prime}}{\alpha_{\lambda\lambda^\prime}^2}\right)^{5/2}
 \exp\left(-\frac{3m_\sigma^2}{2m_\Lambda^2}\frac{p_\Lambda^2}{\alpha_{\lambda\lambda^\prime}^2}\right).
\end{align*}

\vspace{0.5cm}

\subsection{\texorpdfstring{$3/2^{-}$}{}}

\begin{align*}
F_1 & = I_H\frac{3m_\sigma}{\alpha_\lambda}\left[1+\frac{m_\sigma}{\alpha_{\lambda\lambda^\prime}^2}
\left(\frac{\alpha_{\lambda^\prime}^2}{m_s}+\frac{\alpha_\lambda^2}{m_c}\right)\right],\\
F_2 & =-I_H\left[\frac{3m_{\sigma}^2}{m_s}\frac{\alpha_{\lambda}^2}{\alpha_{\lambda\lambda^\prime}^2\alpha_\lambda}
 -\frac{5\alpha_\lambda\alpha_{\lambda^\prime}^2m_{\sigma}}{4\alpha_{\lambda\lambda^\prime}^2m_cm_s}\right],\\
F_3 & = -I_H\left[\frac{3m_{\sigma}^2\alpha_{\lambda}}{m_c\alpha_{\lambda\lambda^{\prime}}^2}+\frac{\alpha_\lambda}{2m_c}\right],\\
F_4 & = I_H\frac{\alpha_\lambda}{m_c},\\
G_1 & = I_H\left[\frac{3m_\sigma}{\alpha_\lambda}- \frac{\alpha_\lambda}{2m_c}
\left(1+\frac{3m_\sigma\alpha_{\lambda^\prime}^2}{2m_s\alpha_{\lambda\lambda^\prime}^2}\right)\right],\\
G_2 & = -I_H\left[\frac{3m_{\sigma}^2}{m_s}\frac{\alpha_{\lambda^\prime}^2}{\alpha_\lambda\alpha_{\lambda\lambda^\prime}^2} 
+ \frac{m_\sigma\alpha_\lambda\alpha_{\lambda^\prime}^2}{4m_cm_s\alpha_{\lambda\lambda^\prime}^4}
(\alpha_{\lambda\lambda^\prime}^2+12 m_\sigma^2)\right],\\
G_3 & = I_H\frac{\alpha_\lambda}{m_c\alpha_{\lambda\lambda^\prime}}\left[\frac{\alpha_{\lambda\lambda^\prime}^2}{2} 
+3m_\sigma^2 + \frac{\alpha_{\lambda^\prime}^2m_\sigma}{m_s\alpha_{\lambda\lambda^\prime}^2}(\alpha_{\lambda\lambda^\prime}^2+6m_\sigma^2)\right],\\
G_4 & = -I_H \left[\frac{\alpha_\lambda}{m_c} 
+ \frac{m_\sigma}{m_cm_s}\frac{\alpha_\lambda\alpha_{\lambda^\prime}^2}{\alpha_{\lambda\lambda^\prime}^2}\right],
\end{align*}

where
\begin{align*}
 I_H=-\frac{1}{\sqrt{3}}\left(\frac{\alpha_\lambda\alpha_{\lambda^\prime}}{\alpha_{\lambda\lambda^\prime}^2}\right)^{5/2}
 \exp\left(-\frac{3m_\sigma^2}{2m_\Lambda^2}\frac{p_\Lambda^2}{\alpha_{\lambda\lambda^\prime}^2}\right).
\end{align*}

\vspace{0.5cm}

\subsection{\texorpdfstring{$3/2^{+}$}{}}

\begin{align*}
F_1 & = -I_H\frac{m_\sigma}{2}\left[\frac{5}{m_s}-\frac{3}{m_c}\right],\\
F_2 & = I_H\frac{m_\sigma}{\alpha_\lambda}\left[\frac{6m_{\sigma}}{\alpha_{\lambda}}-\frac{5\alpha_{\lambda}}{2m_s}
+\frac{6m_{\sigma}^2\alpha_{\lambda}}{m_c\alpha_{\lambda\lambda^\prime}^2}
 -\frac{m_{\sigma}\alpha_\lambda}{2m_cm_s\alpha_{\lambda\lambda^\prime}^2}
 (\alpha_{\lambda}^2-2\alpha_{\lambda^\prime}^2)\right],\\
F_3&= I_H\frac{2m_{\sigma}^2\alpha_{\lambda}}{m_c\alpha_{\lambda\lambda^{\prime}}^2},\\
F_3 & = -I_H\frac{m_\sigma}{m_c}\left[1+\frac{6m_{\sigma}^2}{\alpha_{\lambda\lambda^\prime}^2}\right],\\
F_4 & = I_H\frac{2m_\sigma}{m_c},\\
G_1 & = -I_H\left[\frac{6m_{\sigma}^2}{\alpha_\lambda^2}-\frac{m_\sigma}{2m_c} + \frac{m_\sigma^2}{6\alpha_{\lambda\lambda^\prime}^2m_cm_s}
 \left(11\alpha_\lambda^2 - 6\alpha_{\lambda^\prime}^2\right)\right],\\
G_2 & = I_H\left[\frac{6m_{\sigma}^2}{\alpha_{\lambda}^2} - \frac{5m_\sigma}{2m_s}-\frac{2m_\sigma}{m_c}
+\frac{5\alpha_{\lambda}^2}{12m_cm_s}-\frac{2m_\sigma^2\alpha_\lambda^2}{3\alpha_{\lambda\lambda^\prime}^2m_cm_s}\right],\\
G_3 & = -I_H\left[\frac{m_\sigma}{2m_c}-\frac{5\alpha_\lambda^2}{24m_cm_s} 
- \frac{m_\sigma^2}{4m_cm_s\alpha_{\lambda\lambda^\prime}^2}(5\alpha_{\lambda}^2-2\alpha_{\lambda^\prime}^2)\right],\\
G_4 & = -I_H \frac{5\alpha_\lambda^2}{6m_cm_s}, 
\end{align*}

where
\begin{align*}
 I_H=\frac{1}{\sqrt{5}}\left(\frac{\alpha_\lambda\alpha_{\lambda^\prime}}{\alpha_{\lambda\lambda^\prime}^2}\right)^{7/2}
 \exp\left(-\frac{3m_\sigma^2}{2m_\Lambda^2}\frac{p_\Lambda^2}{\alpha_{\lambda\lambda^\prime}^2}\right).
\end{align*}

\vspace{0.5cm}

\subsection{\texorpdfstring{$5/2^{+}$}{}}

\begin{align*}
F_1 & = I_H\frac{3m_\sigma^2}{\alpha_\lambda^2}\left[1+\frac{m_\sigma}{\alpha_{\lambda\lambda^\prime}^2}
\left(\frac{\alpha_{\lambda^\prime}^2}{m_s}+\frac{\alpha_\lambda^2}{m_c}\right)\right],\\
F_2 & = - I_H\frac{m_\sigma^2}{m_s\alpha_{\lambda\lambda^\prime}^2}
\left[\frac{3m_{\sigma}\alpha_{\lambda^\prime}^2}{\alpha_\lambda^2} -\frac{1}{4m_c}(8\alpha_\lambda^2+7\alpha_{\lambda^\prime}^2)\right],\\
F_3 & = -I_H\frac{m_\sigma}{m_c}\left[1+\frac{3m_{\sigma}^2}{\alpha_{\lambda\lambda^\prime}^2}\right],\\
F_4 & = I_H\frac{2m_\sigma}{m_c},\\
G_1 & = I_H\left[\frac{3m_\sigma^2}{\alpha_\lambda^2}- \frac{m_\sigma}{m_c}-
\frac{m_\sigma^2}{12m_sm_c\alpha_{\lambda\lambda^\prime}^2}(8\alpha_\lambda^2+15\alpha_{\lambda^\prime}^2)\right],\\
G_2 & = -I_H\frac{m_\sigma^2}{m_s\alpha_{\lambda\lambda^\prime}^2}\left[\frac{3m_{\sigma}\alpha_{\lambda^\prime}^2}{\alpha_\lambda^2}
+ \frac{1}{12m_c}(8\alpha_\lambda^2+3\alpha_{\lambda^\prime}^2) 
+ \frac{3m_\sigma^2\alpha_{\lambda^\prime}^2}{m_c\alpha_{\lambda\lambda^\prime}^2}\right],\\
G_3 & = I_H\frac{m_\sigma}{m_c}\left[1+ \frac{3m_\sigma^2}{\alpha_{\lambda\lambda^\prime}^2}
+\frac{m_\sigma\alpha_{\lambda^\prime}^2}{m_s\alpha_{\lambda\lambda^\prime}^2}(1+\frac{6m_\sigma^2}{\alpha_{\lambda\lambda^\prime}^2})\right],\\
G_4 & = -I_H \frac{2m_\sigma}{m_c}\left[1+ \frac{m_\sigma}{m_s}\frac{\alpha_{\lambda^\prime}^2}{\alpha_{\lambda\lambda^\prime}^2}\right],
\end{align*}

where
\begin{align*}
 I_H = \frac{1}{\sqrt{2}}\left(\frac{\alpha_\lambda\alpha_{\lambda^\prime}}{\alpha_{\lambda\lambda^\prime}^2}\right)^{7/2}
 \exp\left(-\frac{3m_\sigma^2}{2m_\Lambda^2}\frac{p_\Lambda^2}{\alpha_{\lambda\lambda^\prime}^2}\right).
\end{align*}

\section{Wave Functions}\label{wavefunctions}
The baryon wave functions are expanded in the harmonic oscillator basis. 
For $\Lambda$ states with spin-parity $J^P=\frac{1}{2}^{+}$, the wave function expansion is
\begin{align}
 \Psi_{\Lambda_{Q},\frac{1}{2}^{+}M}=&\phi_{\Lambda_{Q}}\Biggl(\biggl[\eta_{1}\psi_{000000}(\vec{p}_{\rho},\vec{p}_{\lambda})
 +\eta_{2}\psi_{001000}(\vec{p}_{\rho},\vec{p}_{\lambda})+\eta_{3}\psi_{000010}(\vec{p}_{\rho},\vec{p}_{\lambda})\biggr]
 \chi_{\frac{1}{2}}^{\rho}(M)\nonumber \\
 &+\eta_{4}\psi_{000101}(\vec{p}_{\rho},\vec{p}_{\lambda}) \chi_{\frac{1}{2}}^{\lambda}(M)
 +\eta_{5}\biggl[\psi_{1M_{L}0101}(\vec{p}_{\rho},\vec{p}_{\lambda})\chi_{\frac{1}{2}}^{\lambda}(M-M_L)\biggr]_{1/2,M} \nonumber \\
 &+\eta_{6}\biggl[\psi_{1M_{L}0101}(\vec{p}_{\rho},\vec{p}_{\lambda}) \chi_{\frac{3}{2}}^{S}(M-M_L)\biggr]_{1/2,M}
 +\eta_{7}\biggl[\psi_{2M_{L}0101}(\vec{p}_{\rho},\vec{p}_{\lambda})\chi_{\frac{3}{2}}^{S}(M-M_L)\biggr]_{1/2,M}\Biggr).
\end{align}
where the $\eta_i$'s are  the expansion coefficients and ${[\psi_{LM_L n_\rho m_\rho n_\lambda m_\lambda} \chi_S(M-M_L)]}_{J,M}$ is
 a short-hand notation for the Clebsch-Gordan sum $\sum_{M_L}C^{JM}_{L M_L,SM-M_L}$.

For $J^P=\frac{1}{2}^{-}$ and $J^P = \frac{3}{2}^{-}$, the wave function expansion is
\begin{align}
 \Psi_{\Lambda_{Q},J^{-}M}=&\phi_{\Lambda_{Q}}\biggl(\eta_{1}\biggl[\psi_{1M_{L}0001}(\vec{p}_{\rho},\vec{p}_{\lambda})
 \chi_{\frac{1}{2}}^{\rho}(M-M_L)\biggr]_{J,M}
 +\eta_{2}\biggl[\psi_{1M_{L}0100}(\vec{p}_{\rho},\vec{p}_{\lambda})
 \chi_{\frac{1}{2}}^{\lambda}(M-M_L)\biggr]_{J,M}\nonumber \\
 &
 +\eta_{3}\biggl[\psi_{1M_{L}0100}(\vec{p}_{\rho},\vec{p}_{\lambda})
 \chi_{\frac{3}{2}}^{S}(M-M_L)\biggr]_{J,M}\biggr).
\end{align}

For $J^P=\frac{3}{2}^{+}$, the wave function is
\begin{align}
 \Psi_{\Lambda_{Q},\frac{3}{2}^{+}M}=&\phi_{\Lambda_{Q}}\biggl(
\eta_{1}\biggl[\psi_{2M_{L}0002}(\vec{p}_{\rho},\vec{p}_{\lambda})\chi_{\frac{1}{2}}^{\rho}(M-M_L)\biggr]_{3/2,M}
 +\eta_{2}\biggl[\psi_{2M_{L}0200}(\vec{p}_{\rho},\vec{p}_{\lambda}) \chi_{\frac{1}{2}}^{\rho}(M-M_L)\biggr]_{3/2,M} \nonumber \\
&+\eta_{3}\biggl[\psi_{1M_{L}0101}(\vec{p}_{\rho},\vec{p}_{\lambda})\chi_{\frac{1}{2}}^{\lambda}(M-M_L)\biggr]_{3/2,M}
 +\eta_{4}\biggl[\psi_{2M_{L}0101}(\vec{p}_{\rho},\vec{p}_{\lambda}) \chi_{\frac{1}{2}}^{\lambda}(M-M_L)\biggr]_{3/2,M}\nonumber\\
&+\eta_{5}\psi_{000101}(\vec{p}_{\rho},\vec{p}_{\lambda})\chi_{\frac{3}{2}}^{S}(M)
 +\eta_{6}\biggl[\psi_{1M_{L}0101}(\vec{p}_{\rho},\vec{p}_{\lambda})\chi_{\frac{3}{2}}^{S}(M-M_L)\biggr]_{3/2,M}\nonumber\\
& +\eta_{7}\biggl[\psi_{2M_{L}0101}(\vec{p}_{\rho},\vec{p}_{\lambda}) \chi_{\frac{3}{2}}^{S}(M-M_L)\biggr]_{3/2,M}\biggr).
\end{align}

For $J^P=\frac{5}{2}^{+}$, the wave function is
\begin{align}
 \Psi_{\Lambda_{Q},\frac{5}{2}^{+}M}=&\phi_{\Lambda_{Q}}\Biggl(\eta_{1}\biggl[\psi_{2M_{L}0200}(\vec{p}_{\rho},\vec{p}_{\lambda})
 \chi_{\frac{1}{2}}^{\rho}(M-M_L)\biggr]_{5/2,M}
 +\eta_{2}\biggl[\psi_{2M_{L}0002}(\vec{p}_{\rho},\vec{p}_{\lambda}) \chi_{\frac{1}{2}}^{\rho}(M-M_L)\biggr]_{5/2,M}\nonumber \\
&+\eta_{3}\biggl[\psi_{2M_{L}0101}(\vec{p}_{\rho},\vec{p}_{\lambda})\chi_{\frac{1}{2}}^{\lambda}(M-M_L)\biggr]_{5/2,M} 
+\eta_{4}\psi_{1M_{L}0101}(\vec{p}_{\rho},\vec{p}_{\lambda})\chi_{\frac{3}{2}}^{S}(M)\nonumber \\
& +\eta_{5}\biggl[\psi_{2M_{L}0101}(\vec{p}_{\rho},\vec{p}_{\lambda})
 \chi_{\frac{3}{2}}^{S}(M-M_L)\biggr]_{5/2,M}\Biggr).
\end{align}
No other states are expected to have significant overlap with the decaying ground state $\Lambda_c$ in the spectator approximation.
\section{Hadron Tensors}
\subsection{Hadron Tensor in \texorpdfstring{$\Lambda_c^{+}\to\Lambda^{*} l^{+}\nu_l$}{} transitions}\label{HadTen1}

\subsubsection{\texorpdfstring{$1/2^{+}$}{}}

\vspace{0.5cm}
\begin{equation}
 \alpha=2\left[YF_1^2+XG_1^2\right].
\end{equation}

\begin{equation}
 \beta_{++}=\sum_{i=1,j=1}^{i=3,j=3}(A_{ij}F_iF_j+A_{ij}^\prime G_iG_j),
\end{equation}
where

\begin{minipage}{0.45\textwidth}
\begin{align*}
A_{11} & =2,,\\
A_{22} & = \frac{1}{2m_{\Lambda_c}^2}X,\\
A_{33} & = \frac{1}{2m_{\Lambda}^2}X,\\
A_{12} & = \frac{2}{m_{\Lambda_c}} (m_{\Lambda_c}+m_{\Lambda}),\\
A_{23} & = \frac{1}{m_{\Lambda}m_{\Lambda_c}}X,\\
A_{31} & = \frac{2}{m_{\Lambda}}(m_{\Lambda_c}+m_{\Lambda}),
\end{align*}
\end{minipage}
\begin{minipage}{0.45\textwidth}
\begin{align*}
A_{11}^\prime & =2,\\
A_{22}^\prime & = \frac{1}{2m_{\Lambda_c}^2}Y,\\
A_{33}^\prime & = \frac{1}{2m_{\Lambda}^2}Y,\\
A_{12}^\prime & = \frac{-2}{m_{\Lambda_c}} (m_{\Lambda_c}-m_{\Lambda}),\\
A_{23}^\prime & = \frac{1}{m_{\Lambda}m_{\Lambda_c}}Y,\\
A_{31}^\prime & = \frac{-2}{m_{\Lambda}}(m_{\Lambda_c}-m_{\Lambda}).
\end{align*}
\end{minipage}

\begin{equation*}
 \gamma(1/2^{+}) = 4F_1G_1
\end{equation*}

where $X \equiv [(m_{\Lambda_c}+m_{\Lambda})^2-q^2]$ and $Y \equiv [(m_{\Lambda_c}-m_{\Lambda})^2-q^2]$.

\vspace{0.5cm}

\subsubsection{\texorpdfstring{$1/2^{-}$}{}}

\begin{equation}
 \alpha =2\left[XF_1^2 +YG_1^2\right],
\end{equation}

\begin{equation}
 \beta_{++}=\sum_{i=1,j=1}^{i=3,j=3}(A_{ij}F_iF_j+A_{ij}^\prime G_iG_j),
\end{equation}
where

\begin{minipage}{0.45\textwidth}
\begin{align*}
A_{11} & = 2,\\
A_{22} & = \frac{1}{2m_{\Lambda_c}^2}Y,\\
A_{33} & = \frac{1}{2m_{\Lambda}^2}Y,\\ 
A_{12} & = \frac{-2}{m_{\Lambda_c}} (m_{\Lambda_c}-m_{\Lambda}),\\
A_{23} & = \frac{1}{m_{\Lambda}m_{\Lambda_c}}Y,\\
A_{31} & = \frac{-2}{m_{\Lambda}}
(m_{\Lambda_c}-m_{\Lambda}),
\end{align*}
\end{minipage}
\begin{minipage}{0.45\textwidth}
\begin{align*}
A_{11}^\prime & =2,\\
A_{22}^\prime & = \frac{1}{2m_{\Lambda_c}^2}X,\\
A_{33}^\prime & = \frac{1}{2m_{\Lambda}^2}X,\\ 
A_{12}^\prime & = \frac{2}{m_{\Lambda_c}} (m_{\Lambda_c}+m_{\Lambda}),\\
A_{23}^\prime & = \frac{1}{m_{\Lambda}m_{\Lambda_c}}X,\\
A_{31}^\prime & = \frac{2}{m_{\Lambda}}(m_{\Lambda_c}+m_{\Lambda}),
\end{align*}
\end{minipage}

\begin{equation}
 \gamma(1/2^{+}) = 4F_1G_1.
\end{equation}

\vspace{0.5cm}

\subsubsection{\texorpdfstring{$3/2^{-}$}{}}

\begin{equation}
 \alpha =\sum_{i=1,j=1}^{i=4,j=4}\frac{1}{Z}(B_{ij}F_iF_j+B_{ij}^\prime G_iG_j),
\end{equation}
where $Z = 3m_{\Lambda_c}^2m_{\Lambda}^2$ and the non vanishing coefficients are
\begin{align*}
B_{11} &= XY^2,\,\,\,\,
B_{14} = B_{14}^\prime = -2m_{\Lambda_c}m_{\Lambda}XY,\\
B_{44} &= 4m_{\Lambda_c}^2m_{\Lambda}^2X,\,\,\,\,
B_{11}^\prime = X^2Y,\,\,\,\,
B_{44}^\prime = 4m_{\Lambda_c}^2m_{\Lambda}^2Y.
\end{align*}

\begin{equation}
 \beta_{++}=\sum_{i=1,j=1}^{i=4,j=4}\frac{1}{Z}(A_{ij}F_iF_j+A_{ij}^\prime G_iG_j),
\end{equation}

\begin{minipage}{0.48\textwidth}
\begin{align*}
A_{11}  = & XY,\\
A_{12}  = & \frac{1}{m_{\Lambda_c}} (m_{\Lambda_c}+m_{\Lambda})XY,\\
A_{13}  = & \frac{1}{m_{\Lambda}} (m_{\Lambda_c}+m_{\Lambda})XY,\\
A_{14}  = & 2m_{\Lambda_c}\big[m_{\Lambda_c}(m_{\Lambda_c}^2-m_{\Lambda}^2-q^2)\\
&
+m_{\Lambda}(m_{\Lambda_c}^2-m_{\Lambda}^2)\big],\\
A_{22}  = & \frac{1}{4m_{\Lambda_c}^2}X^2Y,\\
A_{23}  = & \frac{1}{2m_{\Lambda_c}m_\Lambda}X^2Y,\\
A_{24}  = & X(m_{\Lambda_c}^2-m_{\Lambda}^2-q^2),\\
A_{33}  = & \frac{1}{4m_{\Lambda}^2}X^2Y,\\
A_{34}  = & \frac{m_{\Lambda_c}}{m_{\Lambda}}X(m_{\Lambda_c}^2-m_{\Lambda}^2-q^2),\\
A_{44}  = & m_{\Lambda_c}^2X,
\end{align*}
\end{minipage}
\begin{minipage}{0.48\textwidth}
\begin{align*}
A_{11}^\prime  = & XY,\\
A_{12}^\prime  = & \frac{-1}{m_{\Lambda_c}} (m_{\Lambda_c}-m_{\Lambda})XY,\\
A_{13}^\prime  = & \frac{-1}{m_{\Lambda}} (m_{\Lambda_c}-m_{\Lambda})XY,\\
A_{14}^\prime  = & -2m_{\Lambda_c}\big[m_{\Lambda_c}(m_{\Lambda_c}^2-m_{\Lambda}^2-q^2)\\
&
-m_{\Lambda}(m_{\Lambda_c}^2-m_{\Lambda}^2)\big],\\
A_{22}^\prime  = & \frac{1}{4m_{\Lambda_c}^2}XY^2,\\
A_{23}^\prime  = & \frac{1}{2m_{\Lambda_c}m_\Lambda}XY^2,\\
A_{24}^\prime  = & Y(m_{\Lambda_c}^2-m_{\Lambda}^2-q^2),\\
A_{33}^\prime  = & \frac{1}{4m_{\Lambda}^2}XY^2,\\
A_{34}^\prime  = &\frac{m_{\Lambda_c}}{m_{\Lambda}}Y(m_{\Lambda_c}^2-m_{\Lambda}^2-q^2),\\
A_{44}^\prime  = & m_{\Lambda_c}^2Y,
\end{align*}
\end{minipage}

\begin{equation*}
 \gamma=\sum_{i=1,j=1}^{i=4,j=4}\frac{1}{Z}C_{ij}F_iG_j,
\end{equation*}

\begin{align*}
C_{11} & = 2XY,\,\,\,\,
C_{14}  = -2m_{\Lambda_c}m_\Lambda Y,\,\,\,\,
C_{41}  = -2m_{\Lambda_c}m_\Lambda X,\,\,\,\,
C_{44}  = -4m_{\Lambda_c}^2 m_\Lambda^2.
\end{align*}

\subsubsection{\texorpdfstring{$3/2^{+}$}{}}

\begin{equation}
 \alpha =\sum_{i=1,j=1}^{i=4,j=4}\frac{1}{Z}(B_{ij}F_iF_j+B_{ij}^\prime G_iG_j),
\end{equation}
where $Z = 3m_{\Lambda_c}^2m_{\Lambda}^2$ and the non vanishing coefficients are
\begin{align*}
B_{11} &= X^2Y,\,\,\,\,
B_{14} = B_{14}^\prime = -2m_{\Lambda_c}m_{\Lambda}XY,\\
B_{44} &= 4m_{\Lambda_c}^2m_{\Lambda}^2Y,\,\,\,\,
B_{11}^\prime = XY^2,\,\,\,\,
B_{44}^\prime = 4m_{\Lambda_c}^2m_{\Lambda}^2X,
\end{align*}

\begin{equation}
 \beta_{++}=\sum_{i=1,j=1}^{i=4,j=4}\frac{1}{Z}(A_{ij}F_iF_j+A_{ij}^\prime G_iG_j),
\end{equation}

\begin{minipage}{0.48\textwidth}
\begin{align*}
A_{11} & = XY,\\
A_{12} & = \frac{-1}{m_{\Lambda_c}} (m_{\Lambda_c}-m_{\Lambda})XY,\\
A_{13} & = -\frac{1}{m_{\Lambda}}(m_{\Lambda_c}+m_{\Lambda})XY,\\
A_{14} & = -2m_{\Lambda_c}\big[m_{\Lambda_c}(m_{\Lambda_c}^2-m_{\Lambda}^2-q^2)\\
&
\hspace{2cm}-m_{\Lambda}(m_{\Lambda_c}^2-m_{\Lambda}^2)\big],\\
A_{22} & = \frac{1}{4m_{\Lambda_c}^2}XY^2,\\
A_{23} & = \frac{1}{2m_{\Lambda_c}m_\Lambda}XY^2,\\
A_{24} & = Y(m_{\Lambda_c}^2-m_{\Lambda}^2-q^2),\\
A_{33} & = \frac{1}{4m_{\Lambda}^2}XY^2,\\
A_{34} & = \frac{m_{\Lambda_c}}{m_{\Lambda}}Y(m_{\Lambda_c}^2-m_{\Lambda}^2-q^2),\\
A_{44} & = m_{\Lambda_c}^2Y,
\end{align*}
\end{minipage}
\begin{minipage}{0.48\textwidth}
\begin{align*}
A_{11}^\prime & = XY,\\
A_{12}^\prime & = \frac{1}{m_{\Lambda_c}} (m_{\Lambda_c}+m_{\Lambda})XY,\\
A_{13}^\prime & = \frac{1}{m_{\Lambda}} (m_{\Lambda_c}+m_{\Lambda})XY,\\
A_{14}^\prime & = 2m_{\Lambda_c}\big[m_{\Lambda_c}(m_{\Lambda_c}^2-m_{\Lambda}^2-q^2)\\
& \hspace{2cm} +m_{\Lambda}(m_{\Lambda_c}^2-m_{\Lambda}^2)\big],\\
A_{22}^\prime & = \frac{1}{4m_{\Lambda_c}^2}X^2Y,\\
A_{23}^\prime & = \frac{1}{2m_{\Lambda_c}m_\Lambda}X^2Y,\\
A_{24}^\prime & = X(m_{\Lambda_c}^2-m_{\Lambda}^2-q^2),\\
A_{33}^\prime & = \frac{1}{4m_{\Lambda}^2}X^2Y,\\
A_{34}^\prime & = \frac{m_{\Lambda_c}}{m_{\Lambda}}X(m_{\Lambda_c}^2-m_{\Lambda}^2-q^2),\\
A_{44}^\prime & = m_{\Lambda_c}^2X,\\
\end{align*}
\end{minipage}

\begin{equation*}
 \gamma=\sum_{i=1,j=1}^{i=4,j=4}\frac{1}{Z}C_{ij}F_iG_j,
\end{equation*}

\begin{align*}
C_{11} & = 2XY,\\
C_{14} & = -2m_{\Lambda_c}m_\Lambda X,\\
C_{41} & = -2m_{\Lambda_c}m_\Lambda Y,\\
C_{44} & = -4m_{\Lambda_c}^2 m_\Lambda^2.\\
\end{align*}

\subsubsection{\texorpdfstring{$5/2^{+}$}{}}

\begin{equation}
 \alpha =\sum_{i=1,j=1}^{i=4,j=4}\frac{1}{Z}(B_{ij}F_iF_j+B_{ij}^\prime G_iG_j),
\end{equation}
where $Z^\prime = 20m_{\Lambda_c}^4m_{\Lambda}^4$ and the non vanishing coefficients are
\begin{align*}
B_{11} &= -X^2Y^3,\\
B_{14} & = B_{14}^\prime = 2m_{\lambda_c}m_{\Lambda}X^2Y^2,\\
B_{44} &= -3m_{\Lambda_c}^2m_{\Lambda}^2X^2Y,\\
B_{11}^\prime &=-X^3Y^2 ,\\
B_{44}^\prime &= -3m_{\Lambda_c}^2m_{\Lambda}^2XY^2,\\
\end{align*}

\begin{equation}
 \beta_{++}=\sum_{i=1,j=1}^{i=4,j=4}\frac{1}{Z}(A_{ij}F_iF_j+A_{ij}^\prime G_iG_j),
\end{equation}

\begin{minipage}{0.48\textwidth}
\begin{align*}
A_{11}  = & X^2Y^2,\\
A_{12}  = &\frac{1}{m_{\Lambda_c}} (m_{\Lambda_c}+m_{\Lambda})X^2Y^2,\\
A_{13}  = &\frac{1}{m_{\Lambda}}(m_{\Lambda_c}+m_{\Lambda})X^2Y^2,\\
A_{14}  = & 2m_{\Lambda_c}XY\big[m_{\Lambda_c}(m_{\Lambda_c}^2-m_{\Lambda}^2-q^2)\\
& \hspace{1.5cm}-m_{\Lambda}(m_{\Lambda_c}^2-m_{\Lambda}^2)\big],\\
A_{22}  = &\frac{1}{4m_{\Lambda_c}^2}X^3Y^2,\\
A_{23}  = & \frac{1}{2m_{\Lambda_c}m_\Lambda}X^3Y^2,\\
A_{24}  = & X^2Y(m_{\Lambda_c}^2-m_{\Lambda}^2-q^2),\\
A_{33}  = & \frac{1}{4m_{\Lambda}^2}X^3Y^2,\\
A_{34}  = & \frac{m_{\Lambda_c}}{m_{\Lambda}}X^2Y(m_{\Lambda_c}^2-m_{\Lambda}^2-q^2),\\
A_{44}  = & m_{\Lambda_c}^2X\lambda(m_{\Lambda_c}^2,m_\Lambda^2,q^2),
\end{align*}
\end{minipage}
\begin{minipage}{0.48\textwidth}
\begin{align*}
A_{11}^\prime  = & X^2Y^2,\\
A_{12}^\prime  = & \frac{-1}{m_{\Lambda_c}} (m_{\Lambda_c}-m_{\Lambda})X^2Y^2,\\
A_{13}^\prime  = & \frac{-1}{m_{\Lambda}} (m_{\Lambda_c}-m_{\Lambda})X^2Y^2,\\
A_{14}^\prime  = &-2m_{\Lambda_c}XY\big[m_{\Lambda_c}(m_{\Lambda_c}^2-m_{\Lambda}^2-q^2) \\
 &\hspace{2cm} -m_{\Lambda}(m_{\Lambda_c}^2-m_{\Lambda}^2)\big],\\
A_{22}^\prime  = & \frac{1}{4m_{\Lambda_c}^2}X^2Y^3,\\
A_{23}^\prime  = &\frac{1}{2m_{\Lambda_c}m_\Lambda}X^2Y^3,\\
A_{24}^\prime  = & XY^2(m_{\Lambda_c}^2-m_{\Lambda}^2-q^2),\\
A_{33}^\prime  = & \frac{1}{4m_{\Lambda}^2}X^2Y^3,\\
A_{34}^\prime  = & \frac{m_{\Lambda_c}}{m_{\Lambda}}XY^2(m_{\Lambda_c}^2-m_{\Lambda}^2-q^2),\\
A_{44}^\prime  = & m_{\Lambda_c}^2Y\lambda(m_{\Lambda_c}^2,m_\Lambda^2,q^2), 
\end{align*}
\end{minipage}

\vspace{0.5cm}
where $\lambda(m_{\Lambda_c}^2,m_\Lambda^2,q^2) = (m_{\Lambda_c}^4+m_{\Lambda}^4+q^4-2m_{\Lambda_c}^2m_{\Lambda}^2-2m_{\Lambda_c}^2q^2-2m_{\Lambda}^2q^2)$.
\begin{equation*}
 \gamma=\sum_{i=1,j=1}^{i=4,j=4}\frac{1}{Z}C_{ij}F_iG_j,
\end{equation*}

\begin{align*}
C_{11} & = 2X^2Y^2,\\
C_{14} & = -2m_{\Lambda_c}m_\Lambda XY^2,\\
C_{41} & = -2m_{\Lambda_c}m_\Lambda X^2Y,\\
C_{44} & = -2m_{\Lambda_c}^2 m_\Lambda^2 XY.
\end{align*}
\subsection{Hadron tensor in \texorpdfstring{$\Lambda_c^{+}\to\Lambda^{*} l^{+}\nu_l\to\Sigma\pi l^{+}\nu_l/ NK l^{+}\nu_l$}{} transitions} 
 
The most general form of the contribution of the $i$th state to the matrix element for the four-body decay $\Lambda_c^{+}\to \Lambda_il^{+}\nu_l\to BMl^{+}\nu_l$ can be written
\begin{equation*}
M^i_\nu =\bar{u}(p_B)\left(\sum_{j=1}^{16}c^i_j {\cal O}_j\right) u(P+L),
\end{equation*}
where the Lorentz-Dirac operators ${\cal O}_i$ are
\begin{align*}
{\cal O}_1=&\gamma_\nu ,\,\,\,\, {\cal O}_2=\slashed{P}\gamma_\nu,\,\,\,\, {\cal O}_3= P_\nu,\,\,\,\, {\cal O}_4=\slashed{P}P_\nu,\,\,\,\,
{\cal O}_5= L_\nu,\,\,\,\, {\cal O}_6=\slashed{P}L_\nu,\,\,\,\, {\cal O}_7= Q_\nu,\,\,\,\, {\cal O}_8=\slashed{P}Q_\nu,\nonumber\\
{\cal O}_9=&\gamma_\nu\gamma_5 ,\,\,\,\, {\cal O}_{10}=\slashed{P}\gamma_\nu\gamma_5,\,\,\,\, {\cal O}_{11}= P_\nu\gamma_5,\,\,\,\, {\cal O}_{12}=\slashed{P}P_\nu\gamma_5,\,\,\,\,
{\cal O}_{13}= L_\nu\gamma_5,\,\,\,\, {\cal O}_{14}=\slashed{P}L_\nu\gamma_5,\,\,\,\, {\cal O}_{15}= Q_\nu\gamma_5,\,\,\,\, {\cal O}_{16}=\slashed{P}Q_\nu\gamma_5.
\end{align*}
Thus, there are sixteen independent Lorentz-Dirac structures in the amplitude. 
The $c^i_j$ can be written
\begin{equation}
c^i_j =g_{\Lambda_i BM}\sum_{k}(C^{i,F}_{jk}F_k+C^{i,G}_{jk}G_k).
\end{equation}
where $k$ runs from $1$ to $3$ for spin $\frac{1}{2}$ states and from $1$ to $4$ for states with higher spin.

The hadron tensor arising from a single intermediate $\Lambda_i$ can be written
\begin{eqnarray}\label{htens}
H^i_{\mu\nu} =&& \sum_{\text{spins}}M^{i\dagger}_\mu M^i_\nu
= \alpha^i g_{\mu\nu} +\beta_{PP}^iP_\mu P_\nu +\beta_{PQ}^iP_\mu Q_\nu
				+\beta_{QP}^iQ_\mu P_\nu +\beta_{QQ}^iQ_\mu Q_\nu\nonumber\\
				&+&\beta_{QL}^iQ_\mu L_\nu
				+\beta_{LQ}^iL_\mu Q_\nu 
				+\beta_{LL}^iL_\mu L_\nu +\beta_{PL}^iP_\mu L_\nu
				+\beta_{PL}^iP_\mu L_\nu\nonumber\\
				&+& i\gamma_a^i\epsilon^{\mu\nu\rho\delta} P_\rho Q_\delta
				+ i\gamma_b^i\epsilon^{\mu\nu\rho\delta} L_\rho P_\delta
				+ i\gamma_c^i\epsilon^{\mu\nu\rho\delta} L_\rho Q_\delta
				+ i\gamma_d^i\epsilon^{\sigma\mu\rho\delta}L_\sigma P_\rho Q_\delta P_\nu
				+ i\gamma_e^i\epsilon^{\sigma\mu\rho\delta}L_\sigma P_\rho Q_\delta Q_\nu\nonumber\\
				&+& i\gamma_f^i\epsilon^{\sigma\mu\rho\delta}L_\sigma P_\rho Q_\delta L_\nu				
				+ i\gamma_g^i\epsilon^{\sigma\nu\rho\delta}L_\sigma P_\rho Q_\delta P_\mu
				+ i\gamma_h^i\epsilon^{\sigma\nu\rho\delta}L_\sigma P_\rho Q_\delta Q_\mu
				+ i\gamma_k^i\epsilon^{\sigma\nu\rho\delta}L_\sigma P_\rho Q_\delta L_\mu.
\end{eqnarray}
The terms in $\gamma^i$ do not contribute to the decay width. 
Because they are proportional to at least one power of the lepton mass, contributions from $\beta^i_{PL}$, $\beta^i_{QL}$, $\beta^i_{LL}$, $\beta^i_{LP}$, $\beta^i_{LQ}$ are small. 
The $\alpha^i$ from each intermediate state considered takes the form 
\begin{equation}
\alpha^i=\sum_{j,k=1}^{16}a^i_{jk}c^{\dag i}_jc^i_k.
\end{equation}
Similarly, for the $\beta^i_{P_1P_2}$ ($P_1$ and $P_2$ denotes $P$, $Q$ or $L$),
\begin{equation}
\beta^i_{P_1P_2}=\sum_{j,k=1}^{16}b^i_{jk}c^{\dag i}_jc^i_k.
\end{equation}
When we treat the coherent sum of the contributions from all the states we consider, we write
\begin{equation}
M_\nu=\bar{u}(p_B)\left(\sum_{j=1}^{16}{\cal C}_j {\cal O}_j\right)u(p_{\Lambda_c})=\bar{u}(p_B)\sum_{i=1}^6\left(\sum_{j=1}^{16}c^i_j {\cal O}_j\right) u(p_{\Lambda_c}),
\end{equation}
which ultimately leads to
\begin{equation}
{\cal C}_j=\sum_{i=1}^6c^i_j.
\end{equation}
In this case, the hadron tensor takes the same form as in eq. \ref{htens} with the superscripts $i$ removed. The coefficients contributing to the differential decay widths we consider are then
\begin{eqnarray}
\alpha&=&\sum_{j,k=1}^{16}a_{jk}{\cal C}^{\dag}_j{\cal C}_k,\nonumber\\
\beta_{P_1P_2}&=&\sum_{j,k=1}^{16}b_{jk}{\cal C}^{\dag}_j{\cal C}_k.
\end{eqnarray}

For each intermediate $\Lambda_i$ we consider, the $c^i_j$ can be written
\begin{equation}
c^i_j =g_{\Lambda_i BM}\sum_{k}(C^{i,F}_{jk}F_k+C^{i,G}_{jk}G_k).
\end{equation}
where $k$ runs from $1$ to $3$ for spin $\frac{1}{2}$ states and from $1$ to $4$ for states with higher spin.
Here, $g_{\Lambda\Sigma\pi}$ is the strong coupling constant for the decay $\Lambda^i\to BM$.

For future convenience, we define
\begin{align*}
B = \frac{1}{6 m_{\Lambda}^2 m_{\Lambda_c}},\hspace{1cm}
D = \frac{1}{20 m_{\Lambda}^6 m_{\Lambda_c}^2},\hspace{1cm}
C_1  =(P\cdot Q), \hspace{1cm}
C_2 & =(P\cdot L), \hspace{1cm}
C_3  =(Q\cdot L).
\end{align*}

The nonzero coefficients $a_{jk}$s and $b_{jk}$s for $\alpha$ and $\beta$s are listed in the next few subsections. 
\subsubsection{$\alpha$}

\begin{minipage}{0.45\textwidth}
\begin{align*}
a_{1,1}  = & 2 (2 m_{\Lambda_c} m_{\Sigma}-C_2-C_1-C_3-S_{\Sigma\pi}),\\
a_{1,2}  = & 2 (m_{\Lambda_c} (C_1+S_{\Sigma\pi})-2 m_{\Sigma} C_2-2 m_{\Sigma} S_{\Sigma\pi}),\\
\end{align*}
\end{minipage}
\begin{minipage}{0.45\textwidth}
\begin{align*}
a_{9,9}  = & 2 (2 m_{\Lambda_c} m_{\Sigma}+C_2+C_1+C_3+S_{\Sigma\pi}),\\
a_{9,10} = & 2 (m_{\Lambda_c} C_1+m_{\Lambda_c} S_{\Sigma\pi}+2 m_{\Sigma} C_2+2 m_{\Sigma} S_{\Sigma\pi}),\\
\end{align*}
\end{minipage}

\vspace{-0.5cm}

\begin{align*}
a_{2,2}  = & 2 \big((2 m_{\Lambda_c} m_{\Sigma}+C_2-C_1+C_3-S_{\Sigma\pi}) S_{\Sigma\pi} -2C_1C_2 \big),\\
a_{10,10}  = & -2 \big( (2 m_{\Lambda_c} m_{\Sigma}+C_2+C_1-C_3+S_{\Sigma\pi}) S_{\Sigma\pi}+2C_1C_2 \big).
\end{align*}
\subsubsection{$\beta_{PP}$}
\begin{align*}
b_{1,1}  = b_{9,9} = 4, \hspace{0.5cm}
b_{1,2}  =b_{2,1} = -b_{9,10} = -b_{10,9} =  16m_{\Sigma},\hspace{0.5cm}
b_{2,2}  =b_{10,10} = 4(2C_1 + S_{\Sigma\pi}),
\end{align*}

\begin{align*}
b_{3,3} & = 2(2m_{\Lambda_c}m_{\Sigma} + C_2 + C_1 + C_3 + S_{\Sigma\pi}),\\
b_{3,4} & = b_{4,3} = 2(4m_{\Sigma}C_2 + 2m_{\Lambda_c}C_1 + 2m_{\Lambda_c}S_{\Sigma\pi} + 4m_{\Sigma}S_{\Sigma\pi}),\\
b_{4,4} & = 2\big(2C_2C_1 +(2m_{\Lambda_c}m_{\Sigma} + C_2 + C_1 - C_3 + S_{\Sigma\pi})S_{\Sigma\pi}\big),\\
b_{11,11} & = 2(-2m_{\Lambda_c}m_{\Sigma} + C_2 + C_1 + C_3 + S_{\Sigma\pi}),\\
b_{11,12} & = b_{12,11} = 2(-4m_{\Sigma}C_2 + 2m_{\Lambda_c}C_1 + 2m_{\Lambda_c}S_{\Sigma\pi} - 4m_{\Sigma}S_{\Sigma\pi}),\\
b_{12,12} & = 2\big(2C_2C_1 +(- 2m_{\Lambda_c}m_{\Sigma} + C_2 + C_1 - C_3 + S_{\Sigma\pi})S_{\Sigma\pi}\big),
\end{align*}

\begin{minipage}{0.45\textwidth}
\begin{align*}
b_{1,3} & = 2(m_{\Lambda_c} + 2m_{\Sigma}),\\
b_{1,4} & = 2(2m_{\Lambda_c}m_{\Sigma} - C_3 + S_{\Sigma\pi}),\\
b_{2,3} & = 2(2m_{\Lambda_c}m_{\Sigma} + 2C_1 + C_3 + S_{\Sigma\pi}),\\
b_{2,4} & = 2(2m_{\Lambda_c}C_1 + m_{\Lambda_c}S_{\Sigma\pi} + 2m_{\Sigma}S_{\Sigma\pi}),
\end{align*}
\end{minipage}
\begin{minipage}{0.45\textwidth}
\begin{align*}
b_{9,11} & = 2(m_{\Lambda_c} - 2m_{\Sigma}),\\
b_{9,12} & = 2(-2m_{\Lambda_c}m_{\Sigma} - C_3 + S_{\Sigma\pi}),\\
b_{10,11} & = 2(-2m_{\Lambda_c}m_{\Sigma} + 2C_1 + C_3 + S_{\Sigma\pi}),\\
b_{10,12} & = 2(2m_{\Lambda_c}C_1 + m_{\Lambda_c}S_{\Sigma\pi} - 2m_{\Sigma}S_{\Sigma\pi}),
\end{align*}
\end{minipage}
\subsubsection{$\beta_{PQ}$, $\beta_{QP}$}

\begin{align*}
b_{1,1}  & = b_{9,9} = 2, \hspace{0.5cm}b_{2,2} = b_{10,10} = -2S_{\Sigma\pi},\hspace{0.5cm}
b_{1,3}   = b_{9,11}=2m_{\Lambda_c},\\
b_{1,4}  &  =-b_{2,1} = b_{9,12}=- b_{10,11}=2(C_2 + S_{\Sigma\pi}),\hspace{0.5cm}
b_{2,4}   =b_{10,12}=-2m_{\Lambda_c}S_{\Sigma\pi},
\end{align*}

\begin{minipage}{0.45\textwidth}
\begin{align*}
b_{1,7} & = 2(m_{\Lambda_c} + 2m_{\Sigma}),\\
b_{1,8} & = 2(2m_{\Lambda_c}m_{\Sigma} - C_3 + S_{\Sigma\pi}),\\
b_{2,7} & = 2(2m_{\Lambda_c}m_{\Sigma} + 2C_1 + C_3 + S_{\Sigma\pi}),\\
b_{2,8} & = 2(2m_{\Lambda_c}C_1 + m_{\Lambda_c}S_{\Sigma\pi} + 2m_{\Sigma}S_{\Sigma\pi}),\\
b_{3,7} & = 2(2m_{\Lambda_c}m_{\Sigma} + C_2 + C_1 + C_3 + S_{\Sigma\pi}),
\end{align*}
\end{minipage}
\begin{minipage}{0.45\textwidth}
\begin{align*}
b_{9,15} & = 2(m_{\Lambda_c} - 2m_{\Sigma}),\\
b_{9,16} & = 2(-2m_{\Lambda_c}m_{\Sigma} - C_3 + S_{\Sigma\pi}),\\
b_{10,15} & = 2(-2m_{\Lambda_c}m_{\Sigma} + 2C_1 + C_3 + S_{\Sigma\pi}),\\
b_{10,16} & = 2(2m_{\Lambda_c}C_1 + m_{\Lambda_c}S_{\Sigma\pi} - 2m_{\Sigma}S_{\Sigma\pi}),\\
b_{11,15} & = 2(-2m_{\Lambda_c}m_{\Sigma} + C_2 + C_1 + C_3 + S_{\Sigma\pi}),
\end{align*}
\end{minipage}

\begin{align*}
b_{3,8} & =b_{4,7}= 2(2m_{\Sigma}C_2 + m_{\Lambda_c}C_1 + m_{\Lambda_c}S_{\Sigma\pi} + 2m_{\Sigma}S_{\Sigma\pi}),\\
b_{4,8} & =2\big(2C_2C_1 + (2m_{\Lambda_c}m_{\Sigma}+ C_2 + C_1 - C_3 + S_{\Sigma\pi})S_{\Sigma\pi}\big),\\
b_{11,16} & = b_{12,15} = 2(-2m_{\Sigma}C_2 + m_{\Lambda_c}C_1 + m_{\Lambda_c}S_{\Sigma\pi} - 2m_{\Sigma}S_{\Sigma\pi}),\\
b_{12,16} & = 2\big(2C_2C_1 +(- 2m_{\Lambda_c}m_{\Sigma} + C_2 + C_1 - C_3 + S_{\Sigma\pi})S_{\Sigma\pi}\big).
\end{align*}

\subsubsection{$\beta_{QQ}$}

\begin{align*}
b_{1,7}  = b_{9,15} =  4m_{\Lambda_c},\hspace{0.5cm}
b_{1,8}  = -b_{2,7}=b_{9,16}= -b_{10,15} = 2(2C_2 + 2S_{\Sigma\pi}),\hspace{0.5cm}
b_{2,8}  = b_{10,16}= -4m_{\Lambda_c}S_{\Sigma\pi},
\end{align*}

\begin{align*}
b_{7,7} & = 2(2m_{\Lambda_c}m_{\Sigma} + C_2 + C_1 + C_3 + S_{\Sigma\pi}),\\
b_{7,8} & = b_{8,7} = 2(4m_{\Sigma}C_2 + 2m_{\Lambda_c}C_1 + 2m_{\Lambda_c}S_{\Sigma\pi} + 4m_{\Sigma}S_{\Sigma\pi}), \\
b_{8,8} & = 2\big(2C_2C_1 + (2m_{\Lambda_c}m_{\Sigma} + C_2 + C_1 - C_3 + S_{\Sigma\pi})S_{\Sigma\pi}\big), \\
b_{15,15} & = 2(-2m_{\Lambda_c}m_{\Sigma} + C_2 + C_1 + C_3 + S_{\Sigma\pi}),\\
b_{15,16} & = b_{16,15} = 2(-4m_{\Sigma}C_2 + 2m_{\Lambda_c}C_1 + 2m_{\Lambda_c}S_{\Sigma\pi} - 4m_{\Sigma}S_{\Sigma\pi}), \\
b_{16,16} & = 2\big(2C_2C_1 +(- 2m_{\Lambda_c}m_{\Sigma} + C_2 + C_1 - C_3 + S_{\Sigma\pi})S_{\Sigma\pi}\big),
\end{align*}

\subsubsection{$\beta_{QL}$, $\beta_{LQ}$}

\begin{minipage}{0.35\textwidth}
\begin{align*}
b_{1,1} & = b_{9,9} = 2, \hspace{0.5cm}\\
b_{2,2} & = b_{10,10} = -2S_{\Sigma\pi},\\
b_{1,5} & = b_{9,13}=  2m_{\Lambda_c},\\
b_{1,6} & = -b_{2,5} = b_{9,14} = -b_{10,13} = 2(C_2 + S_{\Sigma\pi}),\\ 
b_{2,6} & =b_{10,14} = -2m_{\Lambda_c}S_{\Sigma\pi},\\
b_{1,7} & =  -b_{9,15} =  4m_{\Sigma},\\
b_{1,8} & = b_{2,7}  = b_{9,16} =b_{10,15}=  2(C_1 + S_{\Sigma\pi}), \\
b_{2,8} & = -b_{10,16} = 4m_{\Sigma}S_{\Sigma\pi},
\end{align*}
\end{minipage}
\begin{minipage}{0.6\textwidth}
\begin{align*}
 b_{5,7} & = b_{13,15} = 2(2m_{\Lambda_c}m_{\Sigma} + C_2 + C_1 + C_3 + S_{\Sigma\pi}),\\ 
 b_{5,8} & = b_{6,7} = 2(2m_{\Sigma}C_2 + m_{\Lambda_c}C_1 + (m_{\Lambda_c} + 2m_{\Sigma})S_{\Sigma\pi}), \\
 b_{6,8} & =2(2C_2C_1 + (2m_{\Lambda_c}m_{\Sigma} + C_2 + C_1 - C_3)S_{\Sigma\pi} + S_{\Sigma\pi}^2),\\
 b_{13,16} & = b_{14,15} =2(-2m_{\Sigma}C_2 + m_{\Lambda_c}C_1 + (m_{\Lambda_c} - 2m_{\Sigma})S_{\Sigma\pi}),\\ 
 b_{14,16} & =2(2C_2C_1 + (-2m_{\Lambda_c}m_{\Sigma} + C_2 + C_1 - C_3)S_{\Sigma\pi} + S_{\Sigma\pi}^2).
\end{align*}
\end{minipage}

\subsubsection{$\beta_{LL}$}

\begin{align*}
b_{1,5}  = -b_{9,13} = 8m_{\Sigma}, \hspace{0.5cm}
b_{1,6}  = b_{2,5}  =b_{9,14}= b_{10,13}=2(2C_1 + 2S_{\Sigma\pi}),\hspace{0.5cm}
b_{2,6}  = -b_{10,14} = 8m_{\Sigma}S_{\Sigma\pi}.
\end{align*}

\begin{align*}
 b_{5,5} & = 2(2m_{\Lambda_c}m_{\Sigma} + C_2 + C_1 + C_3 + S_{\Sigma\pi}),\\
 b_{5,6} & = b_{6,5}= 2(4m_{\Sigma}C_2 + 2m_{\Lambda_c}C_1 + 2m_{\Lambda_c}S_{\Sigma\pi} + 4m_{\Sigma}S_{\Sigma\pi}),\\ 
 b_{6,6} & =  2\big(2C_2C_1 + (2m_{\Lambda_c}m_{\Sigma} + C_2 + C_1 - C_3 + S_{\Sigma\pi})S_{\Sigma\pi}\big),\\
 b_{13,13} & = 2(-2m_{\Lambda_c}m_{\Sigma} + C_2 + C_1 + C_3 + S_{\Sigma\pi}),\\
 b_{13,14} & = b_{14,13} = 2(-4m_{\Sigma}C_2 + 2m_{\Lambda_c}C_1 + 2m_{\Lambda_c}S_{\Sigma\pi} - 4m_{\Sigma}S_{\Sigma\pi}),\\ 
 b_{14,14} & = 2\big(2C_2C_1 -(2m_{\Lambda_c}m_{\Sigma} - C_2 - C_1 + C_3 - S_{\Sigma\pi})S_{\Sigma\pi}\big).
\end{align*}

\subsubsection{$\beta_{PL}$, $\beta_{LP}$}
\begin{align*}
b_{1,1} & = b_{9,9} = 2, \hspace{0.5cm}
b_{1,2} = b_{2,1} = - b_{9,10} = -b_{10,9} = 8m_{\Sigma}, \hspace{0.5cm}
b_{2,2} = b_{10,10} =  2(2C_1 + S_{\Sigma\pi}),\\
b_{1,3} & = -b_{9,11} = 4m_{\Sigma},\hspace{0.5cm}
b_{1,4} =  b_{2,3} = b_{9,12} =  b_{10,11} = 2(C_1 + S_{\Sigma\pi}), \hspace{0.5cm}
b_{2,4} =-b_{10,12}=  4m_{\Sigma}S_{\Sigma\pi},
\end{align*}

\begin{align*}
b_{3,5}  & = 2(2m_{\Lambda_c}m_{\Sigma} + C_2 + C_1 + C_3 + S_{\Sigma\pi}), \\
b_{3,6} & =  b_{4,5} =  2(2m_{\Sigma}C_2 + m_{\Lambda_c}C_1 + m_{\Lambda_c}S_{\Sigma\pi} + 2m_{\Sigma}S_{\Sigma\pi}),\\ 
b_{4,6} & =  2\big(2C_1C_2+ (2m_{\Lambda_c}m_{\Sigma} + C_2 - C_3 + S_{\Sigma\pi}  +C_1) S_{\Sigma\pi}\big),\\
b_{11,13} & = 2(-2m_{\Lambda_c}m_{\Sigma} + C_2 + C_1 + C_3 + S_{\Sigma\pi}), \\
b_{11,14} & = b_{12,13} = 2(-2m_{\Sigma}C_2 + m_{\Lambda_c}C_1 + m_{\Lambda_c}S_{\Sigma\pi} - 2m_{\Sigma}S_{\Sigma\pi}), \\
b_{12,14} & = 2\big(2C_1C_2 -(2m_{\Lambda_c}m_{\Sigma} - C_2 + C_3 - C_1- S_{\Sigma\pi} ) S_{\Sigma\pi}\big), 
\end{align*}

\begin{minipage}{0.45\textwidth}
\begin{align*}
b_{1,5} & =  2(m_{\Lambda_c} + 2m_{\Sigma}), \\
b_{1,6} & =  2(2m_{\Lambda_c}m_{\Sigma} - C_3 + S_{\Sigma\pi}), \\
b_{2,5} & =  2(2m_{\Lambda_c}m_{\Sigma} + 2C_1 + C_3 + S_{\Sigma\pi}), \\
b_{2,6} & =  2(2m_{\Lambda_c}C_1 + m_{\Lambda_c}S_{\Sigma\pi} + 2m_{\Sigma}S_{\Sigma\pi}), 
\end{align*}
\end{minipage}
\begin{minipage}{0.45\textwidth}
\begin{align*}
b_{9,13} & =  2(m_{\Lambda_c} - 2m_{\Sigma}),\\
b_{9,14} & =  2(-2m_{\Lambda_c}m_{\Sigma} - C_3 + S_{\Sigma\pi}), \\
b_{10,13} & =  2(-2m_{\Lambda_c}m_{\Sigma} + 2C_1 + C_3 + S_{\Sigma\pi}), \\
b_{10,14} & =  2(2m_{\Lambda_c}C_1 + m_{\Lambda_c}S_{\Sigma\pi} - 2m_{\Sigma}S_{\Sigma\pi}). 
\end{align*}
\end{minipage}

\vspace{0.5cm}

The nonzero terms in the coefficients $c^i_j$  are listed in the following subsections.

\subsubsection{\texorpdfstring{$\Lambda_1=\Lambda\frac{1}{2}^{+}(1115)$, $\Lambda_2=\Lambda\frac{1}{2}^{+}(1600)$}{}}

\begin{minipage}{0.3\textwidth}
\begin{align*}
C^{G}_{1,1} & =  -C^{F}_{9,1} = M_{\Gamma},\\
C^{G}_{2,1} & =  - C^{F}_{10,1} = -1, 
\end{align*}
\end{minipage}
\begin{minipage}{0.3\textwidth}
\begin{align*}
C^{G}_{3,2} & =  -C^{F}_{11,2} = C^{G}_{5,2} = -C^{F}_{13,2} =-\frac{M_{\Gamma}}{m_{\Lambda_c}},\\
C^{G}_{4,2} & =  -C^{F}_{12,2}  = C^{G}_{6,2} =  -C^{F}_{14,2} = \frac{1}{ m_{\Lambda_c}},
\end{align*}
\end{minipage}
\begin{minipage}{0.3\textwidth}
\begin{align*}
C^{G}_{3,3} & = - C^{F}_{11,3} =  - \frac{M_{\Gamma}}{m_{\Lambda}},\\
C^{G}_{4,3} & = -C^{F}_{12,3} = \frac{1}{m_{\Lambda}}.
\end{align*}
\end{minipage}

\subsubsection{\texorpdfstring{$\Lambda_3=\Lambda\frac{1}{2}^{-}(1405)$}{}}
\begin{minipage}{0.3\textwidth}
\begin{align*}
C^{G}_{1,1} & = -C^{F}_{9,1} = -M_{\Gamma}, \\
C^{G}_{2,1} & = -C^{F}_{10,1} = -1, 
\end{align*}
\end{minipage}
\begin{minipage}{0.3\textwidth}
\begin{align*}
C^{G}_{3,2} & = - C^{F}_{11,2} = C^{G}_{5,2} = - C^{F}_{13,2} = -\frac{M_{\Gamma}}{m_{\Lambda_c}},\\
C^{G}_{4,2} & = C^{F}_{12,2}= -C^{G}_{6,2} = C^{F}_{14,2}  = \frac{1}{m_{\Lambda_c}},
\end{align*}
\end{minipage}
\begin{minipage}{0.3\textwidth}
\begin{align*}
C^{G}_{3,3} & = - C^{F}_{11,3} = -\frac{M_{\Gamma}}{m_{\Lambda}},\\
C^{G}_{4,3} & = - C^{F}_{12,3} = -\frac{1}{m_{\Lambda}}.
\end{align*}
\end{minipage}

\subsubsection{\texorpdfstring{$\Lambda_4=\Lambda\frac{3}{2}^{-}(1520)$}{}}

\begin{align*}
C^{G}_{1,1} & = -C^{F}_{9,1} =B \bigg[\big(2 M_{\Gamma}-m_{\Lambda}\big)S_{\Sigma\pi}^2 +\big[M_{\Gamma} \big[m_{\Lambda}(-m_{\Lambda}+2m_{\Sigma})-2 (C_1-C_2)\big]-m_{\Lambda}(2 m_{\Lambda}m_{\Sigma}+C_1)\big] S_{\Sigma\pi}\\
&
 + M_{\Gamma}\big[3 m_{\Lambda}^2 (C_1-C_2+C_3)+2C_2(m_{\Lambda}m_{\Sigma} -C_1)\big]-2 m_{\Lambda} C_2 C_1\bigg]\\
%
%
C^{G}_{1,4} & =-C^{F}_{9,4} =m_{\Lambda} m_{\Lambda_c}B\bigg[- M_{\Gamma} S_{\Sigma\pi} 
+ M_{\Gamma} (2 m_{\Lambda}m_{\Sigma}+C_1)-2 m_{\Lambda}C_1\bigg] \\
%
%
C^{G}_{2,1} & =-C^{F}_{10,1}= B\bigg[-2S_{\Sigma\pi}^2+\bigg[-M_{\Gamma} m_{\Lambda}+ m_{\Lambda}(3m_{\Lambda}+2m_{\Sigma})
+2 (C_1-C_2)\bigg]S_{\Sigma\pi}\\
&
+m_{\Lambda}\big[-M_{\Gamma}(2m_{\Lambda} m_{\Sigma}+2 C_2+C_1)+ m_{\Lambda}(3C_2-C_1-3C_3)+2
m_{\Sigma} C_2\big] +2 C_2 C_1\bigg]\\
%
%
C^{G}_{2,4} &= -C^{F}_{10,4}= m_{\Lambda} m_{\Lambda_c}B\bigg[S_{\Sigma\pi}-2 M_{\Gamma} m_{\Lambda}+2 m_{\Lambda}
m_{\Sigma}-C_1\bigg]\\
%
%
C^{G}_{3,2} & = \frac{m_{\Lambda}}{m_{\Lambda_c}}C^{G}_{3,3}=C^{G}_{5,2}
=\frac{B}{m_{\Lambda_c}}\bigg[-2 M_{\Gamma}S_{\Sigma\pi}^2
+ \big[M_{\Gamma}\big[m_{\Lambda}(3 m_{\Lambda}-2m_{\Sigma})-m_{\Lambda}m_{\Lambda_c}+2(C_1-C_2)\big]+2 m_{\Lambda}C_1\big]S_{\Sigma\pi}\\
& 
+\Big[ M_{\Gamma}\big[3 m_{\Lambda}^2 (C_2-C_1-C_3)+m_{\Lambda_c}m_{\Lambda}(2 m_{\Lambda}m_{\Sigma}+C_1)
 -2C_2( m_{\Lambda}m_{\Sigma} -C_1)\big]-2m_{\Lambda} C_1(m_{\Lambda}m_{\Lambda_c}-C_2)\Big]\bigg]\\
%
%
C^{G}_{3,4} & =-C^{F}_{11,4}= m_{\Lambda_c}B\bigg[-2 M_{\Gamma} S_{\Sigma\pi}
+M_{\Gamma} \big[m_{\Lambda}(3 m_{\Lambda}-2m_{\Sigma})+2C_1\big]+2m_{\Lambda}C_1\bigg]\\
%
%
C^{G}_{4,2} & =\frac{m_{\Lambda}}{m_{\Lambda_c}}C^{G}_{4,3}=C^{G}_{6,2}
= \frac{B}{m_{\Lambda_c}} \bigg[2S_{\Sigma\pi}^2
+\big[ m_{\Lambda}(2 M_{\Gamma}-3m_{\Lambda}-2m_{\Sigma}+m_{\Lambda_c})-2 (C_1-C_2)\big]S_{\Sigma\pi}\\
& 
+m_{\Lambda}\big[-2 (M_{\Gamma}+m_{\Sigma})( m_{\Lambda_c}m_{\Lambda}-C_2)-3 m_{\Lambda}(C_2- C_1- C_3)-m_{\Lambda_c}C_1\big]-2 C_2 C_1\bigg]\\
%
%
C^{G}_{4,4} & = -C^{F}_{12,4} = m_{\Lambda_c}B\bigg[2 S_{\Sigma\pi}+ m_{\Lambda}\big[2 M_{\Gamma} -3 m_{\Lambda}-2m_{\Sigma}\big]-2 C_1\bigg]
\\
%
%
C^{F}_{11,2} & =\frac{m_{\Lambda}}{m_{\Lambda_c}}C^{F}_{11,3} =C^{F}_{13,2}
= \frac{B}{m_{\Lambda_c}}\bigg[2 M_{\Gamma}S_{\Sigma\pi}^2
- \Big[M_{\Gamma}\big[m_{\Lambda}(3 m_{\Lambda}-2m_{\Sigma})+m_{\Lambda}m_{\Lambda_c}+2(C_1-C_2)\big]-2 m_{\Lambda}C_1\Big]S_{\Sigma\pi}\\
& 
- M_{\Gamma}\big[3 m_{\Lambda}^2 (C_2-C_1-C_3)-m_{\Lambda_c}m_{\Lambda}(2 m_{\Lambda}m_{\Sigma}+C_1)
 -2C_2( m_{\Lambda}m_{\Sigma} +C_1)\big]-2m_{\Lambda} C_1(m_{\Lambda}m_{\Lambda_c}+C_2)\bigg]\\
%
C^{F}_{12,2} & = \frac{m_{\Lambda}}{m_{\Lambda_c}}C^{F}_{12,3}= C^{F}_{14,2} 
=-\frac{B}{m_{\Lambda_c}} \bigg[2S_{\Sigma\pi}^2
+\big[ m_{\Lambda}(2 M_{\Gamma}-3m_{\Lambda}-2m_{\Sigma}-m_{\Lambda_c})-2 (C_1-C_2)\big]S_{\Sigma\pi}\\
& 
+m_{\Lambda}\big[2 (M_{\Gamma}-m_{\Sigma})( m_{\Lambda_c}m_{\Lambda}+C_2)-3 m_{\Lambda}(C_2- C_1- C_3)
+m_{\Lambda_c}C_1\big]-2 C_2 C_1\bigg]\\
%
C^{G}_{7,4} & =  -C^{F}_{15,4} =  -\frac{M_{\Gamma}}{2},\hspace{1cm}
C^{G}_{8,4}   = -C^{F}_{16,4} =\frac{1}{2}.
\end{align*}

\subsubsection{\texorpdfstring{$\Lambda_5=\Lambda\frac{3}{2}^{+}(1890)$}{}}

\begin{align*}
C^{G}_{1,1} & = -C^{F}_{9,1} = B\bigg[\big[m_{\Lambda}-2 M_{\Gamma}\big]S_{\Sigma\pi}^2 +\big[M_{\Gamma} \big[m_{\Lambda}(m_{\Lambda}+2m_{\Sigma})+2 (C_1-C_2)\big]-m_{\Lambda}(2 m_{\Lambda}m_{\Sigma}-C_1)\big] S_{\Sigma\pi}\\
& 
 + M_{\Gamma}\big[3 m_{\Lambda}^2 (C_2-C_1-C_3)+2C_2(m_{\Lambda}m_{\Sigma} +C_1)\big]+2 m_{\Lambda} C_2 C_1\bigg],\\
%
%
C^{G}_{1,4} & = -C^{F}_{9,4} =m_{\Lambda} m_{\Lambda_c}B\bigg[ M_{\Gamma} S_{\Sigma\pi} 
+ M_{\Gamma} (2 m_{\Lambda}m_{\Sigma}-C_1)+2 m_{\Lambda}C_1\bigg], \\
%
%
C^{G}_{2,1} & =-C^{F}_{10,1}= B\bigg[-2S_{\Sigma\pi}^2+\bigg[-M_{\Gamma} m_{\Lambda}+ m_{\Lambda}(3m_{\Lambda}-2m_{\Sigma})+2 (C_1-C_2)\bigg]S_{\Sigma\pi}\\
& 
+m_{\Lambda}\big[M_{\Gamma}(2m_{\Lambda} m_{\Sigma}-2 C_2- C_1)+ m_{\Lambda}(3C_2-C_1-3C_3)-2
m_{\Sigma} C_2\big] +2 C_2 C_1\bigg],\\
%
%
C^{G}_{2,4} &= -C^{F}_{10,4}= m_{\Lambda} m_{\Lambda_c}B\bigg[S_{\Sigma\pi}-2 M_{\Gamma} m_{\Lambda}-2 m_{\Lambda}
m_{\Sigma}-C_1\bigg],\\
%
%
C^{G}_{3,2} & = \frac{m_{\Lambda}}{m_{\Lambda_c}}C^{G}_{3,3} = C^{G}_{5,2}
=\frac{B}{m_{\Lambda_c}}\bigg[-2 M_{\Gamma}S_{\Sigma\pi}^2 + \Big[M_{\Gamma}\big[m_{\Lambda}(3 m_{\Lambda}+2m_{\Sigma})+m_{\Lambda}m_{\Lambda_c}+2(C_1-C_2)\big]+2 m_{\Lambda}C_1\Big]S_{\Sigma\pi}\\
& 
+\Big[ M_{\Gamma}\big[3 m_{\Lambda}^2 (C_2-C_1-C_3)+m_{\Lambda_c}m_{\Lambda}(2 m_{\Lambda}m_{\Sigma}-C_1)
 +2C_2( m_{\Lambda}m_{\Sigma} +C_1)\big]+2m_{\Lambda} C_1(m_{\Lambda}m_{\Lambda_c}+C_2)\Big]\bigg],\\
%
%
C^{G}_{3,4} & =-C^{F}_{11,4}= m_{\Lambda_c}B\bigg[-2 M_{\Gamma} S_{\Sigma\pi}+M_{\Gamma} \big[m_{\Lambda}(3 m_{\Lambda}+2m_{\Sigma})+2C_1\big]+2m_{\Lambda}C_1\bigg],\\
%
%
C^{G}_{4,2} & =\frac{m_{\Lambda}}{m_{\Lambda_c}}C^{G}_{4,3}=C^{G}_{6,2}= \frac{B}{m_{\Lambda_c}} \bigg[-2S_{\Sigma\pi}^2+\big[ m_{\Lambda}(-2 M_{\Gamma}+3m_{\Lambda}-2m_{\Sigma}+m_{\Lambda_c})+2 (C_1-C_2)\big]S_{\Sigma\pi}\\
& 
+m_{\Lambda}\big[-2 (M_{\Gamma}+m_{\Sigma})( m_{\Lambda_c}m_{\Lambda}+C_2)+3 m_{\Lambda}(C_2- C_1- C_3)-m_{\Lambda_c}C_1\big]+2 C_2 C_1\bigg],\\
%
%
C^{G}_{4,4} & = -C^{F}_{12,4} = m_{\Lambda_c}B\bigg[-2 S_{\Sigma\pi}+ m_{\Lambda}\big[-2 M_{\Gamma} +3 m_{\Lambda}-2m_{\Sigma}\big]+2 C_1\bigg],
\\
%
%
C^{F}_{11,2} & =\frac{m_{\Lambda}}{m_{\Lambda_c}}C^{F}_{11,3} =C^{F}_{13,2}= \frac{B}{m_{\Lambda_c}}\bigg[2 M_{\Gamma}S_{\Sigma\pi}^2- \Big[M_{\Gamma}\big[m_{\Lambda}(3 m_{\Lambda}+2m_{\Sigma})-m_{\Lambda}m_{\Lambda_c}-2(C_1-C_2)\big]-2 m_{\Lambda}C_1\Big]S_{\Sigma\pi}\\
& 
- M_{\Gamma}\big[3 m_{\Lambda}^2 (C_2-C_1-C_3)-m_{\Lambda_c}m_{\Lambda}(2 m_{\Lambda}m_{\Sigma}-C_1)
 +2C_2( m_{\Lambda}m_{\Sigma} +C_1)\big]+2m_{\Lambda} C_1(m_{\Lambda}m_{\Lambda_c}-C_2)\bigg],\\
%
C^{F}_{12,2} & =\frac{m_{\Lambda}}{m_{\Lambda_c}} C^{F}_{12,3}=C^{F}_{14,2} =-\frac{B}{m_{\Lambda_c}} \bigg[-2S_{\Sigma\pi}^2+\big[ m_{\Lambda}(-2 M_{\Gamma}+3m_{\Lambda}-2m_{\Sigma}-m_{\Lambda_c})+2 (C_1-C_2)\big]S_{\Sigma\pi}\\
& 
+m_{\Lambda}\big[2 (M_{\Gamma}+m_{\Sigma})( m_{\Lambda_c}m_{\Lambda}-C_2)+3 m_{\Lambda}(C_2- C_1- C_3)
+m_{\Lambda_c}C_1\big]+2 C_2 C_1\bigg],\\
%
C^{G}_{7,4} & =  - C^{F}_{15,4} =  -\frac{M_{\Gamma}}{2},\hspace{1cm}
C^{G}_{8,4}   = -C^{F}_{16,4} =-\frac{1}{2}.
\end{align*}

\subsubsection{\texorpdfstring{$\Lambda_6=\Lambda\frac{5}{2}^{+}(1820)$}{}}
The coefficients take the form
\begin{align*}
C^{F(G)}_{ij} =D\Bigg[ T_5S_{\Sigma\pi}^5+T_4S_{\Sigma\pi}^4+T_3S_{\Sigma\pi}^3+T_2S_{\Sigma\pi}^2+T_1S_{\Sigma\pi}+T_0\bigg],
\end{align*}

\begin{align*}
C^{G}_{1,1} & = - C^{F}_{9,1}=D\bigg[2 M_{\Gamma}S_{\Sigma\pi}^5
 -4\Big[ m_{\Lambda}^2 (M_{\Gamma}-m_{\Sigma})+ M_{\Gamma}(C_1-C_2)\Big]S_{\Sigma\pi}^4 
+ T_3 S_{\Sigma\pi}^3 + T_2 S_{\Sigma\pi}^2 +T_1 S_{\Sigma\pi} +T_0 \bigg],\\
&\hspace{0.5cm}\text{where}\\
&\hspace{0.5cm}
T_3=m_{\Lambda}^4 (M_{\Gamma}-8m_{\Sigma})+m_{\Lambda}^2 (2M_{\Gamma}(2 C_1-3C_2+C_3)+4m_{\Sigma}(2 C_2- C_1)+2M_{\Gamma}(C_1^2+C_2^2-4C_1C_2),
\\
&\hspace{0.5cm}
T_2=m_{\Lambda}^6 (M_{\Gamma}+4m_{\Sigma})+m_{\Lambda}^4\Big[M_{\Gamma}\big[4m_{\Sigma}^2+2(C_1-2 C_2+2 C_3)+ q^2\big]+4m_{\Sigma}\big[2C_1-3C_2+C_3\big]\Big]\\
&\hspace{0.8cm} 
+m_{\Lambda}^2\Big[2 M_{\Gamma}\big[C_2 (C_1 -C_2+C_3)-C_1 C_3\big]+4m_{\Sigma}C_2\big[ C_2-2C_1\big]\Big]+4 M_{\Gamma} C_2C_1(C_1-C_2),
\\
&\hspace{0.5cm}
T_1
=-m_{\Lambda}^6\Big[2M_{\Gamma}\big[2 m_{\Sigma}^2-3 C_2+ C_1+3 C_3\big]+4
m_{\Sigma}\big[C_1-C_2+C_3\big]\Big]\\
&\hspace{0.8cm}
+m_{\Lambda}^4\Big[M_{\Gamma}\big[8 m_{\Sigma}^2
C_2-6 C_2^2+8 C_2C_1+6  C_2 C_3-7C_1^2-8 C_1 C_3-2 C_1 q^2\big]
+4m_{\Sigma} C_2 \big[2C_1-C_2+C_3\big]\Big]\\
&\hspace{0.8cm}
+2m_{\Lambda}^2C_2 C_1\Big[ M_{\Gamma} \big[ 2C_1-C_2-C_3\big]-2
m_{\Sigma} C_2\Big]+2 M_{\Gamma} C_2^2C_1^2,
\\
&\hspace{0.5cm}
T_0
=M_{\Gamma}m_{\Lambda}^2 \Big[
m_{\Lambda}^4 \Big(-4m_{\Sigma}^2\big[2C_2+ q^2\big]+5\big[C_1^2+ C_2^2+C_3^2\big]
-2C_1\big[C_2 -2q^2\big]+10\big[C_1-C_2\big] C_3\Big)\\
&\hspace{0.8cm}
+m_{\Lambda}^2\Big(4 m_{\Sigma}^2 C_2^2+2C_2 C_1\big[3C_2-4C_1-5 C_3\big]+C_1^2 q^2\Big)+2C_2^2C_1^2\Big].
\\
%
C^{G}_{1,4} & = -C^{F}_{9,4}= m_{\Lambda}^2 m_{\Lambda_c}D\bigg[ -S_{\Sigma\pi}^4 + \big[m_{\Lambda}^2 +(2C_1- C_2)\big] S_{\Sigma\pi}^3 
+T_2 S_{\Sigma\pi}^2 
+T_1 S_{\Sigma\pi} +T_0 \bigg],\\
&\hspace{0.5cm}\text{where}\\
&\hspace{0.5cm}
T_2 = \big[m_{\Lambda}^2 (2 M_{\Gamma} m_{\Sigma}+C_2-4  C_1-C_3)+C_1(2  C_2- C_1)\big],\\
&\hspace{0.5cm}
T_1=2 m_{\Lambda}^4 ( C_1- M_{\Gamma}m_{\Sigma})+m_{\Lambda}^2\Big[2 M_{\Gamma} 
m_{\Sigma}\big[ C_2-C_1\big]+C_1\big[3C_1-3C_2+ C_3\big]\Big]- C_2 C_1^2,
\\
&\hspace{0.5cm}
T_0=m_{\Lambda}^4\big[M_{\Gamma} m_{\Sigma}-C_1\big]\big[C_1-C_2+ C_3\big]+m_{\Lambda}^2C_2
C_1\big[C_1-M_{\Gamma} m_{\Sigma}\big].
\\
C^{G}_{2,1} & = -C^{F}_{10,1}=D\bigg[-2S_{\Sigma\pi}^5 -4 (C_2-C_1) S_{\Sigma\pi}^4 +T_3 S_{\Sigma\pi}^3
 + T_2 S_{\Sigma\pi}^2 +T_1 S_{\Sigma\pi} +T_0 \bigg],
\\
&\hspace{0.5cm} {\text{where}}\\
&\hspace{0.5cm}
T_3 = \Big[7 m_{\Lambda}^4+m_{\Lambda}^2 (4 M_{\Gamma}m_{\Sigma}-2(C_2+2 C_1C_3)-2 (C_1^2+C_2^2-4C_1C_2)\Big],\\
&\hspace{0.5cm}
T_2=-5 m_{\Lambda}^6+m_{\Lambda}^4\Big[-4 m_{\Sigma}\big[2M_{\Gamma}+m_{\Sigma}\big]+2(C_2- C_1-4 C_3)-q^2\Big]\\
&\hspace{0.8cm}
+m_{\Lambda}^2\Big[4M_{\Gamma}m_{\Sigma}\big[2 C_2-C_1\big] -2 C_2(C_1+C_2+C_3)+2C_1(2C_1+C_3)\Big]
+4 C_2C_1\big[C_2-C_1\big],
\\
&\hspace{0.5cm}
T_1
=m_{\Lambda}^6\Big[4 m_{\Sigma}\big[M_{\Gamma}+m_{\Sigma}\big]+2\big[ C_1-5C_2+5 C_3\big]\Big]\\
&\hspace{0.8cm}
+m_{\Lambda}^4\Big[4M_{\Gamma}m_{\Sigma}\big[2C_1-3 C_2+ C_3\big]-8m_{\Sigma}^2
C_2+2 C_2\big[5C_2-2 C_1-5C_3\big]-C_1\big[C_1-4 C_3+2 q^2\big]\Big]\\
&\hspace{0.8cm}
+m_{\Lambda}^2\Big[4 M_{\Gamma}m_{\Sigma} C_2\big[C_2-2C_1\big]+2 C_2C_1\big[C_2
+2 C_1+C_3\big]\Big]-2 C_2^2 C_1^2,
\\
&\hspace{0.5cm}
T_0
=m_{\Lambda}^6\Big[4 M_{\Gamma}m_{\Sigma}\big[ C_2-C_1-C_3\big]+4m_{\Sigma}^2\big[2C_2+q^2\big]-\big[C_1^2+5C_2^2+5C_3^2\big]-2C_2\big[C_1-5C_3\big]\\
&\hspace{0.8cm}
-2 C_1\big[3C_3-2 q^2\big]\Big]
+m_{\Lambda}^4
\Big[4 M_{\Gamma}m_{\Sigma}C_2 \big[2 C_1-C_2+C_3\big]-4m_{\Sigma}^2 C_2^2\\
&\hspace{0.8cm}
+C_1\big[2 C_2(3C_3-C_2)-C_1q^2\big]\Big]
-2 M_{\Gamma}m_{\Lambda}^2m_{\Sigma} C_2^2 C_1.
\\
%
C^{G}_{24} & = -C^{F}_{10,4} = m_{\Lambda}^2 m_{\Lambda_c}D\bigg[ M_{\Gamma} S_{\Sigma\pi}^3 + \Big[m_{\Lambda}^2 (2 m_{\Sigma}-3 M_{\Gamma})+M_{\Gamma}(C_2-2 C_1)\Big] S_{\Sigma\pi}^2 + T_1 S_{\Sigma\pi}+T_0\bigg],\\
&\hspace{0.5cm}\text{where}\\
&\hspace{0.5cm}
T_1 = 2m_{\Lambda}^4 (M_{\Gamma} - m_{\Sigma})+m_{\Lambda}^2 M_{\Gamma}(4C_1-3C_2+C_3)
+2 m_{\Sigma} (C_2-C_1)+M_{\Gamma}(C_1-2 C_2), \\
&\hspace{0.5cm}
T_0 = 2m_{\Lambda}^6 ( M_{\Gamma}-m_{\Sigma})( C_2-C_1-C_3)
+m_{\Lambda}^4\bigg[M_{\Gamma} C_1(3 C_2 -C_1-C_3)-2 m_{\Sigma} C_2C_1\bigg]+M_{\Gamma} m_{\Lambda}^2 
C_2 C_1^2.\\
%
C^{G}_{3,2} & =\frac{m_\Lambda}{m_{\Lambda_c}}C^{G}_{3,3} = C^{G}_{5,2} =\frac{D}{m_{\Lambda_c}} \bigg[-2 M_{\Gamma}S_{\Sigma\pi}^5 
+ T_4 S_{\Sigma\pi}^4 + T_3 S_{\Sigma\pi}^3 + T_2 S_{\Sigma\pi}^2 +T_1 S_{\Sigma\pi} +T_0 \bigg],
\\
&\hspace{0.5cm} {\text{where}}\\
& \hspace{0.5cm}
T_4 = \Big[2 m_{\Lambda}^2(M_{\Gamma}-2m_{\Sigma}-m_{\Lambda_c})+4M_{\Gamma}(C_1-C_2)\Big],\\
& \hspace{0.5cm}
T_3=m_{\Lambda}^4\Big[5 M_{\Gamma}+4
m_{\Sigma}+2m_{\Lambda_c}\Big]+ 2m_{\Lambda}^2
\Big[M_{\Gamma} (2C_2-C_3)+2m_{\Sigma}(C_1-2C_2)+m_{\Lambda_c}(2C_1-C_2)\Big]\\
&\hspace{0.8cm}
-2 M_{\Gamma}(C_1^2+C_2^2-4C_1C_2),\\
&\hspace{0.5cm}
T_2=-5 M_{\Gamma} m_{\Lambda}^6-m_{\Lambda}^4
\Big[M_{\Gamma}\big[4 m_{\Sigma}^2+2(5C_1-5C_2-3C_3)+q^2-4m_{\Sigma}m_{\Lambda_c}\big]+4m_{\Sigma}
(C_1-2C_2+C_3)\\
&\hspace{0.8cm}
+2(4 C_1-C_2+ C_3)\Big]
+m_{\Lambda}^2\Big[2M_{\Gamma} \big[C_2(C_2+C_1-C_3)-C_1(C_1-C_3)\big]\\
&\hspace{0.8cm}
+4m_{\Sigma}C_2(2C_1-C_2) - 2 m_{\Lambda_c}C_1(C_1-2C_2)\Big]-4 M_{\Gamma} C_1C_2(C_1-C_2),
\\
&\hspace{0.5cm}
T_1
=2m_{\Lambda}^6\Big[M_{\Gamma}\big[2m_{\Sigma}^2-5 C_2+3 C_1+5  C_3-2m_{\Sigma}m_{\Lambda_c}\big]
+2 C_1m_{\Lambda_c}\Big]\\
&\hspace{0.8cm}
+m_{\Lambda}^4\Big[M_{\Gamma}\big[-8 m_{\Sigma}^2C_2+2C_2(3C_2-7C_1-3C_3)
+C_1(9C_1+10C_3)+2C_1q^2
+4m_{\Sigma}m_{\Lambda_c} (C_2-C_1)\big]\\
&\hspace{0.8cm}
+4m_{\Sigma}C_2[C_2-C_1-C_3]
+2m_{\Lambda_c}C_1(-3 C_2 +3 C_1+C_3)\Big]\\
&\hspace{0.8cm}
+2m_{\Lambda}^2C_1C_2\Big[ M_{\Gamma}(C_2-3C_1+C_3)+2m_{\Sigma} C_2-m_{\Lambda_c}C_1\Big]-2
M_{\Gamma} C_2^2 C_1^2,
\\
&\hspace{0.5cm}
T_0
=m_{\Lambda}^6\Big[ M_{\Gamma}\big[4m_{\Sigma}^2
(2C_2+q^2)+C_2(2 C_1-5C_2+10C_3)-C_1(5 C_1+10C_3+4 q^2)-5C_3^2\\
&\hspace{0.8cm}
+4m_{\Lambda_c}m_{\Sigma}( 
C_1-C_2+C_3)\big]
+4 C_1(C_2- C_1-C_3)\Big]\\
&\hspace{0.8cm}
+m_{\Lambda}^4\Big[M_{\Gamma}\big[-4 m_{\Sigma}^2
C_2^2+2C_1C_2(-3 C_2+4 C_1+5 C_3)- C_1^2q^2-4m_{\Sigma}m_{\Lambda_c} C_2 C_1\big]
+4m_{\Lambda_c}C_2 C_1^2\Big]
\\
&\hspace{0.8cm}
-4 M_{\Gamma} m_{\Lambda}^2C_2^2 C_1^2.
\\
%
%
C^{G}_{3,4} & = C^{F}_{11,4} = m_{\Lambda_c}D\bigg[-2 M_{\Gamma}S_{\Sigma\pi}^4 + \Big[m_{\Lambda}^2(2 M_{\Gamma} -m_{\Lambda_c}-4
m_{\Sigma}+2 M_{\Gamma}(2C_1-C_2)\Big] S_{\Sigma\pi}^3 + T_2 S_{\Sigma\pi}^2 +T_1 S_{\Sigma\pi} +T_0\bigg],
\\
&\hspace{0.5cm} {\text{where}}\\
& \hspace{0.5cm}
T_2=m_{\Lambda}^4\Big[5 M_{\Gamma} +m_{\Lambda_c}+4m_{\Sigma}\Big]+m_{\Lambda}^2\Big[ M_{\Gamma}\big[2
 C_2+C_1- C_3\big]+2 m_{\Lambda_c} C_1+4 m_{\Sigma} (C_1-C_2)\Big]\\
&\hspace{0.8cm}
+4M_{\Gamma} C_1(2C_2-C_1),
\\
&
\hspace{0.5cm}
T_1=-5 M_{\Gamma} m_{\Lambda}^6 +m_{\Lambda}^4\Big[
M_{\Gamma}\big[2 m_{\Sigma}(m_{\Lambda_c}-2 m_{\Sigma})+6C_2-7 C_1-3 C_3\big]
-3m_{\Lambda_c} C_1+2 m_{\Sigma}(2 C_2-C_1-C_3)\Big]\\
&\hspace{0.8cm}
+m_{\Lambda}^2\Big[ M_{\Gamma}C_1\big[2C_2-3C_1 +C_3\big]-m_{\Lambda_c}C_1^2+4 m_{\Sigma} C_2 C_1\Big]
-2M_{\Gamma}  C_2 C_1^2,
\\
&\hspace{0.5cm}
T_0
=m_{\Lambda}^6\Big[M_{\Gamma}\big[-2 m_{\Sigma}(m_{\Lambda_c}-2m_{\Sigma})-5  C_2+C_1+5C_3\big]+2 m_{\Lambda_c}
C_1\Big]\\
&\hspace{0.8cm}
+m_{\Lambda}^4\Big[M_{\Gamma}\big[-2m_{\Sigma}( m_{\Lambda_c} C_1+2m_{\Sigma} C_2)+C_1(4 C_1-6C_2+5 C_3)\big]
+2m_{\Lambda_C} C_1^2\Big]-4M_{\Gamma} m_{\Lambda}^2  C_2 C_1^2.
\\
C^{G}_{4,2} & = \frac{m_{\Lambda}}{m_{\Lambda_c}}C^{G}_{4,3}  = C^{G}_{6,2} =
\frac{D}{m_{\Lambda_c}}\bigg[2S_{\Sigma\pi}^5 + \Big[2 m_{\Lambda}^2+4(C_2-C_1)\Big]S_{\Sigma\pi}^4 
+ T_3 S_{\Sigma\pi}^3 + T_2 S_{\Sigma\pi}^2 +T_1 S_{\Sigma\pi} +T_0 \bigg],
\\
&\hspace{0.5cm} {\text{where}}\\
&\hspace{0.5cm}
T_3=-9m_{\Lambda}^4+2m_{\Lambda}^2\Big[ M_{\Gamma}(m_{\Lambda_c}-2m_{\Sigma})+(2C_2+C_3)\Big]+2(C_1^2+C_2^2-4C_1C_2),
\\
&\hspace{0.5cm}
T_2=5m_{\Lambda}^6+m_{\Lambda}^4\Big[2 M_{\Gamma}(2m_{\Sigma}-3m_{\Lambda_c})
+4m_{\Sigma}(m_{\Sigma}+m_{\Lambda_c})+2(5 C_1-9C_2+5C_3)+q^2\Big]\\
&\hspace{0.8cm}
+2m_{\Lambda}^2\Big[ M_{\Gamma}\big[2m_{\Sigma}(C_1-2C_2)
+m_{\Lambda_c}(C_2-2C_1)\big]+ C_2(C_2-C_1+ C_3)- C_1(C_1+ C_3)\Big]\\
&\hspace{0.8cm}
+4 C_1C_2 (C_1- C_2),
\\
&\hspace{0.5cm}
T_1=m_{\Lambda}^6\Big[4 M_{\Gamma}m_{\Lambda_c}-4
m_{\Sigma}(m_{\Sigma}+m_{\Lambda_c})+2(5 C_2-3C_1-5 C_3)\Big]\\
&\hspace{0.8cm}
+m_{\Lambda}^4\Big[M_{\Gamma}\big[4 m_{\Sigma}(2 C_2-C_1-C_3)+2m_{\Lambda_c}(4C_1-3 C_2+C_3)\big]
+8m_{\Sigma}^2C_2+10 C_2(C_1- C_2+C_3)\\
&\hspace{0.8cm}
-C_1(5 C_1-6C_3-2 q^2)
+4m_{\Sigma}m_{\Lambda_c}( C_2-C_1)\Big]\\
&\hspace{0.8cm}
+2m_{\Lambda}^2
\Big[ M_{\Gamma}\big[2m_{\Sigma}C_2(2C_1-C_2)+m_{\Lambda_c}C_1(C_1- 2C_2 )\big]- C_2C_1(C_2+C_1+C_3)\Big]+2 C_2^2C_1^2,
\\
&\hspace{0.5cm}
T_0
=m_{\Lambda}^6\Big[4 M_{\Gamma}m_{\Lambda_c}( C_2-C_1-C_3)-4m_{\Sigma}^2(2C_2-q^2)+5(C_2^2+C_1^2+C_3^2)
-2 C_1(C_2-2q^2)\\
&\hspace{0.8cm}
+10(C_1-C_2)C_3
+4m_{\Sigma}m_{\Lambda_c} (C_1-C_2+C_3)\Big]\\
&\hspace{0.8cm}
+m_{\Lambda}^4\Big[2M_{\Gamma}\big[2m_{\Sigma} C_2(C_2-C_1-C_3)+ m_{\Lambda_c}C_1(3C_2-C_1- C_3)
+4m_{\Sigma}C_2(m_{\Sigma}C_2-m_{\Lambda_c}C_1)\\
&\hspace{0.8cm}
+2C_1 C_2(C_2-2 C_1-3 C_3)+C_1^2 q^2\Big]
+ 2m_{\Lambda}^2 M_{\Gamma} C_2 C_1\Big[2m_{\Sigma} C_2+m_{\Lambda_c}C_1\Big].
\\
C^{G}_{4,4} & =m_{\Lambda_c}D \bigg[2 S_{\Sigma\pi}^4 + 2 \Big[m_{\Lambda}^2 + ( C_2-2 C_1)\Big] S_{\Sigma\pi}^3 
+S_{\Sigma\pi}^2+T_1 S_{\Sigma\pi} +T_0\bigg],\\
&\hspace{0.5cm}\text{where}\\
&\hspace{0.5cm}
T_2 =  -9 m_{\Lambda}^4+m_{\Lambda}^2\big[M_{\Gamma} (m_{\Lambda_c}-4 m_{\Sigma})+2C_2-C_1+C_3\big]+2C_1(C_1-2C_2), \\
&\hspace{0.5cm}
T_1 = m_{\Lambda}^4\big[M_{\Gamma}(-3  m_{\Lambda_c}+4m_{\Sigma})+2 m_{\Sigma}(m_{\Lambda_c}+2m_{\Sigma})+5(C_1-2C_2+C_3)\big]\\
&\hspace{0.8cm}
+m_{\Lambda}^2\big[2M_{\Gamma}\big(2 m_{\Sigma} (C_1-C_2)- m_{\Lambda_c} C_1\big)
-C_1(2 C_2 + C_1+ C_3)\big]+5 m_{\Lambda}^6 +2 C_2 C_1^2,
\\
&\hspace{0.5cm}
T_0=m_{\Lambda}^6\big[2 M_{\Gamma} m_{\Lambda_c}-2
 m_{\Sigma}(m_{\Lambda_c}+2 m_{\Sigma})+5 C_2- C_1-5C_3\big]
 +m_{\Lambda}^4\big[ M_{\Gamma}\big(
3m_{\Lambda_c} C_1+ 2m_{\Sigma}(2C_2-C_1-C_3)\big)\\
&\hspace{0.8cm}
-2m_{\Sigma}( m_{\Lambda_c} C_1-2 m_{\Sigma} C_2)
+C_1(2C_2 -2  C_1-3 C_3)\big]+m_{\Lambda}^2
M_{\Gamma}C_1 (m_{\Lambda_c} C_1+4 m_{\Sigma} C_2).
\\
C^{F}_{11,2} &  = C^{F}_{13,2}=\frac{m_{\Lambda}}{m_{\Lambda_c}}C^{F}_{11,3} = \frac{D}{m_{\Lambda_c}}\bigg[2 M_{\Gamma}S_{\Sigma\pi}^5  +T_4S_{\Sigma\pi}^4
+ T_3 S_{\Sigma\pi}^3 + T_2 S_{\Sigma\pi}^2 +T_1 S_{\Sigma\pi} +T_0 \bigg],
\\
&\hspace{0.5cm} {\text{where}}\\
&\hspace{0.5cm}
T_4 = -2\Big[ m_{\Lambda}^2(M_{\Gamma}-2m_{\Sigma}+m_{\Lambda_c})+2 M_{\Gamma}(C_1-C_2)\Big],\\
&\hspace{0.5cm}
T_3=-m_{\Lambda}^4\Big[5 M_{\Gamma}+4m_{\Sigma}-2m_{\Lambda_c}\Big]-2m_{\Lambda}^2\Big[
M_{\Gamma} (2C_2-C_3)+2m_{\Sigma}(C_1-2C_2)+m_{\Lambda_c}(C_2-2 C_1)\Big]\\
&\hspace{0.8cm}
+2 M_{\Gamma}(C_1^2+C_2^2-4C_1C_2),\\
&\hspace{0.5cm}
T_2
=5 M_{\Gamma} m_{\Lambda}^6+m_{\Lambda}^4
\Big[ M_{\Gamma}\big[4m_{\Sigma}(m_{\Sigma}+m_{\Lambda_c})+10 (C_1-C_2)+6 C_3+q^2\big]\\
&\hspace{0.8cm}
+4m_{\Sigma}(C_1-2C_2+ C_3)
+2m_{\Lambda_c}(C_2-4 C_1- C_3)\Big]\\
&\hspace{0.8cm}
+2m_{\Lambda}^2\Big[-M_{\Gamma}\big[ C_2(C_2+C_1-C_3)-C_1(C_1-C_3)\big]
+2m_{\Sigma}C_2(C_2-2C_1)
+m_{\Lambda_c}C_1(2 C_2 -C_1)\Big]\\
&\hspace{0.8cm}
+4 M_{\Gamma} C_2C_1(C_1-C_2),
\\
&\hspace{0.5cm}
T_1
=2m_{\Lambda}^6\Big[M_{\Gamma}\big[-2 m_{\Sigma}(m_{\Sigma}+m_{\Lambda_c})+5(C_2-C_3)-3C_1\big]
+2 m_{\Lambda_c}C_1\Big]\\
&\hspace{0.8cm}
+m_{\Lambda}^4\Big[M_{\Gamma}\big[8 m_{\Sigma}^2C_2-2 C_2(3C_2-7 C_1-3 C_3)
-C_1(9 C_1+10 C_3)-2C_1q^2
+4 m_{\Sigma}m_{\Lambda_c}(C_2-C_1)\big]\\
&\hspace{0.8cm}
+4m_{\Sigma}C_2(C_1-C_2+C_3)+2m_{\Lambda_c}C_1(3 C_1-3 C_2 +2 C_3)\Big]\\
&\hspace{0.8cm}
+2m_{\Lambda}^2\Big[M_{\Gamma}(-
C_2+3C_1-C_3)-2m_{\Sigma} C_2-m_{\Lambda_c}C_1\Big]
+2M_{\Gamma} C_2^2 C_1^2,
\\
&\hspace{0.5cm}
T_0=m_{\Lambda}^6\Big[ M_{\Gamma}\big[-m_{\Sigma}^2(2C_2+q^2)
+5(C_3^2+C_2^2+C_3^2)+10(C_1- C_2)C_3-2 C_1(C_2-2q^2)\\
&\hspace{0.8cm}
+4m_{\Sigma}(C_1-C_2+ C_3)\big]+4 m_{\Lambda_c}(C_2-C_1-C_3)\Big]\\
&\hspace{0.8cm}
+m_{\Lambda}^4\Big[ M_{\Gamma}\big[4m_{\Sigma}C_2(m_{\Sigma}C_2-m_{\Lambda_c}C_1)
+2C_2C_1(3C_2-4C_1-5C_3)+C_1^2q^2\big]+4m_{\Lambda_c}C_2 C_1^2\Big]\\
&\hspace{0.8cm}
+4 M_{\Gamma} m_{\Lambda}^2C_2^2 C_1^2,
\\
C^{F}_{12,2} & =\frac{m_{\Lambda}}{m_{\Lambda_c}}C^{F}_{12,3} = C^{F}_{14,2}= -\frac{D}{m_{\Lambda_c}} \bigg[2S_{\Sigma\pi}^5 +2 \Big[m_{\Lambda}^2+2(C_2-C_1)\Big] S_{\Sigma\pi}^4 +S_{\Sigma\pi}^3 + T_2 S_{\Sigma\pi}^2 +T_1 S_{\Sigma\pi} +T_0 \bigg],
\\
&\hspace{0.5cm} {\text{where}}\\
&\hspace{0.5cm}
T_3 = -9m_{\Lambda}^4+2m_{\Lambda}^2\big[- M_{\Gamma}(2m_{\Sigma}+m_{\Lambda_c})+(2 C_2+C_3)\big]+2(C_1^2+C_2^2-4C_1C_2),\\
&\hspace{0.5cm}
T_2	
=5m_{\Lambda}^6+m_{\Lambda}^4\Big[2 M_{\Gamma}(2m_{\Sigma}+3m_{\Lambda_c})+4m_{\Sigma}(m_{\Sigma}-m_{\Lambda_c})
+2(5 C_1+5C_3-9C_2)+q^2\Big]\\
&\hspace{0.8cm}
+m_{\Lambda}^2\Big[M_{\Gamma}\big[4 m_{\Sigma}(C_1-2 C_2)+2m_{\Lambda_c}(2C_1- C_2)\big]+2 C_2(C_2-C_1+C_3)-2 C_1(C_1+C_3)\Big]\\
&\hspace{0.8cm}
+4 C_2 C_1(C_1-C_2),
\\
&\hspace{0.5cm}
T_1
=m_{\Lambda}^6\Big[-4 M_{\Gamma}m_{\Lambda_c}+4
m_{\Sigma}(m_{\Lambda_c}-m_{\Sigma})+2 (5C_2-3C_1-5 C_3)\Big]\\
&\hspace{0.8cm}
+m_{\Lambda}^4\Big[ M_{\Gamma}\big[4m_{\Sigma} (2C_2-C_1-C_3)
+2m_{\Lambda_c}(3C_2-4 C_1-C_3)\big]\\
&\hspace{0.8cm}
+8m_{\Sigma}^2C_2+10C_2( C_1-C_2+C_3)-C_1(5 C_1+6C_3-2 q^2)+4m_{\Sigma}m_{\Lambda_c}( C_1-C_2)\Big]\\
&\hspace{0.8cm}
+2m_{\Lambda}^2\Big[M_{\Gamma}\big[2m_{\Sigma}C_2(2C_1-C_2)+m_{\Lambda_c}C_1(2C_2-C_1)\big]
- C_2C_1(C_1+C_2+C_3)\Big]+2 C_2^2C_1^2,
\\
&\hspace{0.5cm}
T_0
=m_{\Lambda}^6\Big[4 M_{\Gamma}m_{\Lambda_c}(C_1-C_2+C_3)-4m_{\Sigma}^2(2C_2+q^2)
+5 (C_1^2+C_2^2+ C_3^2)+10(C_1- C_2) C_3\\
&\hspace{0.8cm}
-2 C_1(C_2-2 q^2)+4m_{\Sigma}(C_2-C_1-C_3)\Big]\\
&\hspace{0.8cm}
+m_{\Lambda}^4\Big[2 M_{\Gamma}\big[2m_{\Sigma} C_2(C_2-C_1-C_3)+m_{\Lambda_c}C_1(C_1-3C_2 +C_3)\big]\\
&\hspace{0.5cm}
+4m_{\Sigma}C_2(m_{\Sigma}C_2+m_{\Lambda_c}C_1)+2 C_2C_1(C_2-2 C_1-3 C_3)+C_1^2 q^2\Big]+2m_{\Lambda}^2
M_{\Gamma}C_2C_1\Big[2m_{\Sigma} C_2-m_{\Lambda_c}C_1\Big],
\\
C^{F}_{12,4} & = - m_{\Lambda_c}D\bigg[2 S_{\Sigma\pi}^4 + 2 \Big[m_{\Lambda}^2 + ( C_2-2 C_1)\Big] S_{\Sigma\pi}^3 
+T_2S_{\Sigma\pi}^2 +T_1 S_{\Sigma\pi} +T_0 \bigg],
\\
&\hspace{0.5cm} {\text{where}}\\
&\hspace{0.5cm}
T_2=
-9 m_{\Lambda}^4-m_{\Lambda}^2\big[M_{\Gamma} (m_{\Lambda_c}+4m_{\Sigma})-2  C_2+C_1- C_3\big]+2  C_1(C_1 -2 C_2),\\
&\hspace{0.5cm}
T_1
=5 m_{\Lambda}^6+
m_{\Lambda}^4\big[ M_{\Gamma} (3m_{\Lambda_c}+4m_{\Sigma})+2m_{\Sigma}(2m_{\Sigma}- m_{\Lambda_c})+5( C_1-2C_2+C_3\big]\\
&\hspace{0.8cm}
+m_{\Lambda}^2\big[2 M_{\Gamma}\big(m_{\Lambda_c} C_1+m_{\Sigma}(2 C_1-C_2)\big)- C_1(2 C_2+ C_1+C_3)\big]
 +2 C_2 C_1^2,
\\
&\hspace{0.5cm}
T_0=m_{\Lambda}^6\Big[-2 M_{\Gamma} m_{\Lambda_c}+2m_{\Sigma}(m_{\Lambda_c}-2 m_{\Sigma})
+5 C_2- C_1-5 C_3\Big]\\
&\hspace{0.8cm}
+m_{\Lambda}^4\Big[M_{\Gamma}\big[-3 m_{\Lambda_c} C_1+2 m_{\Sigma}(2C_2-C_1-C_3)\big]
+2m_{\Sigma}(m_{\Lambda_c} C_1+2 m_{\Sigma} C_2)+C_1(2 C_2 -2  C_1-3 C_3)\Big]\\
&\hspace{0.8cm}
+m_{\Lambda}^2M_{\Gamma} C_1\Big[4 m_{\Sigma} C_2-m_{\Lambda_c} C_1\Big],
\\
C^{G}_{5,4} & = -C^{G}_{13,4}=m_{\Lambda}^4m_{\Lambda_c}D\bigg[-M_{\Gamma}  S_{\Sigma\pi}^2
+2 M_{\Gamma} C_1S_{\Sigma\pi} 
+ 4m_{\Lambda}^2 M_{\Gamma} (m_{\Sigma}^2-C_1)-M_{\Gamma} C_1^2
\bigg],\\
C^{G}_{6,4} & = -C^{F}_{14,4} = m_{\Lambda}^4m_{\Lambda_c}D\bigg[S_{\Sigma\pi}^2
-2 C_1S_{\Sigma\pi}+\Big[4m_{\Lambda}^2( C_1- m_{\Sigma}^2)+C_1^2\Big]
\bigg], \\
C^{G}_{7,4} & =m_{\Lambda}^2 m_{\Lambda_c}D\bigg[-M_{\Gamma} S_{\Sigma\pi}^3 
-\Big[m_{\Lambda}^2(3 M_{\Gamma}+m_{\Lambda_c}+2m_{\Sigma})- M_{\Gamma}( C_1-C_2)\Big]S_{\Sigma\pi}^2\\
&
+\Big[5 M_{\Gamma} m_{\Lambda}^4 -m_{\Lambda}^2\big[
M_{\Gamma}(3 C_2-5C_1)
-m_{\Lambda_c} C_1+2 m_{\Sigma}C_2\big]+M_{\Gamma} C_2C_1\Big] S_{\Sigma\pi} \\
&
+m_{\Lambda}^4\big(M_{\Gamma}\big[2  m_{\Lambda_c}m_{\Sigma}+5 (C_2-C_1- C_3)\big]
-2m_{\Lambda_c} C_1\big)+5M_{\Gamma} m_{\Lambda}^2C_2 C_1 \bigg],
\\
C^{G}_{8,4} & =m_{\Lambda}^2 m_{\Lambda_c}D\bigg[S_{\Sigma\pi}^3 +
\Big[5 m_{\Lambda}^2+ (C_2-C_1)\Big]S_{\Sigma\pi}^2\\
&
+\Big[-5 m_{\Lambda}^4 +m_{\Lambda}^2\big[M_{\Gamma} (m_{\Lambda_c}-2 m_{\Sigma})+5 C_2-3 C_1\big]-C_2 C_1\Big] S_{\Sigma\pi} \\
&
+m_{\Lambda}^4\big[-2 M_{\Gamma} m_{\Lambda_c}
+2m_{\Lambda_c}m_{\Sigma}+5 (C_1-C_2+ C_3)\big]-m_{\Lambda}^2
\big[M_{\Gamma} (m_{\Lambda_c} C_1+2 m_{\Sigma} C_2)+3 C_2 C_1\big]\bigg],
\\
C^{F}_{15,4} & =m_{\Lambda}^2 m_{\Lambda_c}D\bigg[M_{\Gamma} S_{\Sigma\pi}^3 
+\Big[m_{\Lambda}^2(3 M_{\Gamma}-m_{\Lambda_c}+2m_{\Sigma})- M_{\Gamma}( C_1-C_2)\Big]S_{\Sigma\pi}^2\\
&
-\Big[5 M_{\Gamma} m_{\Lambda}^4 -m_{\Lambda}^2\big[M_{\Gamma}(3 C_2-5C_1)
+m_{\Lambda_c} C_1+2 m_{\Sigma}C_2\big]+M_{\Gamma} C_2C_1\Big] S_{\Sigma\pi}\\
& 
+\Big[m_{\Lambda}^4\big(M_{\Gamma}\big[2  m_{\Lambda_c}m_{\Sigma}-5 (C_2-C_1- C_3)\big]-2
m_{\Lambda_c} C_1\big)-5M_{\Gamma} m_{\Lambda}^2C_2 C_1\Big] \bigg],
\\
C^{F}_{16,4} & = - m_{\Lambda}^2 m_{\Lambda_c}D\bigg[S_{\Sigma\pi}^3 +
\Big[5 m_{\Lambda}^2+ (C_2-C_1)\Big]S_{\Sigma\pi}^2\\
&
+\Big[-5 m_{\Lambda}^4 +m_{\Lambda}^2\big[-M_{\Gamma} (m_{\Lambda_c}+2 m_{\Sigma})+5 C_2-3 C_1\big]-C_2 C_1\Big] S_{\Sigma\pi} \\
&
+m_{\Lambda}^4\big[2 M_{\Gamma} m_{\Lambda_c}-2
m_{\Lambda_c}m_{\Sigma}+5 (C_1-C_2+ C_3)\big]+m_{\Lambda}^2
\big[M_{\Gamma} (m_{\Lambda_c} C_1-2 m_{\Sigma} C_2)-3 C_2 C_1\big]
 \bigg].
\\
\end{align*}

%
\end{document}